\documentclass[twocolumn,floatfix,aps]{revtex4}
\usepackage{amsmath}
\usepackage{makeidx}
\usepackage{amssymb}
\usepackage{graphicx}

\usepackage[usenames]{color}
\usepackage[normalem]{ulem}

\usepackage{hyperref}
\usepackage[T1]{fontenc} 

\begin{document}

\title{Vortex-lattice formation in a spin-orbit coupled rotating spin-1   condensate}

\author{S. K. Adhikari\footnote{s.k.adhikari@unesp.br, \\
          https://professores.ift.unesp.br/sk.adhikari}}
\affiliation{Instituto de F\'{\i}sica Te\'orica, Universidade Estadual Paulista - UNESP, 01.140-070 S\~ao Paulo, S\~ao Paulo, Brazil}
      

\date{\today}

\begin{abstract}

We study the vortex-lattice formation in a rotating {Rashba} spin-orbit (SO) coupled quasi-two-dimensional (quasi-2D) hyper-fine 
spin-1 
spinor Bose-Einstein condensate (BEC) in the $x-y$ plane using a numerical solution of the underlying mean-field Gross-Pitaevskii equation. 
In this case, the non-rotating {Rashba} SO-coupled spinor BEC can have topological excitation
in the form of vortices of different angular momenta in the three components, e.g. the $(0,+1,+2)$- and $(-1,0,+1)$-type states in ferromagnetic and anti-ferromagnetic spinor BEC: the numbers in the parenthesis denote the intrinsic angular momentum 
of the vortex states of the three components with the negative sign denoting an anti-vortex. The presence of these states with intrinsic vorticity breaks  the  symmetry between rotation  with vorticity 
along the $z$ and $-z$ axes and thus generates a rich variety of vortex-lattice and anti-vortex-lattice states in a rotating quasi-2D spin-1 spinor ferromagnetic and anti-ferromagnetic  BEC, not possible in a scalar BEC. {For weak SO coupling, } we find two types of symmetries of these states $-$ 
hexagonal and ``square''.  The hexagonal (square) symmetry state has vortices arranged in closed concentric orbits with a maximum  of $6, 12, 18...$  ($8,12,16...$) vortices in successive orbits. 
Of these two symmetries, the square vortex-lattice state is found to have the smaller energy. 

\end{abstract}

 \maketitle

\newpage

\section{Introduction}
\label{sec:intro}
 
 After the observation of a trapped Bose-Einstein condensate (BEC) in $^{87}$Rb and $^{23}$Na alkali metal atoms at ultra-low  temperature in a laboratory \cite{expt}, rotating
trapped condensates   hosting quantized vortices \cite{vortex} and large vortex lattices \cite{vorlat} were created,
for small and large angular frequencies of rotation, respectively, 
 under   controlled conditions and
studied experimentally. As suggested by Onsager \cite{onsager},
Feynman \cite{feynman} and Abrikosov \cite{abrikosov} these vortices have quantized circulation as in liquid
He II  \cite{fetter}
\begin{equation} \label{cir}
\frac{\widetilde m}{2\pi\hbar }\oint_{\cal C}  {\bf v}.d{\bf r}= \pm l,
\end{equation}
where  ${\bf v}({\bf r}, t)$ is the super-fluid velocity field at a space point ${\bf r}$ and time $t$, $\cal C$  
is a generic closed path, $ l$ is the
quantized angular momentum of an atom in units of $\hbar $ in the rotating BEC and
$\widetilde m$ is the mass of an atom. For notational simplicity,  the circulation (\ref{cir}) is scaled 
by a factor of $2\pi\hbar/\widetilde m$, so that  its absolute value is equal to angular momentum $l$:
 a  positive (negative) circulation corresponds to a vortex (anti-vortex).
 If $l\ne 0$,  there are topological defects inside the closed path $\cal C$, which
manifests in the form of  a quantized vortex line \cite{fetter}.
Quantized vortices of unit
angular momentum were first observed in a uniform
super-fluid He II in a rotating bucket \cite{heii}.
 Vorticity is the curl of the velocity field $\nabla _{\bf r} \times {\bf v}({\bf r}, t)$ and determines
the direction of the angular momentum vector. London gave a qualitative explanation of quantization of circulation in He II \cite{london}. In addition,
 if we assume that the dynamics of the super-fluid is governed by
a  complex scalar field $\phi({\bf r}, t)\equiv  | \psi({\bf r},t)|\exp [i\delta ({\bf r}, t) ]$ with 
${\bf v}({\bf r}, t)  = \nabla_{\bf r} \delta( {\bf r}, t)$, then $\psi({\bf r},t)$ is known to satisfy the mean-field Gross-Pitaevskii (GP) equation \cite{fetter} which has been used successfully to study the formation of vortex and vortex lattice \cite{vor-lat} in a BEC.
 
A  spinor BEC of $^{23}$Na atoms \cite{exptspinor} with  hyper-fine spin $F=1$  has  also been observed and, more recently,   it has been  possible to introduce an artificial synthetic SO coupling by
Raman lasers that coherently couple the spin-component
states in a spinor BEC \cite{thso}. 
Two common  SO couplings are due to Rashba \cite{SOras}
  and Dresselhaus \cite{SOdre}. 
 An equal
mixture of these SO couplings has been realized in pseudo
spin-1/2 $^{87}$Rb \cite{exptso} and $^{23}$Na \cite{exptsona} BECs containing only
two spin components $F_z = 0, -1$ of total spin $F = 1$. 
Later, an equal mixture of Rashba and 
Dresselhaus SO couplings has also been created
 in a spin-1 ferromagnetic $^{87}$Rb BEC, containing all
three spin components $F_z=\pm 1,0$ \cite{bosespin1}.

Spinor BECs can show a rich variety of topological excitation \cite{thspinor,topo2,kita,kita2} not possible in a scalar BEC. 
The predicted 
Mermin-Ho \cite{mh} and Anderson-Toulouse  \cite{at} vortices in $^3$He with a non-singular angular momentum structure, although not observed in $^3$He,  might  appear in a spinor BEC. It was later demonstrated \cite{kita} that, in a trapped slowly rotating ferromagnetic spinor BEC, the Mermin-Ho  and Anderson-Toulouse vortices are thermodynamically stable. Such stable vortices appear in a ferromagnetic spinor BEC in the form of a state of type $(0,+1,+2)$ \cite{kita,kita2}, where the numbers in the parenthesis denote 
the circulation (angular momentum) of vortices in components $F_z=+1,0,-1$,   respectively. For certain values of magnetization,  $(+1,+1,+1)$ and $(+1,0,-1)$-type states  are also demonstrated to appear \cite{kita}
in a ferromagnetic spinor BEC. The $(+1,0,-1)$-type state hosts a vortex (anti-vortex of negative vorticity)  in the component $F_z=+1$  ($F_z=-1$), {whereas the component $F_z=0$ remains vortex free}.
However, such states in an anti-ferromagnetic BEC were not found to be stable \cite{kita}.  The analogue of these states in a pseudo spin-1/2 system is the $(0,+1)$-type   half-quantum state 
\cite{yy,yy2,yy1}.

{
In bosonic spin
systems, SO-coupling  leads to  a variety of novel phenomena
  that are not possible in spinor
BECs without SO coupling \cite{xyz}, for example \cite{rash}, the stripe phase \cite{stripe,st2}, the Rashba pairing bound states
(Rashbons) \cite{rashbon},  spin Hall effect \cite{hall},  spintronics \cite{spint}, as well as the super-fluidity and Mott-insulator phases of SO-coupled quantum gases in optical lattice \cite{AB}.}
A three-component SO-coupled spin-1 BEC is  known to exhibit a rich variety of physical phenomena not 
possible in a two-component pseudo spin-1/2 BEC \cite{thspinor,stripe,prop}. A spin-1 spinor BEC is controlled by two interaction
strengths, e.g., $c_0 \propto (a_0 + 2a_2)/3$ and $c_2 \propto (a_2 - a_0)/3,$
with $a_0$ and $a_2$ the scattering lengths in total spin  0
and 2 channels, respectively, and   appears in two distinct phases:
ferromagnetic ($c_2<0$) and anti-ferromagnetic or polar ($c_2>0$).

In view of this, we investigate in this paper the formation of vortex lattice in a rapidly rotating Rashba SO-coupled ferromagnetic and anti-ferromagnetic trapped spin-1 spinor BEC, to see the effect of the above topological excitation in the generated vortex lattice, if any.  
Previously, the formation of vortex and vortex lattice in a Rashba SO-coupled trapped pseudo spin-1/2 spinor BEC was studied numerically \cite{yy,yy2} and analytically \cite{yy1}.
Different ways of realizing a rotating SO-coupled spinor BEC have been 
suggested \cite{gai}. 
{We find that for a weakly Rashba SO-coupled quasi-two-dimensional (quasi-2D) non-rotating ferromagnetic spin-1 spinor BEC, the { lowest-energy circularly-symmetric state is a}
     $(0,+1,+2)$-type state \cite{cpc}. For a weakly Rashba 
SO-coupled quasi-2D non-rotating anti-ferromagnetic spinor BEC, the { lowest-energy circularly-symmetric} state is of the type
$(-1,0,+1)$ \cite{cpc}. }
 In a rotating quasi-2D scalar BEC in the $x-y$ plane, the generated vortex-lattice structure 
is the same for vorticity of rotation along $z$ or $-z$ axis. 
However, the generated vortex lattice for  vorticity of rotation along $z$ or $-z$ axis will be different for  {a Rashba} SO-coupled spin-1 spinor BEC because of the above symmetry-breaking  $(0,+1,+2)$ and $(-1,0,+1)$-type states.
When subject to rotation, 
both   ferromagnetic and anti-ferromagnetic spinor BECs
form  vortex lattices with a hexagonal  or an {{\it approximate} ``square''  symmetry.
Although, the square symmetry is often distorted, 
from a study of energies, distinct from a rotating scalar BEC, it was found that
the lattice structure with square symmetry has the smaller energy. In case of a scalar BEC 
the vortex lattice with hexagonal symmetry has the smallest energy.}
 In the case of  a ferromagnetic  {Rashba}
SO-coupled spin-1
spinor BEC, for rotation with 
vorticity along $z$ direction, the 
hexagonal or square vortex-lattice structure
 is built around the  $(0,+1,+2)$-type state at the center: the central site of the three components
  $F_z=+1,0,-1$, respectively, host vortices of circulation $0,+1,+2$. 
For rotation with vorticity along $-z$ direction, an anti-vortex lattice is generated in both ferromagnetic and anti-ferromagnetic spinor BEC around a complex anti-vortex structure at the center.

In Sec. \ref{II} we present the mean-field GP equation for a rotating quasi-2D SO-coupled spin-1 spinor condensate in the rotating frame. 
In Sec. \ref{III} we present {the  numerical details for the solution of the GP equation as well as the}
numerical results obtained from its solution   for weak SO coupling using the 
split-time-step    Crank-Nicolson discretization scheme. For rotation with angular momentum along $z$ and $-z$ directions, the generated vortex- and anti-vortex-lattice structures with hexagonal and square symmetries  were studied for ferromagnetic and anti-ferromagnetic spinor BECs.   
Finally,  in Sec. \ref{IV} we present a summary of our study.

\section{The Gross-Pitaevskii equation for a rotating spin-1 condensate}
\label{II}

{We will consider a Rashba SO-coupled BEC with coupling between  the spin and  momentum  given by
$\gamma  ( \Sigma_x p_y - \Sigma_y p_x)$ \cite{exptso}, where $\gamma$ is the strength of SO coupling, 
$p_x$ and $p_y$ are the $x$ and $y$ components of the momentum operator 
 and 
 $\Sigma_x$ and $\Sigma_y$ are the irreducible representations of the $x$ 
and $y$ components of the spin matrix, respectively,
\begin{eqnarray}
\Sigma_x=\frac{1}{\sqrt 2} \begin{pmatrix}
0 & 1 & 0 \\
1 & 0  & 1\\
0 & 1 & 0
\end{pmatrix}, \quad  \Sigma_y=\frac{i}{\sqrt 2 } \begin{pmatrix}
0 & -1 & 0 \\
1 & 0  & -1\\
0 & 1 & 0
\end{pmatrix}.
\end{eqnarray}
}

For the study of vortex-lattice formation in  a rotating  SO-coupled quasi-2D  spin-1 spinor BEC, we consider a 
harmonic trap $V({\bf r})=  \widetilde m\omega^2(x^2+y^2)/2 + \widetilde m\omega_z^2 z^2/2$  with tighter binding in the $z$ direction 
$(\omega_z\gg \omega)$, where   $\omega_z$ is the angular frequency of the  trap in the $z$ direction and $\omega$ that in the $x-y$ plane. The single-particle Hamiltonian 
of the condensate without atomic interaction and 
with Rashba \cite{SOras} SO coupling in this quasi-2D trap, in dimensionless  variables,
is \cite{exptso,zhai}
\begin{equation}
H_0 = -\frac{1}{2}\nabla^2_{\bf r}+\frac{x^2+y^2}{2}+\frac{\omega_z^2z^2}{2\omega^2}  + \gamma( \Sigma_x p_y- \Sigma_y p_x ),
\label{sph} 
\end{equation}
where ${\bf r}=\{x,y,z\}$, { $\nabla^2_{\bf r}=-(p_x^2+p_y^2+p_z^2),$}
$p_x = -i\partial_x, p_y = -i\partial_y$ and $p_z = -i\partial_z$ 
are the momentum operators along $x,y$ and $z$ axes, respectively, $\partial_x,   \partial_y, \partial_z$
are partial space derivatives.  All quantities in (\ref{sph}) and in the following are dimensionless; this is achieved by expressing length ($x,y,z$) in units of harmonic oscillator length $l_0\equiv \sqrt{\hbar/\widetilde m \omega}$, and energy in units of $\hbar \omega$.


The formation of a vortex lattice in a rapidly rotating spinor BEC can be conveniently studied in the rotating frame,
where the generated vortex-lattice state is a stationary one, that can be obtained by the imaginary-time propagation
method \cite{fetter}. Such a dynamical equation in the rotating frame can be written if we note that the Hamiltonian in the rotating
frame is given by $H =H_0 - \Omega_0   L_z$ \cite{landau}, 
where $H_0$ is the laboratory frame Hamiltonian, $\Omega_0$ is the angular frequency of
rotation around the $z$ axis, and $L_z\equiv (xp_y-yp_x) = i(y\partial _x - x\partial _y)$ is the $z$ component of the angular momentum. In a trapped condensate, for rotation around $z$ axis, ordered vortex-lattice formation
is possible for $\Omega_0<\omega$ \cite{fetter}. {As $\Omega_0$  is increased above $\omega$, the whole  super-fluid moves away from the center  towards the boundary because of an excess of centrifugal force, and the super-fluidity of the condensate breaks down \cite{fetter}.
 This 
was verified in  our numerical calculation.}

For tight harmonic binding along $z$ direction,  assuming a Gaussian density distribution in the $z$ direction,
 after integrating out the $z$ coordinate following the procedure of  \cite{quasi12d}, 
in the mean-field approximation, a quasi-2D rotating SO-coupled  spin-1 spinor  BEC is described by the 
 following set of three coupled GP equations for $N$ atoms in dimensionless form for the hyper-fine
spin components $F_z = \pm 1, 0$ \cite{thspinor,thspinorb,GA}
\begin{align}\label{EQ1} 
i \partial_t & \psi_{\pm 1}({\boldsymbol 
\rho})= \left[{\cal H}+{c_2}
\left(n_{\pm 1} -n_{\mp 1} +n_0\right) -\Omega L_z \right] \psi_{\pm 1}({\boldsymbol 
\rho})\nonumber \\
+&\left\{c_2 \psi_0^2({\boldsymbol \rho})\psi_{\mp 1}^*({\boldsymbol \rho})\right\} 
-i {\widetilde \gamma} (\partial_y\psi_{0}  ({\boldsymbol \rho})\pm i \partial_x\psi_{0}({\boldsymbol \rho}) ) \, , 
\\ \label{EQ2}
i \partial_t & \psi_0({\boldsymbol \rho})=\left[ {\cal H}+{c_2}
\left(n_{+ 1}+n_{- 1}\right) -\Omega L_z\right] \psi_{0}({\boldsymbol \rho}) \nonumber \\
+&\left \{ {2} c_2 \psi_{+1}({\boldsymbol \rho})\psi_{-1}({\boldsymbol \rho})\psi_{0}^* ({\boldsymbol \rho})\right\}   
-i{\widetilde \gamma} [-i  \partial_x \{\psi_{+1}  ({\boldsymbol \rho})   \nonumber \\
-&\psi_{-1}({\boldsymbol \rho})\}
+  \partial_y \{\psi_{+1} ({\boldsymbol \rho}) +  \psi_{-1}({\boldsymbol \rho})\}]   \, , \\
{\cal H}=&-\frac{1}{2}\nabla^2_{\boldsymbol \rho}+V({\boldsymbol \rho})+c_0 n,    \\
c_0 =& \frac{2N\sqrt{2\pi\kappa}(a_0+{2}a_2)}{3}, \quad c_2 
= \frac{2N\sqrt{2\pi\kappa}(a_2-a_0)}{3}, \label{EQ4}
\end{align}
where ${\boldsymbol \rho}\equiv \{x,y   \}$, $\nabla_{\boldsymbol \rho}^2\equiv (\partial_x^2+\partial_y^2)$, 
$\kappa=\omega_z/\omega$ $(\gg 1)$, $\Omega=\Omega_0/\omega$ $ (<1)$, $\widetilde \gamma = \gamma/\sqrt 2 $,   $n_j = |\psi_j|^2$, $j= \pm 1 , 0$, are the component densities, in units of $l_0^{-2}$, of hyper-fine spin components $F_z=\pm 1,0$ and $n ({\boldsymbol \rho})= \sum_j n_j({\boldsymbol \rho})$ is the total density, $V({\boldsymbol \rho})\equiv 
(x^2+y^2)/2$ is the  circularly symmetric confining trap in the $x-y$ plane,  $\partial_t $   is the partial time derivative with time in units of $\omega^{-1}$,  $a_0$ and
$a_2$ are the s-wave scattering lengths, in units of $l_0$, in the total spin 0
and 2 channels, respectively, and the asterisk denotes complex conjugate.  The maximum allowed value of angular frequency  in 
 (\ref{EQ1})-(\ref{EQ2}) for the formation of an ordered super-fluid vortex lattice is $|\Omega|=1$ \cite{fetter}.
For notational compactness, the time dependence of the wave functions is not explicitly shown in ~(\ref{EQ1}) and (\ref{EQ2}). The normalization condition is
$ \int n({\boldsymbol \rho})\, d{\boldsymbol \rho}=1. $ 
  Equations (\ref{EQ1})-(\ref{EQ2}) can be derived from the energy functional \cite{thspinor,fetter}
\begin{align}\label{energy}
E[\psi(\Omega)] &=  \frac{1}{2} \int d{\boldsymbol \rho} \Big\{ \sum_j |\nabla_{\boldsymbol \rho}\psi_j|^ 2+
2 Vn+c_0n^2\nonumber \\  &+ c_2\big[n_{+1}^2 +n_{-1}^2
+2(n_{+1}n_0+n_{-1}n_0-n_{+1}n_{-1}\nonumber \\ &  +\psi_{-1}^*\psi_0^2\psi_{+1}^*  
+ \psi_{-1}\psi_0^{*2}\psi_{+1})  \big] -2i\widetilde \gamma\big[ \psi_0^* \partial_y(\psi_{+1}\nonumber \\
&+\psi_{-1} )
+ (\psi_{+1}^*+\psi_{-1}^* )  \partial_y \psi_0   -i \psi_0^* \partial_x(\psi_{+1}-\psi_{-1} )\nonumber \\
&+i (\psi_{+1}^*-\psi_{-1}^* )\partial_x \psi_0 \big] - 2 \Omega \sum_j \psi_j^* L_z \psi_j  \Big\},
\end{align}
 where the space dependence of different variables is not explicitly shown. The $\Omega$ dependence of the 
wave function is shown to recall that the energy functional is a function of the angular frequency of rotation.

\section{Numerical Results}
 
\label{III}

To solve  (\ref{EQ1}) and (\ref{EQ2}) numerically, we propagate
these equations in time by the split-time-step Crank-Nicolson discretization scheme \cite{cpc,bec2009,bec2012,bec2017x} using a space
step of 0.1 and a time step $\Delta$ of 0.001  to obtain the stationary state  by imaginary-time simulation. There are different C and FORTRAN programs for solving the GP equation \cite{bec2009,bec2012}  and one
should use the appropriate one. These programs have
recently been adapted to simulate the vortex lattice in
a rapidly rotating BEC \cite{vor-lat} and we use these in this
study.

The imaginary-time propagation is started with an appropriate initial state consistent with symmetry for quick convergence. { In the ferromagnetic and anti-ferromagnetic phases, the circularly-symmetric ground states 
of the Rashba SO-coupled spin-1 BEC 
are of types $(0,+1,+2)$  and $(-1,0,+1)$ \cite{cpc}, respectively, with vortices in  components. In numerical simulation of vortex lattice in a rotating SO-coupled spin-1 BEC  we will include   these vortices in the initial state. }
 For  a final  localized state  without vorticity, a Gaussian initial state in each component is adequate:
$\psi _j(\boldsymbol \rho) \sim \exp (-\rho^2/\alpha_j^2)$, where $\alpha_j$ is the width. However, for  a 
$(0,+1,+2)$-type  state with vorticity,  we will take the initial functions $\ \psi _{+1}(\boldsymbol \rho) \sim \exp (-\rho^2/\alpha_j^2)$, 
$\ \psi _{0}(\boldsymbol \rho) \sim (x+iy)\exp (-\rho^2/\alpha_j^2)$, 
$\ \psi _{-1}(\boldsymbol \rho) \sim (x+iy)^2 \exp (-\rho^2/\alpha_j^2)$, 
Similarly, 
 for  a 
$(-1,0,+1)$-type  state,  we take the initial functions $\ \psi _{+1}(\boldsymbol \rho) \sim (x-iy) \exp (-\rho^2/\alpha_j^2)$, 
$\ \psi _{0}(\boldsymbol \rho) \sim \exp (-\rho^2/\alpha_j^2)$, 
$\ \psi _{-1}(\boldsymbol \rho) \sim (x+iy) \exp (-\rho^2/\alpha_j^2)$.  With these initial states, with proper vorticity,
the convergence of the imaginary-time propagation is quick.

The parameters of the GP equation $c_0$ and $c_2$ are taken from the following realistic experimental situations.  
For the quasi-2D ferromagnetic 
BEC we use the following parameters of $^{87}$Rb atoms: $N=100,000, a_0=101.8a_B, a_2=100.4a_B,  $ \cite{a02rb} $l_{z}\equiv l_0/\sqrt \kappa=2.0157$ $\mu$m, where $a_B$ is the Bohr radius. Consequently, 
$c_0\equiv   2N\sqrt{2\pi}(a_0+2a_2)/3l_{z}  \approx 1327$ and $c_2\equiv  2N\sqrt{2\pi}(a_2-a_0)/3l_{z}  \approx -6.15$.   
For the quasi-2D anti-ferromagnetic 
BEC we use the following parameters of $^{23}$Na atoms: $N=100,000, a_0=50.00a_B, a_2=55.01a_B$, \cite{baox} $l_{z}=2.9369$ $\mu$m. Consequently, $c_0  \approx 482$ and $c_2 \approx 15$.  
{ The $^{87}$Rb and  $^{23}$Na atoms naturally appear in ferromagnetic ($c_2<0$) and anti-ferromagnetic ($c_2>0$) phases, respectively. Hence,  to simulate a ferromagnetic (anti-ferromagnetic) BEC we use the parameters of  $^{87}$Rb ($^{23}$Na).  Using a Feshbach resonance it is possible to change the sign of $c_2$, thus turning, for example,  a ferromagnetic $^{87}$Rb BEC to an anti-ferromagnetic BEC.  However, we will not consider this possibility in this paper.}

\subsection{Classification of states and symmetries}

{
We will consider   Rashba SO-coupled ferromagnetic and anti-ferromagnetic spinor BECs  for weak  SO coupling ($\gamma \lessapprox 0.75$) and an initial circularly-symmetric solution for the 
  quasi-2D  non-rotating spinor 
 BEC.  For large values of $\gamma$, the non-rotating SO-coupled  spinor BEC  has 
 stripe  (and other) pattern in density \cite{stripe,st2},   breaking  circular symmetry. The rotation of such a state should lead to a complex vortex-lattice structure and will not be considered in this paper.}   

The formation of vortex lattice in a scalar BEC, without intrinsic vorticity in the absence of rotation, is different from that in an SO-coupled spin-1 spinor 
BEC  with states   of type $(0,+1,+2)$ or  $(-1,0,+1)$ with intrinsic vorticity  \cite{kita}.  
For a    {weakly} SO-coupled     spin-1 ferromagnetic
spinor BEC   of type $(0,+1,+2)$,  rotating with the angular momentum vector parallel to the vorticity direction, 
a vortex lattice with hexagonal symmetry can be
 generated maintaining the  states  of circulation   $+1$ and $+2$ at the center in components $j=0$ and $-1$, respectively, while the center of the $j=+1$ component is maintained vortex free. However, in this case, for rotation with the angular  momentum vector anti-parallel to the vorticity direction,  for small angular frequency of rotation, the state 
of type $(0,+1,+2)$  becomes one of type  $(-2,-1,0)$.   The $(-2,-1,0)$-type  state results 
upon a superposition of the $(0,+1,+2)$-type state with a  $(-2,-2,-2)$-type state: the latter is  generated by  rotation.   { This is a two-step process. First a $(-1,0,+1)$-type state is formed upon the superposition of a  $(-1,-1,-1)$-type state, generated by rotation, with the $(0,+1,+2)$-type state. Later another   $(-1,-1,-1)$-type state superposed on the $(-1,0,+1)$-type state yields the  $(-2,-1,0)$-type  state. 
} 
With the increase of angular frequency of rotation,  
an anti-vortex lattice with hexagonal symmetry
can be generated 
maintaining  these central anti-vortices of circulation $-2$ and $-1$ in   components
$j=+1,$ and $0$, respectively. The anti-vortex lattice is really a vortex lattice with opposite vorticity.

For a  {weakly}  SO-coupled    spin-1 anti-ferromagnetic 
spinor BEC  of type $(-1,0,+1)$, rotating with the angular momentum vector parallel to the vorticity $z$ direction, for small angular frequency of rotation, the state $(-1,0,+1)$ transforms into the state $(0,+1,+2)$. The $(0,+1,+2)$-type  state results 
upon a superposition of the $(-1,0,+1)$-type state {with a  $(+1,+1,+1)$-type state:} the latter is  generated by  rotation.   
In this case, upon the increase of angular frequency of rotation, a vortex lattice with hexagonal symmetry can be generated maintaining the central vortices of circulation $+1$ and $+2$ in components $j=0$ and $-1$, respectively. 
However, for a spin-1 SO-coupled 
 spinor BEC  of type $(-1,0,+1)$, rotating with the angular momentum vector anti-parallel to the vorticity direction, for small angular frequency of rotation, the state $(-1,0,+1)$ transforms into the state $(-2,-1,0)$.  With the increase of angular frequency of rotation,  
an anti-vortex lattice with hexagonal symmetry can   again be  generated maintaining  these central anti-vortices of circulation $-2$ and $-1$ in the respective components.

In the case of a rotating SO-coupled spin-1 spinor condensate, we find that there are many different vortex-lattice states
 with different symmetry properties lying close to each other. Hence it is often difficult to 
find the vortex-lattice state with minimum energy by  imaginary-time propagation and 
it is possible that, in some cases, the imaginary-time approach   converges to a nearby excited vortex-lattice state, instead of the lowest-energy  state for certain initial states. To circumvent this problem, we  repeated 
the calculation with different initial states, so as to be sure that the converged vortex-lattice state is indeed the lowest-energy  state.  The use of an analytic initial function modulated by a random phase at different space points
also increases the possibility of the convergence to the minimum-energy state \cite{vor-lat}.
Unlike in a rotating scalar BEC, where the vortex lattice in the lowest-energy state always has a hexagonal symmetry, 
in  the case of a rotating SO-coupled spin-1 spinor BEC,
    vortex- and anti-vortex-lattice states with an
approximate square symmetry can also appear, in addition to those with the usual hexagonal symmetry.   For hexagonal symmetry, vortices are arranged in concentric orbits containing a maximum of 6, 12, 18 ... vortices;  for square symmetry these numbers are 8, 12, 16 ...
 Often, for the same angular frequency of rotation,  it is possible to obtain both types of vortex-lattice states with 
all orbits containing the maximum  number of vortices. When this happens, the vortex- and anti-vortex-lattice  states for both ferromagnetic and anti-ferromagnetic SO-coupled spin-1 { BECs
with square symmetry are  found to have  the smaller numerical energy, although we could not establish this fact theoretically. } The general scenario of vortex-lattice states remain unchanged for different  numerical values of the non-linearity parameters $c_0$ and $c_2$.

\begin{figure}[!t] 
\centering
\includegraphics[trim = 6mm 0mm 11mm 0mm,clip,width=.32\linewidth]{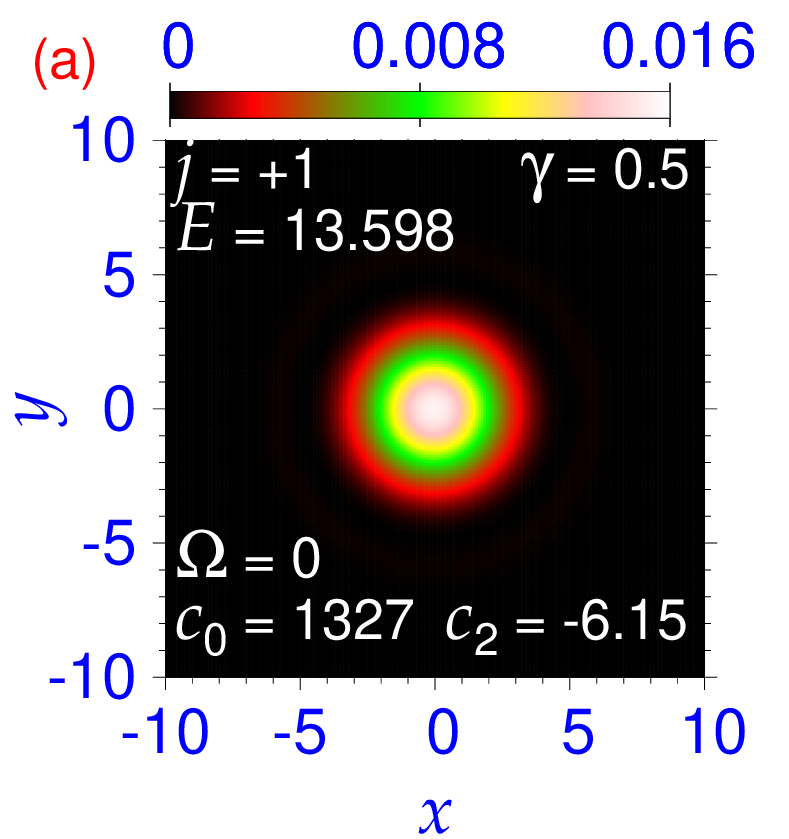} 
\includegraphics[trim = 6mm 0mm 11mm 0mm,clip,width=.32\linewidth]{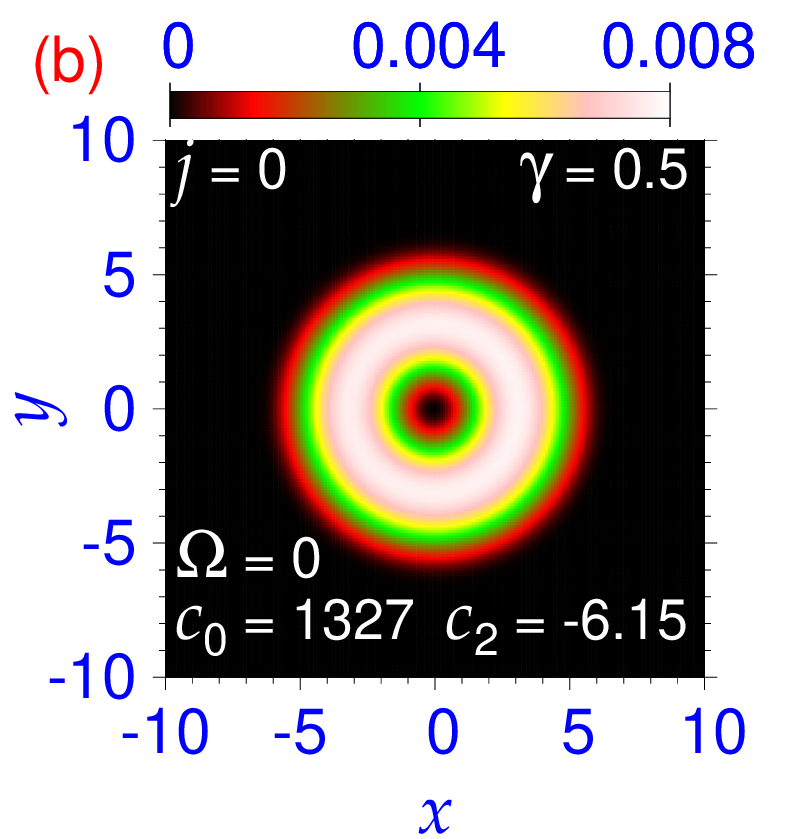}
\includegraphics[trim = 6mm 0mm 11mm 0mm,clip,width=.32\linewidth]{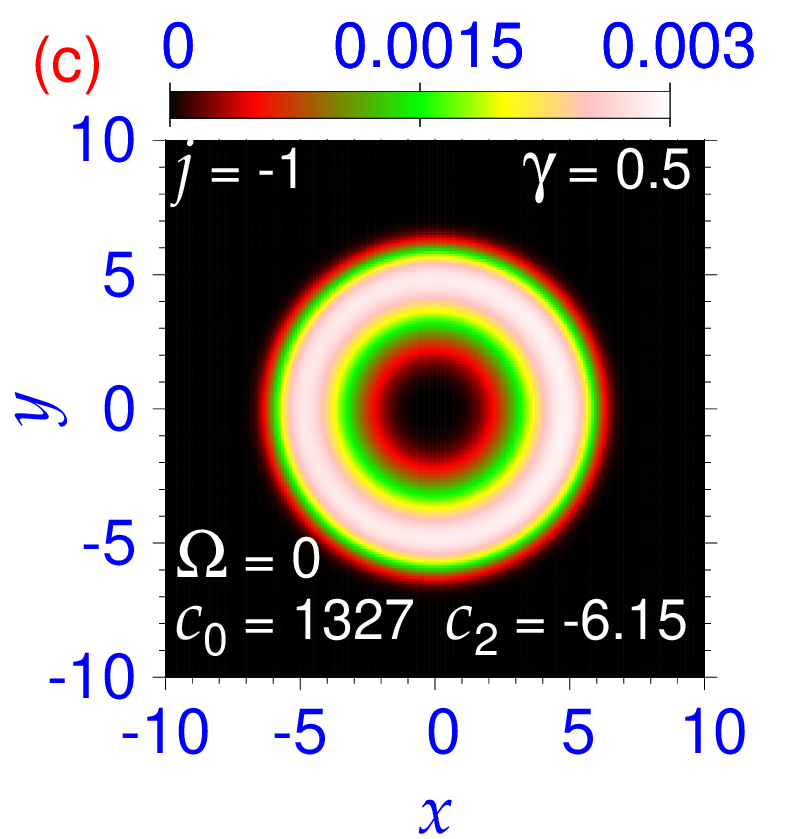}
 \includegraphics[trim = 6mm 0mm 11mm 0mm,clip,width=.32\linewidth]{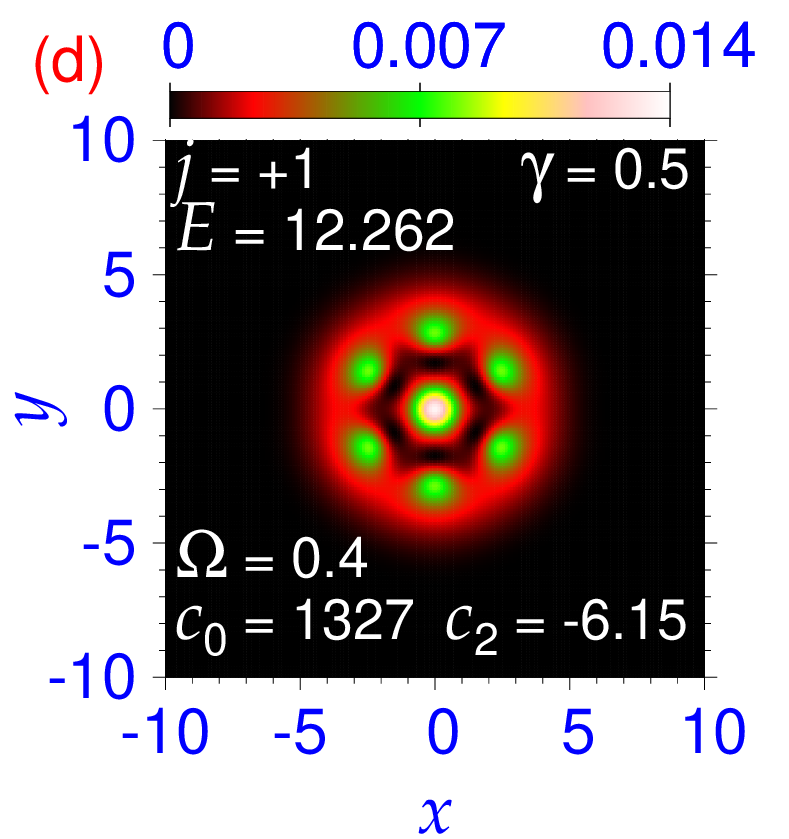} 
\includegraphics[trim = 6mm 0mm 11mm 0mm,clip,width=.32\linewidth]{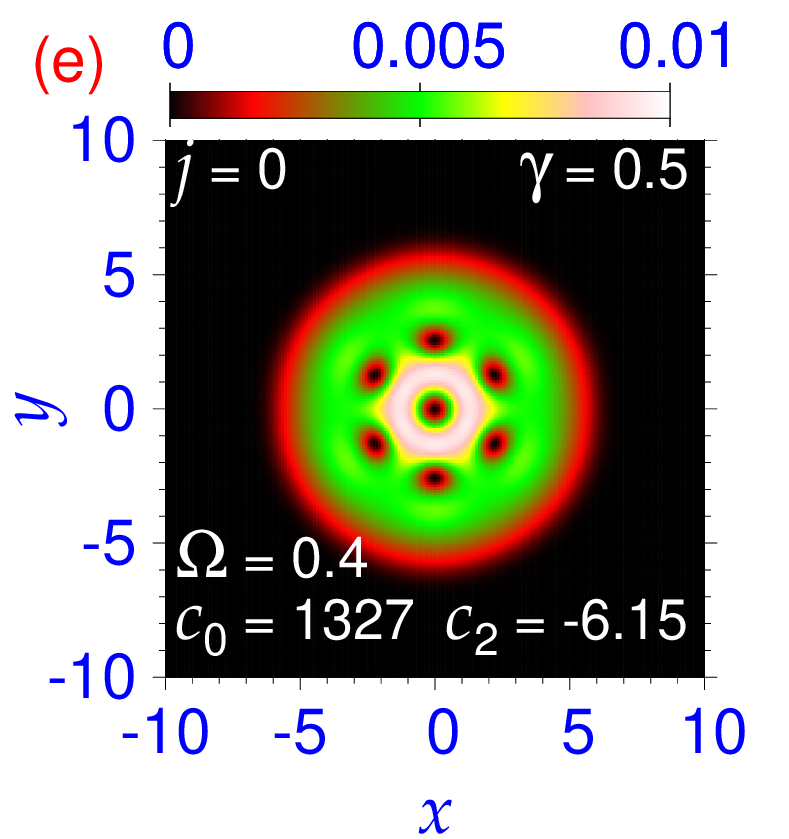}
\includegraphics[trim = 6mm 0mm 11mm 0mm,clip,width=.32\linewidth]{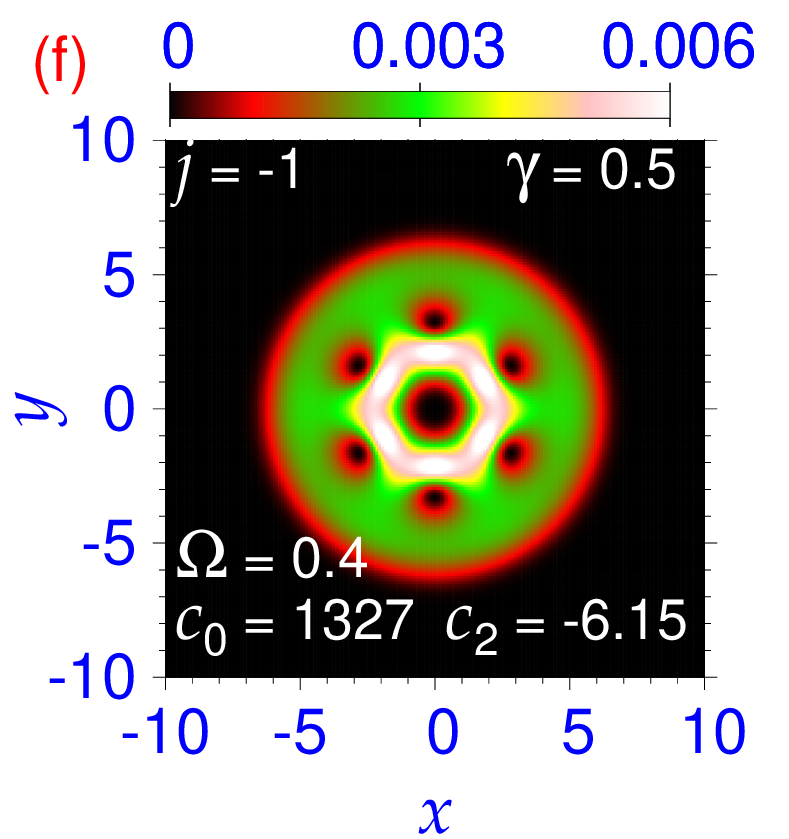}
 \includegraphics[trim = 6mm 0mm 11mm 0mm,clip,width=.32\linewidth]{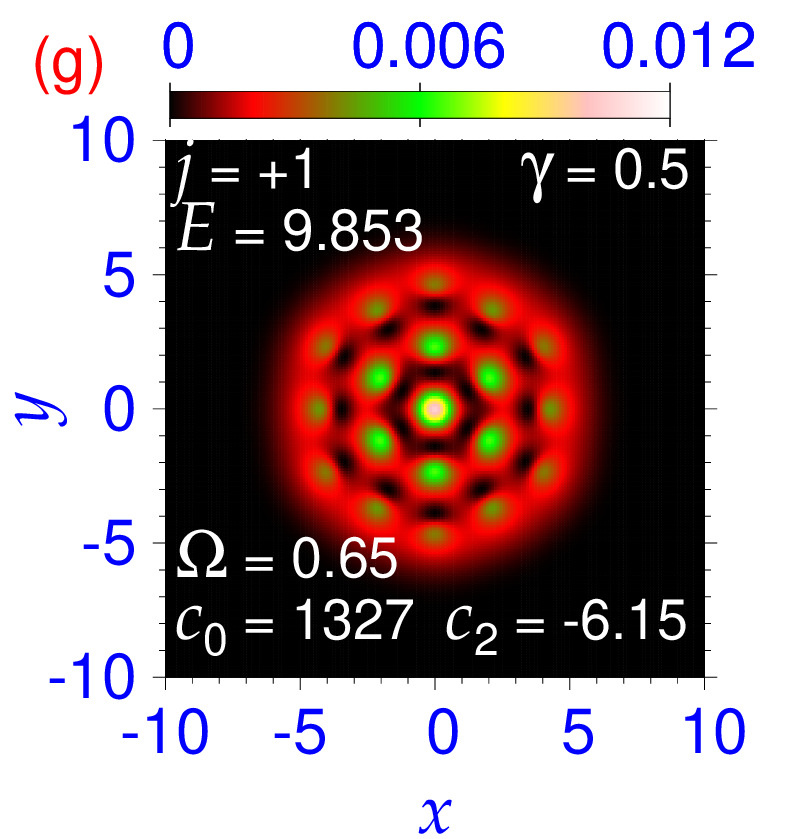} 
\includegraphics[trim = 6mm 0mm 11mm 0mm,clip,width=.32\linewidth]{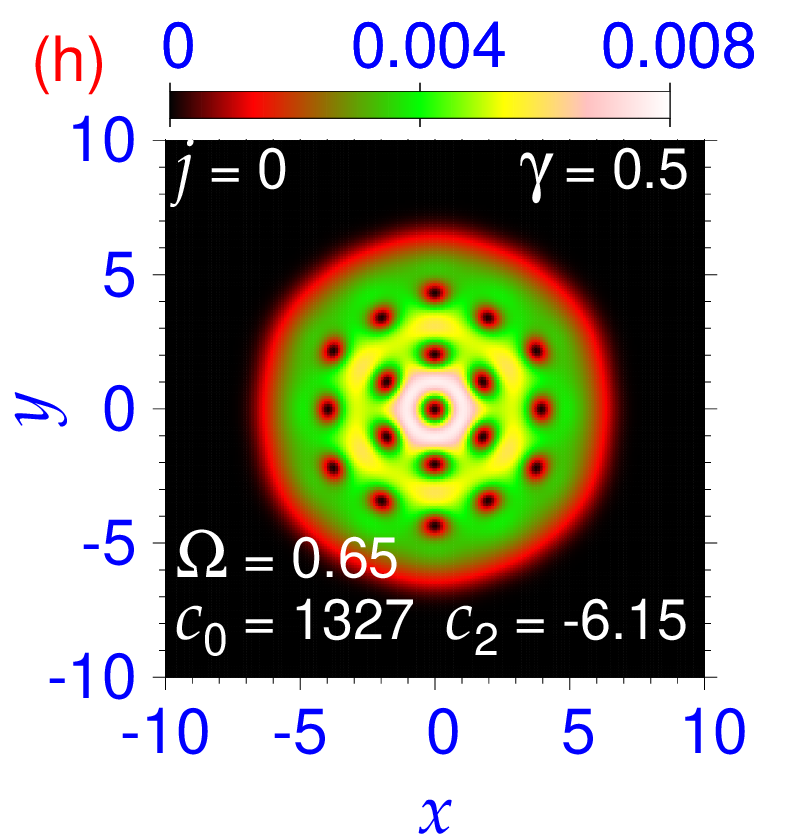}
\includegraphics[trim = 6mm 0mm 11mm 0mm,clip,width=.32\linewidth]{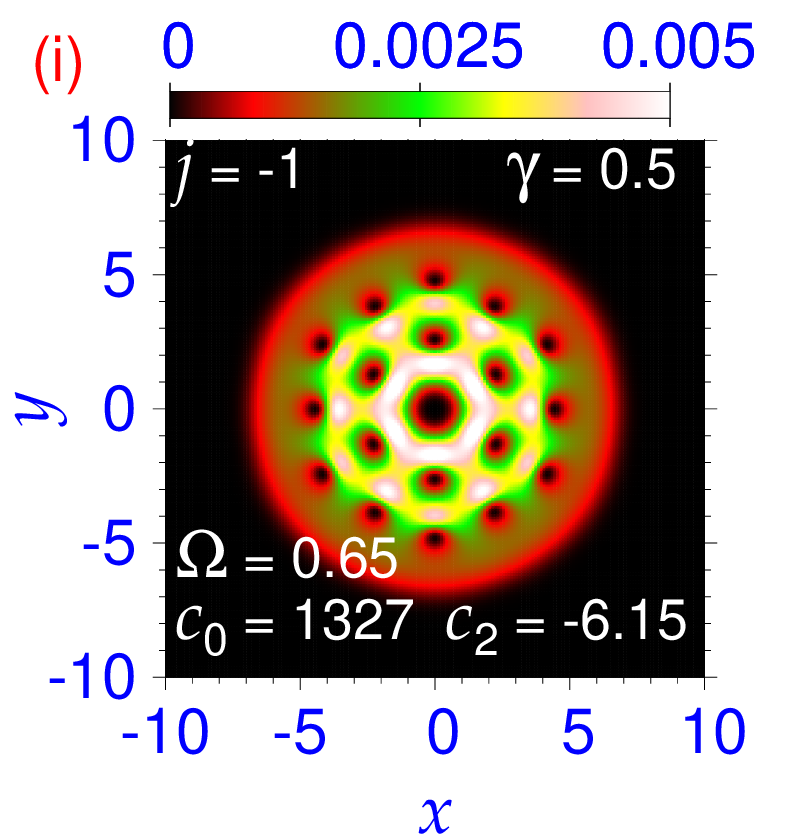}
 \includegraphics[trim = 6mm 0mm 11mm 0mm,clip,width=.32\linewidth]{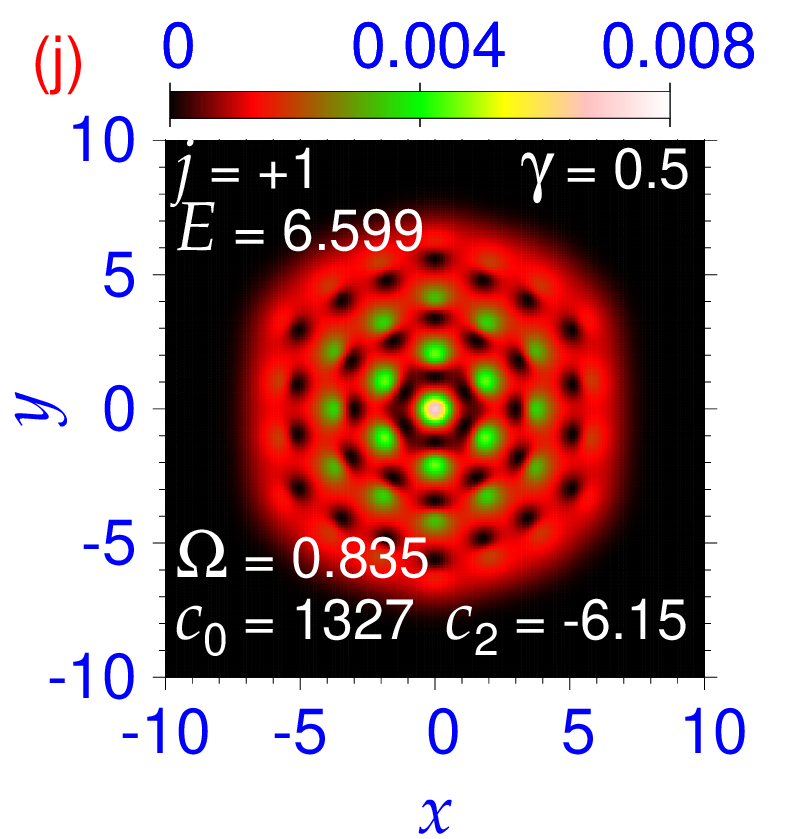} 
\includegraphics[trim = 6mm 0mm 11mm 0mm,clip,width=.32\linewidth]{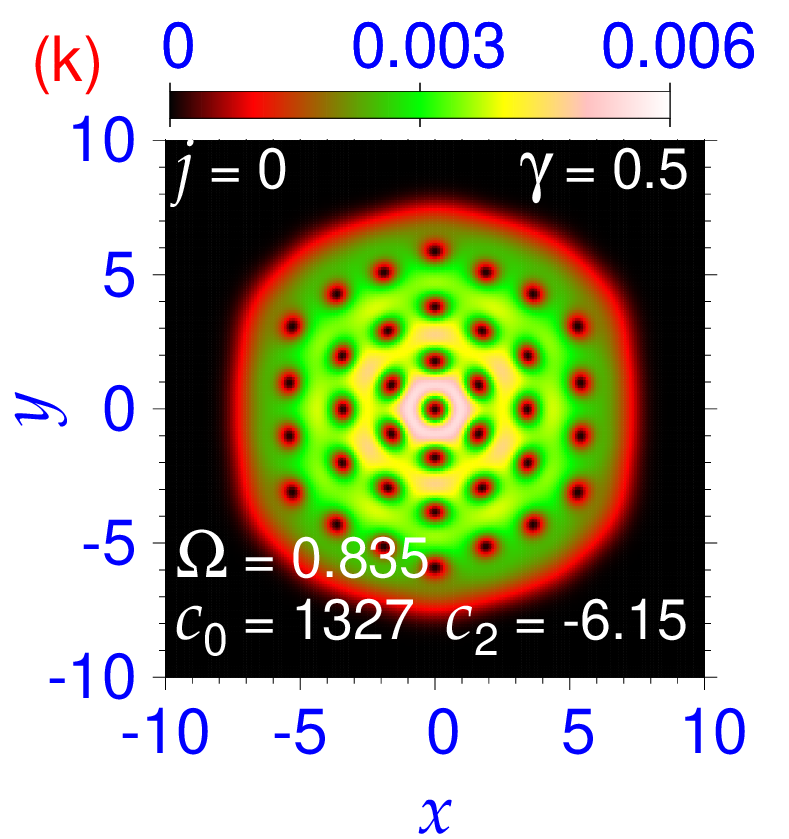}
\includegraphics[trim = 6mm 0mm 11mm 0mm,clip,width=.32\linewidth]{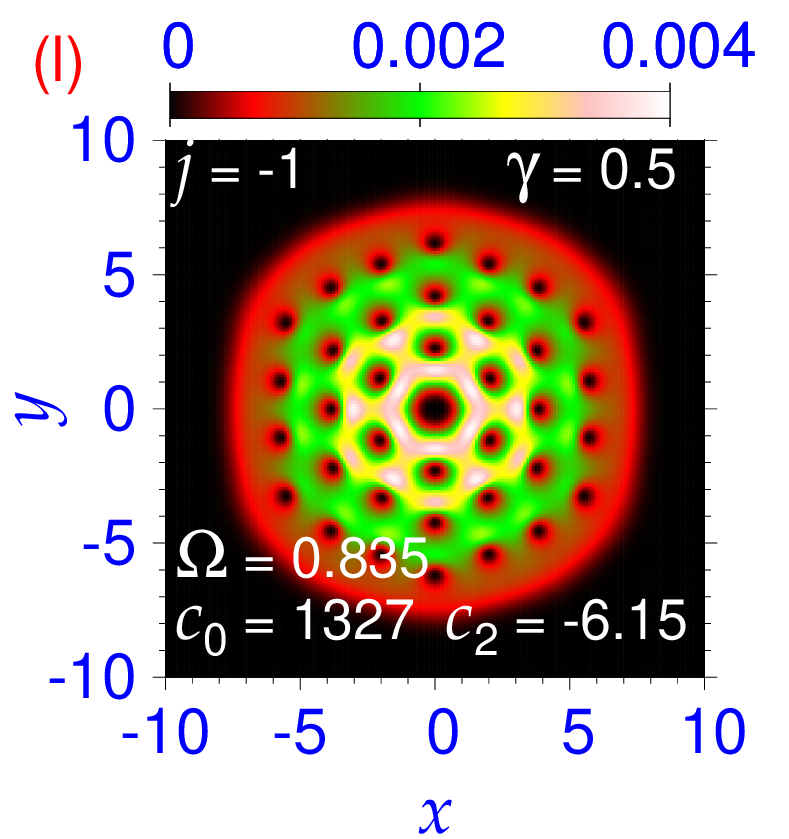} 
 
\caption{(Color online)  Contour plot of component densities $n_j(\boldsymbol \rho)\equiv | \psi _j(\boldsymbol \rho)|^2$ of vortex-lattice states with hexagonal symmetry of a rotating  Rashba SO-coupled ferromagnetic spin-1 quasi-2D spinor BEC   for angular 
frequencies $\Omega =0, 0.4, 0.65,$ and 0.835 in plots (a)-(c), (d)-(f), (g)-(i), and (j)-(l), respectively.  
The angular 
momentum of rotation is parallel to the vorticity direction of the non-rotating state in (a)-(c). The non-linearity parameters $c_0=1327, c_2=-6.15$,  and SO-coupling strength $\gamma =0.5$.
In all density plots of this paper, energy values, viz.  (\ref{energy}), are displayed in the density of the $j=+1$ component.
  All results reported in this paper are in dimensionless units, as outlined in Section~\ref{II}.}
\label{fig1}

\end{figure}

\begin{figure}[!t] 
\centering
\includegraphics[width=.32\linewidth]{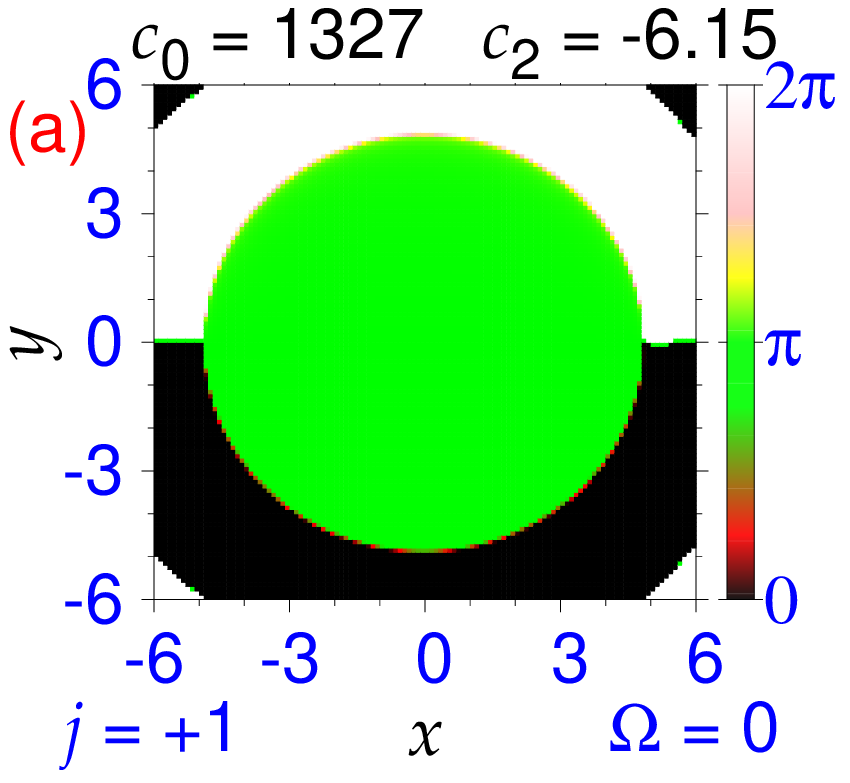} 
\includegraphics[width=.32\linewidth]{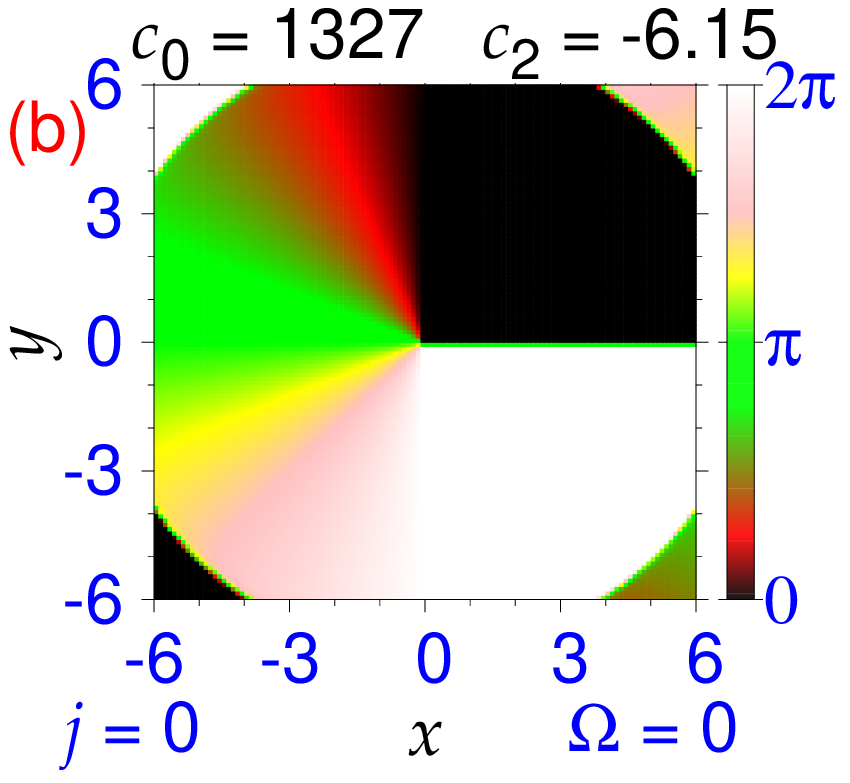}
\includegraphics[width=.32\linewidth]{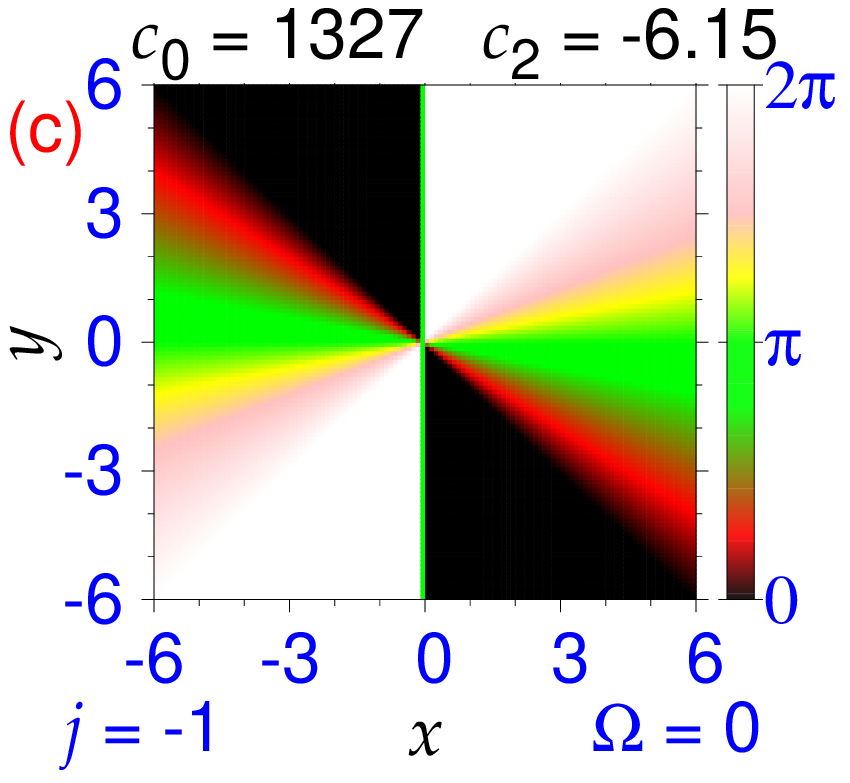}
 \includegraphics[width=.32\linewidth]{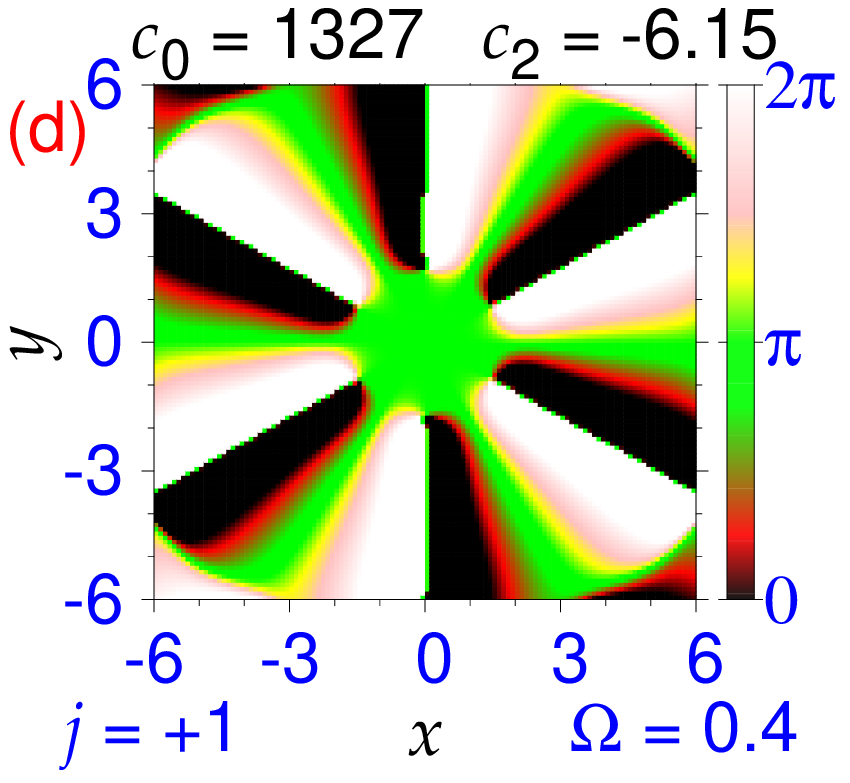} 
\includegraphics[width=.32\linewidth]{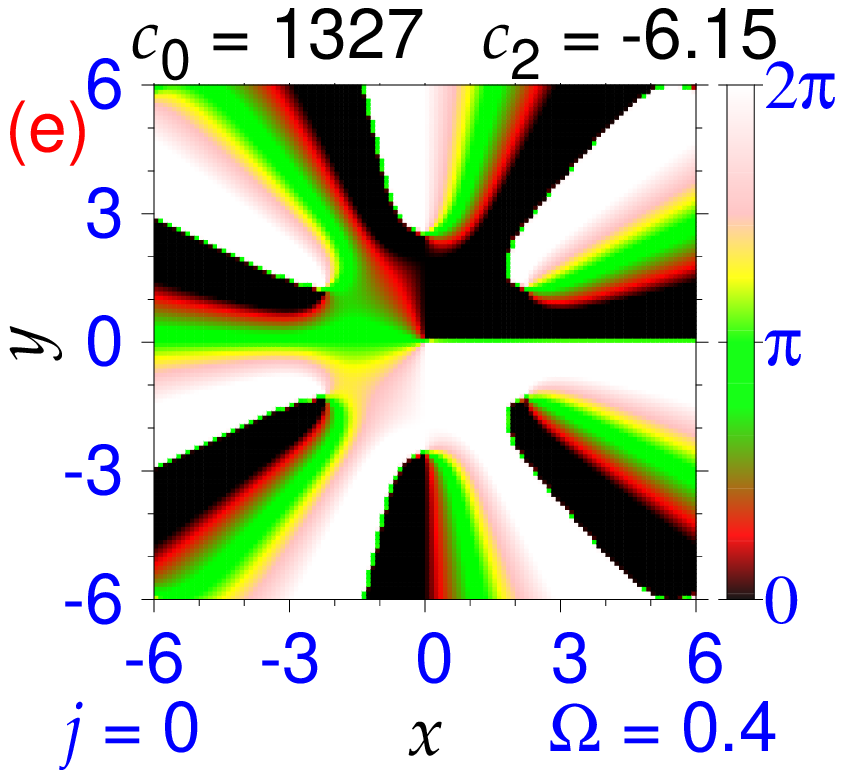}
\includegraphics[width=.32\linewidth]{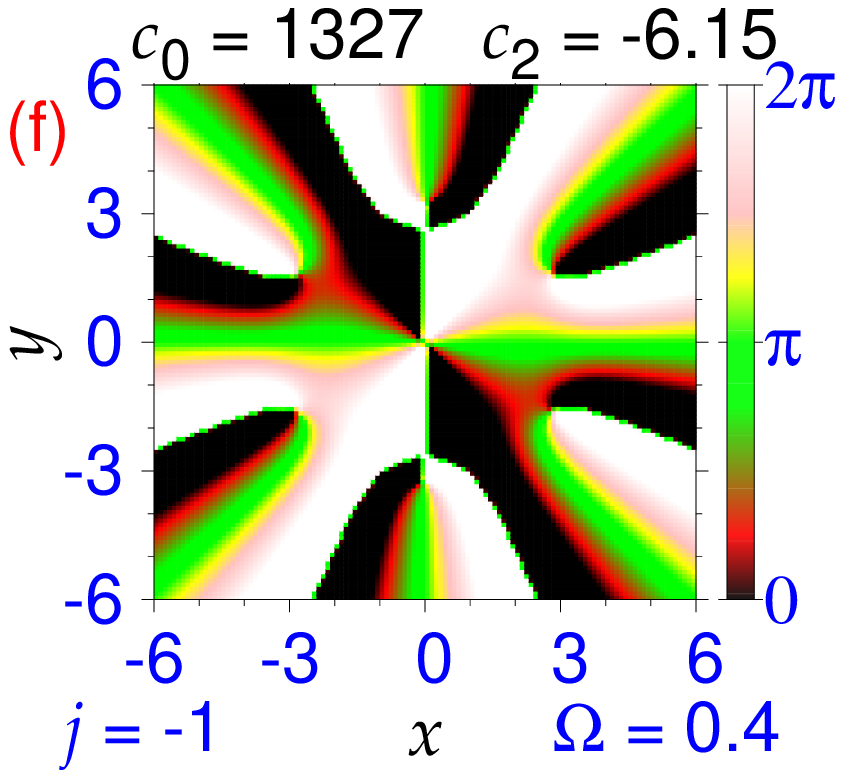}
 \includegraphics[width=.32\linewidth]{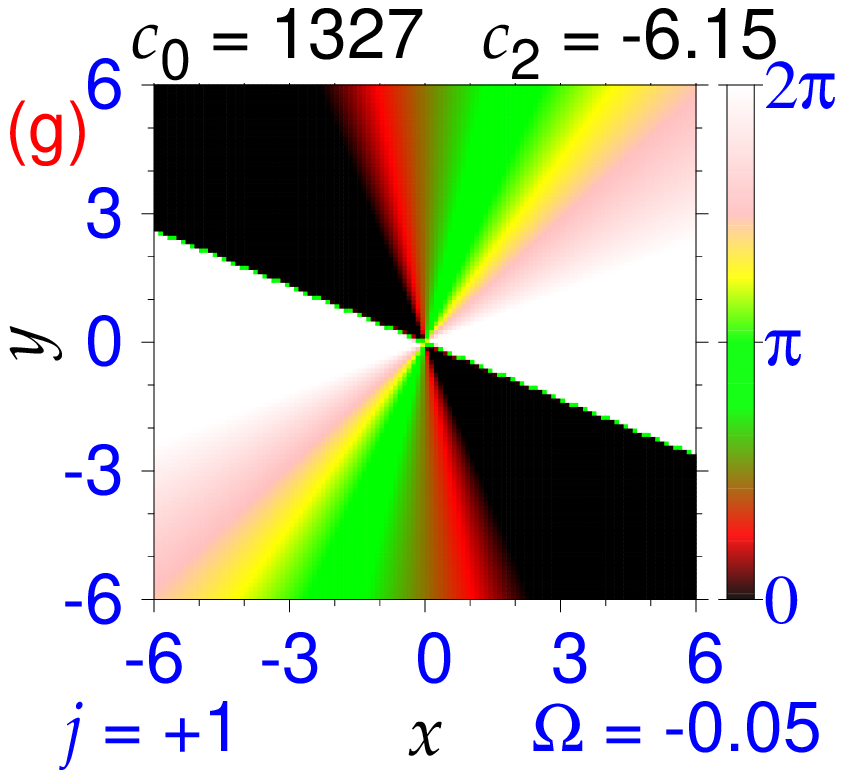} 
\includegraphics[width=.32\linewidth]{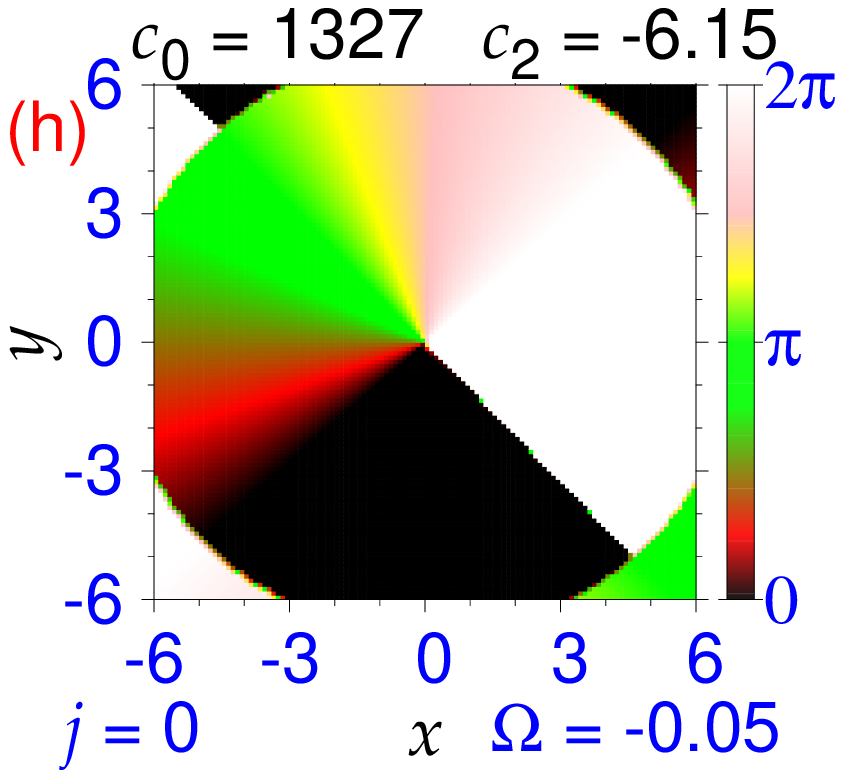}
\includegraphics[width=.32\linewidth]{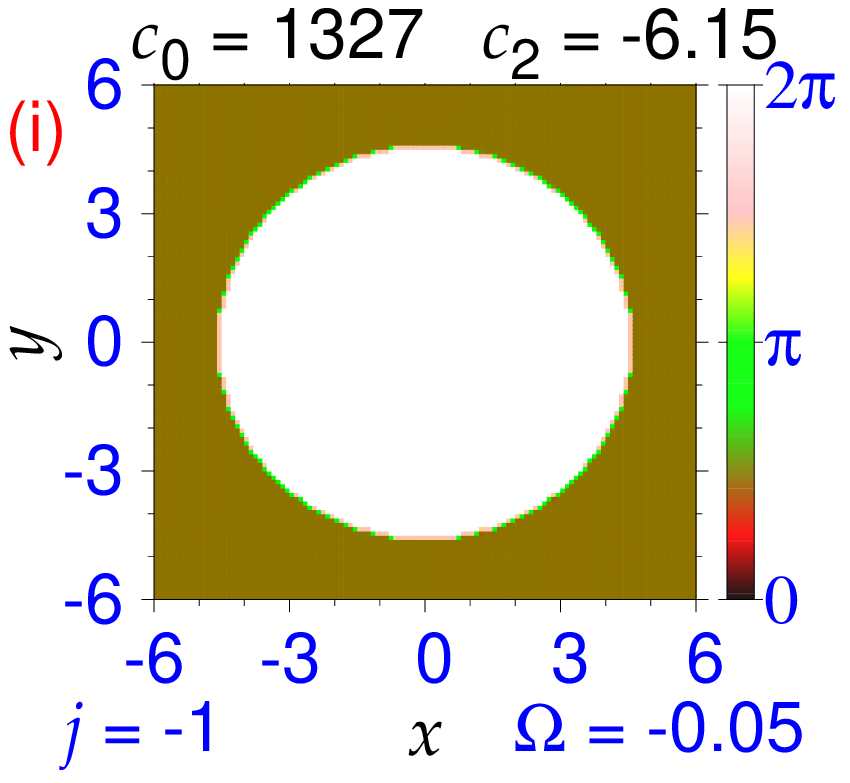} 
\includegraphics[width=.32\linewidth]{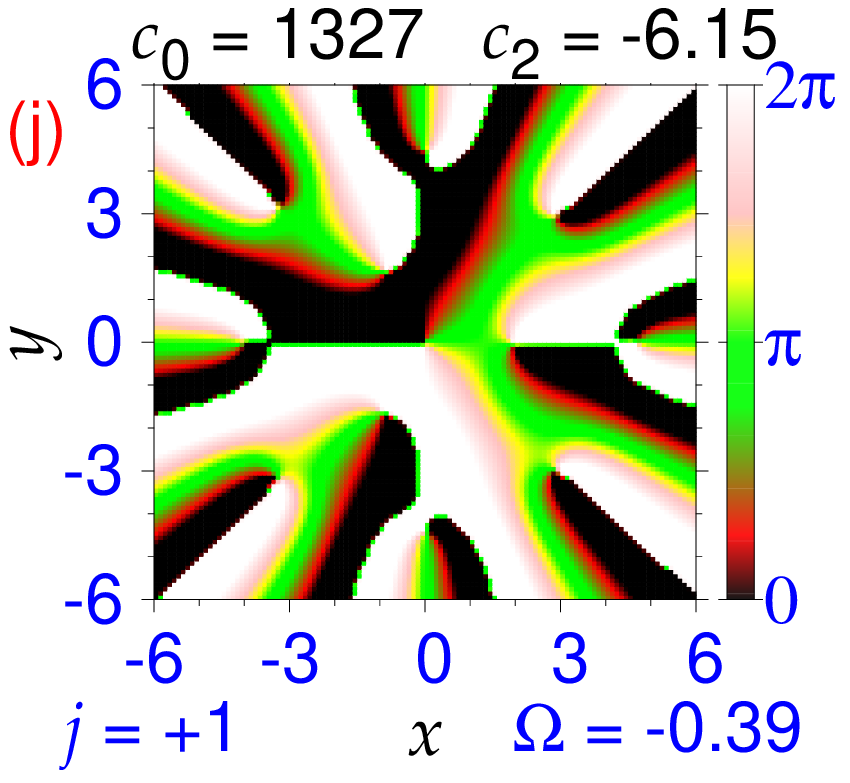} 
\includegraphics[width=.32\linewidth]{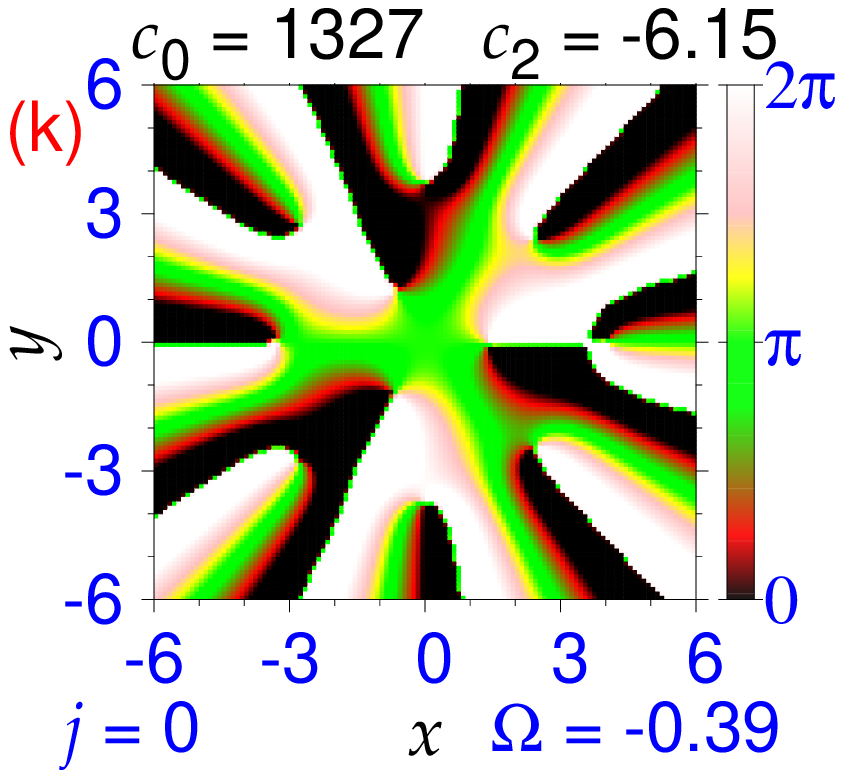}
\includegraphics[width=.32\linewidth]{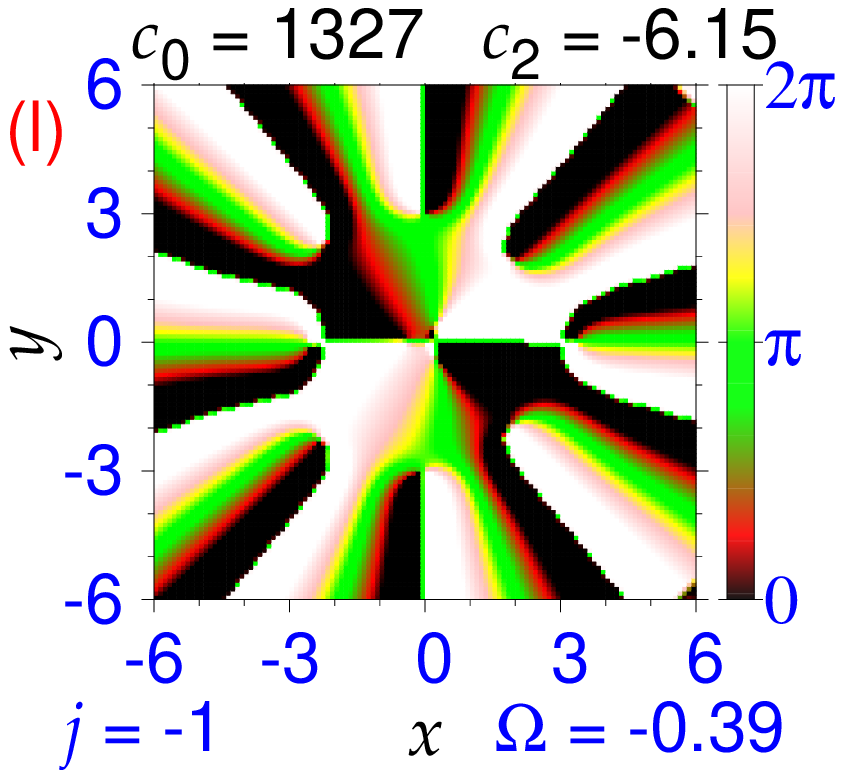} 
 
\caption{(Color online) (a)-(c) Contour plot  of the phase $\delta({\boldsymbol \rho})$ of the wave function of the non-rotating  Rashba SO-coupled ferromagnetic spin-1 quasi-2D spinor BEC    of figures \ref{fig1}(a)-(c). 
(d)-(f) The same of the rotating ferromagnetic spinor BEC, with angular frequency $\Omega=0.4$,
of  figures \ref{fig1}(d)-(f). 
(g)-(i) The same of the rotating ferromagnetic spinor BEC, with angular frequency $\Omega=-0.05$,
of  figures \ref{fig4}(a)-(c). (j)-(l) The same of the rotating ferromagnetic spinor BEC, { with angular frequency $\Omega=-0.39$,}
 of  figures \ref{fig5}(a)-(c).
 }
\label{fig2}

\end{figure}

\subsection{Ferromagnetic condensate}

We study the  formation of vortex-lattice states with hexagonal symmetry
 in a rotating Rashba SO-coupled quasi-2D $^{87}$Rb
ferromagnetic spin-1 BEC with
SO-coupling strength $\gamma=0.5$,  and  
non-linearities $c_0=1327$ and $c_2=-6.15$ 
for different angular frequency $\Omega$ through  a plot of contour density  {$n_j({\boldsymbol \rho})=|\psi_{j} ({\boldsymbol \rho})|^2$}
 of different components.
In figures \ref{fig1}(a)-(c)  we plot the  densities  of components $j=+1,0,-1$ of the  non-rotating { lowest-energy circularly-symmetric} state of type
$(0,+1,+2)$.
We  checked the vorticity and circulation of the components analyzing the phase plot of the wave function displayed in 
figures \ref{fig2}(a)-(c). The phase drop upon a clockwise rotation of $2\pi$ in figure \ref{fig2}(b)  (c) is $2\pi$  ($4\pi$) indicating a
circulation of $+1$ ($+2$). 
 The $j=-1$ component with circulation $+2$ has a larger vortex core than the $j=0$ component with 
circulation $+1$. In figures \ref{fig1}(d)-(l) we display  the vortex lattice with hexagonal symmetry
for increasing angular frequencies $\Omega=0.4, 0.65,$ and $0.835$. { The vortex structure of the BEC with $\Omega =0.4$ can be  found from the  phase plot of the corresponding wave function in figures \ref{fig2}(d)-(f) for components $j=+1,0,-1.$}
The direction of generated angular momentum upon rotation is parallel to the intrinsic vorticity 
of the non-rotating state ($z$ direction).
 For all angular frequencies, a clean vortex lattice is generated as in the case of a scalar BEC. The only difference from a scalar BEC  is that the central spot is vortex free for component $j=+1$ and hosts a vortex of circulation $+2$ in component $j=-1$. In the component $j=0$, the central spot has a vortex of circulation 
$+1$ as in a scalar BEC. A vortex of circulation $+2$ (greater than unity) in a scalar BEC should break into two of unit circulation from an energetic consideration \cite{fetter}. However, in this SO-coupled spin-1 spinor BEC, the vortex of circulation $+2$ at the center  of component $j=-1$ is found to be energetically   stable. 

 \begin{figure}[!t]
\centering
\includegraphics[trim = 6mm 0mm 11mm 0mm,clip,width=.32\linewidth]{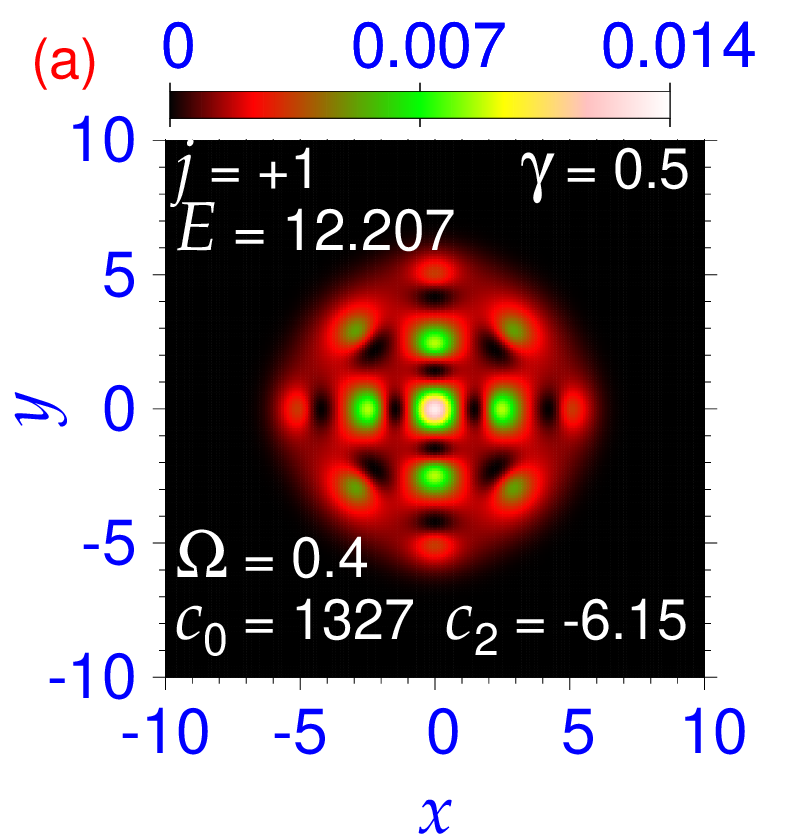}
\includegraphics[trim = 6mm 0mm 11mm 0mm,clip,width=.32\linewidth]{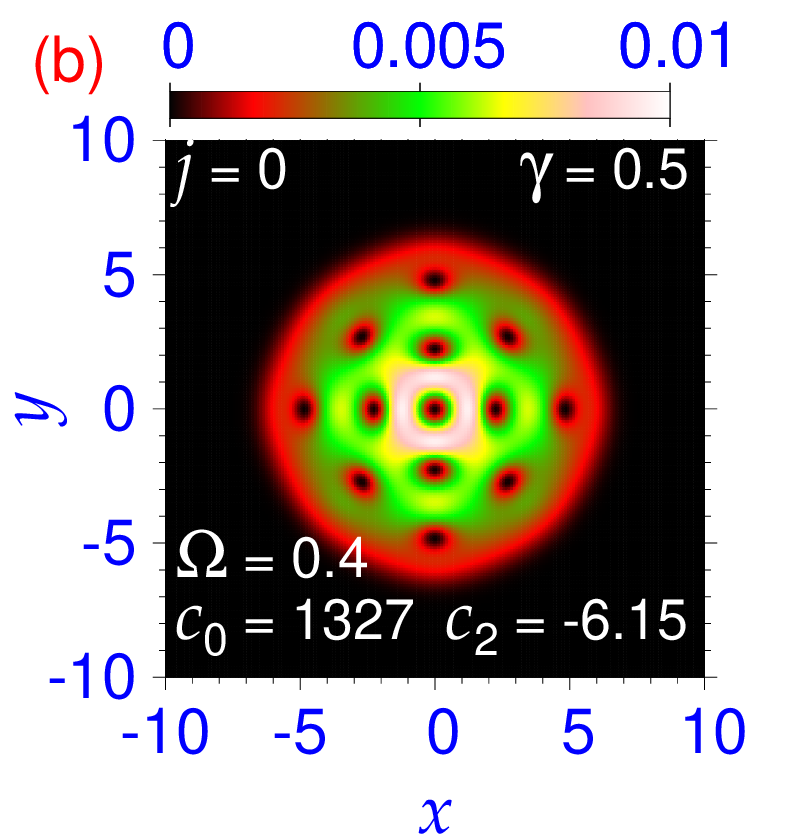}
 \includegraphics[trim = 6mm 0mm 11mm 0mm,clip,width=.32\linewidth]{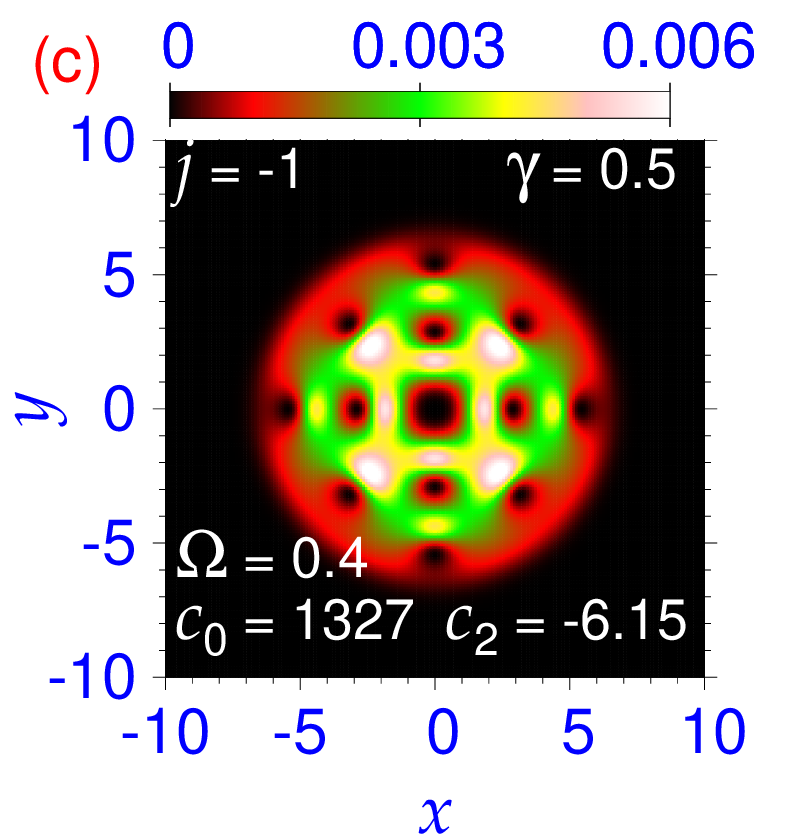}
\includegraphics[trim = 6mm 0mm 11mm 0mm,clip,width=.32\linewidth]{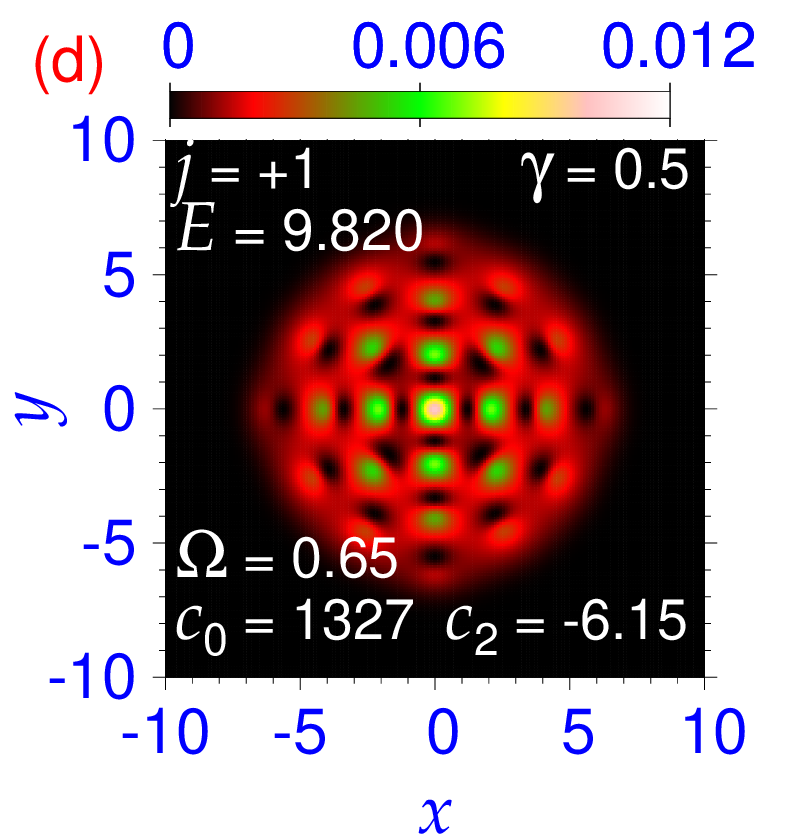}
\includegraphics[trim = 6mm 0mm 11mm 0mm,clip,width=.32\linewidth]{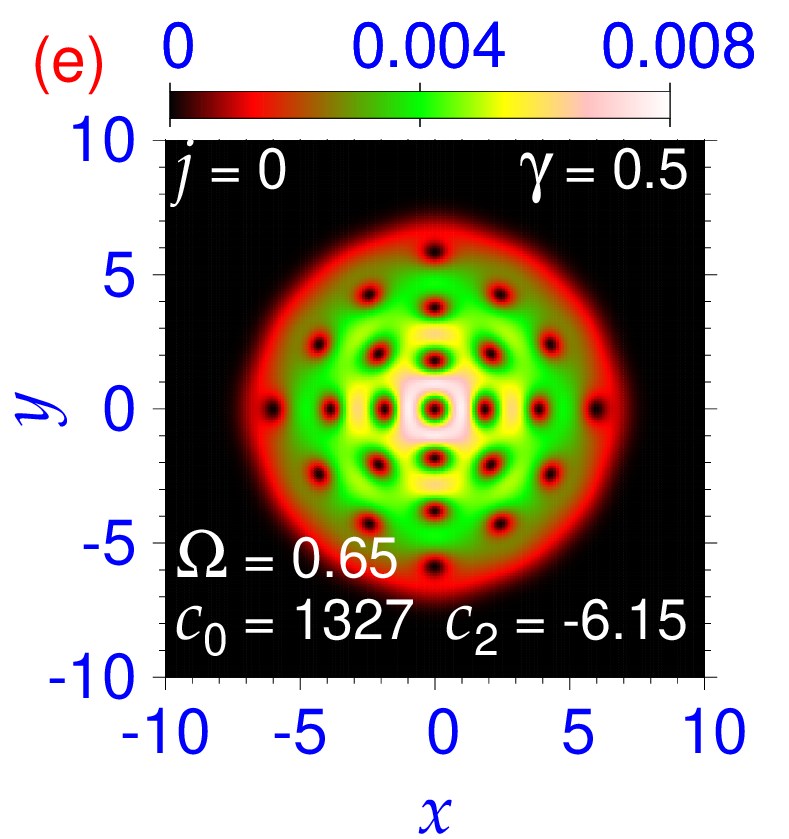}
 \includegraphics[trim = 6mm 0mm 11mm 0mm,clip,width=.32\linewidth]{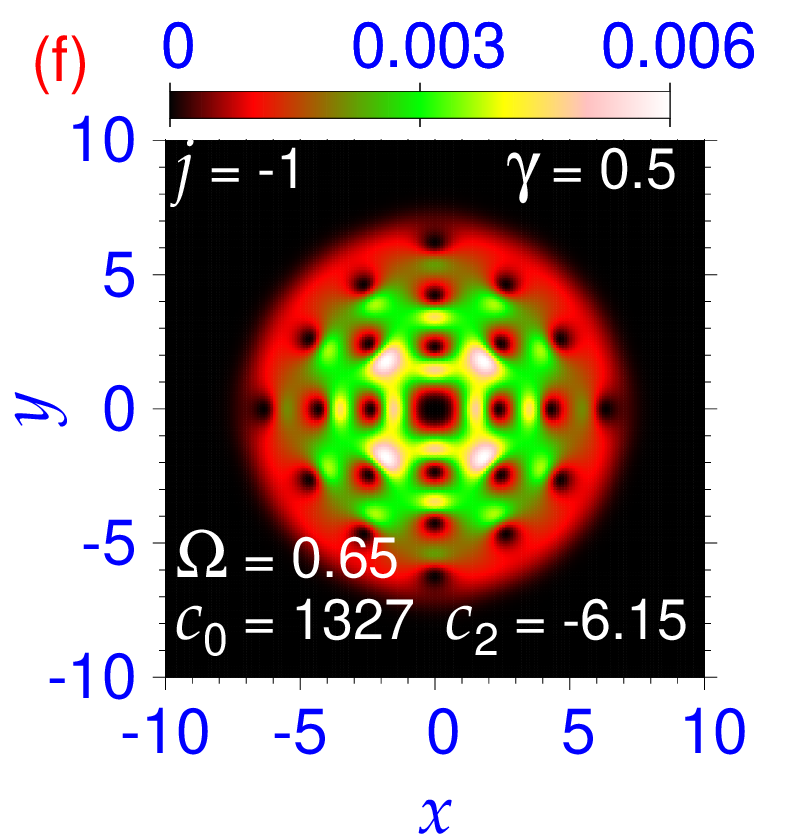}
\includegraphics[trim = 6mm 0mm 11mm 0mm,clip,width=.32\linewidth]{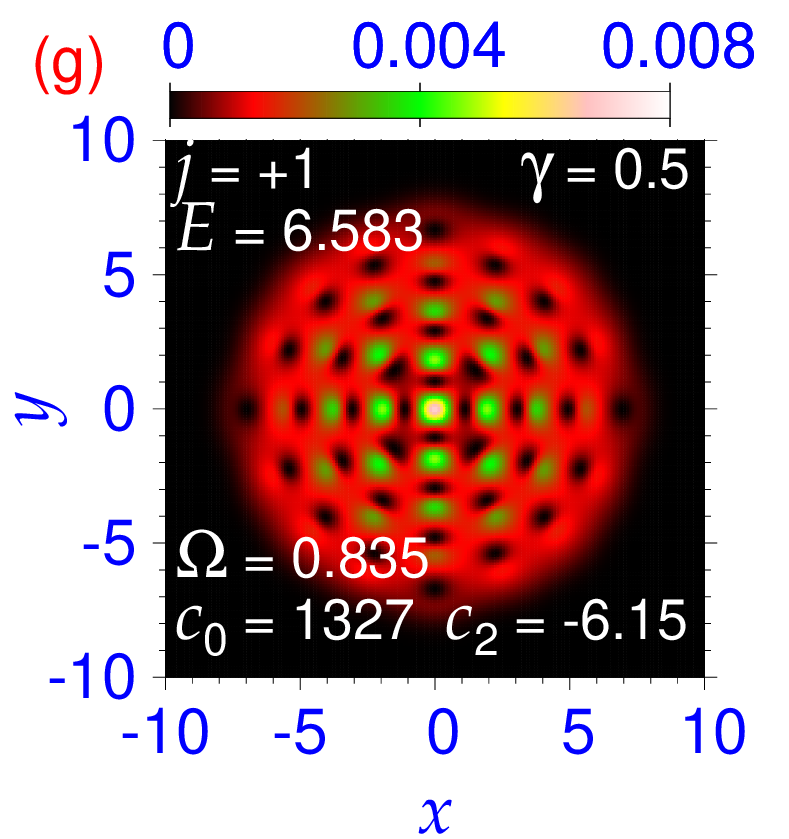}
\includegraphics[trim = 6mm 0mm 11mm 0mm,clip,width=.32\linewidth]{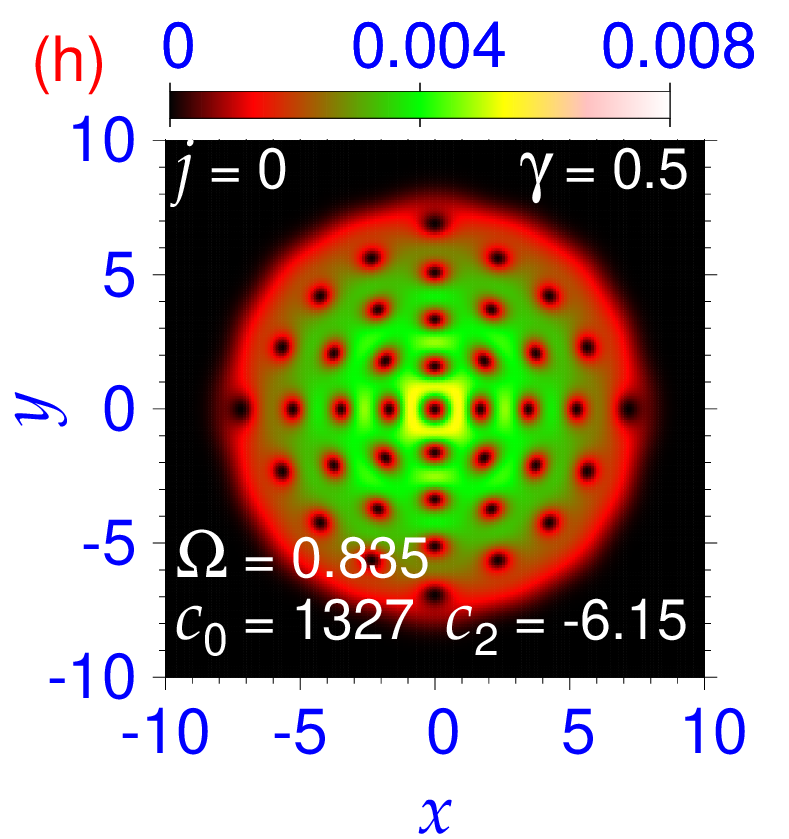}
 \includegraphics[trim = 6mm 0mm 11mm 0mm,clip,width=.32\linewidth]{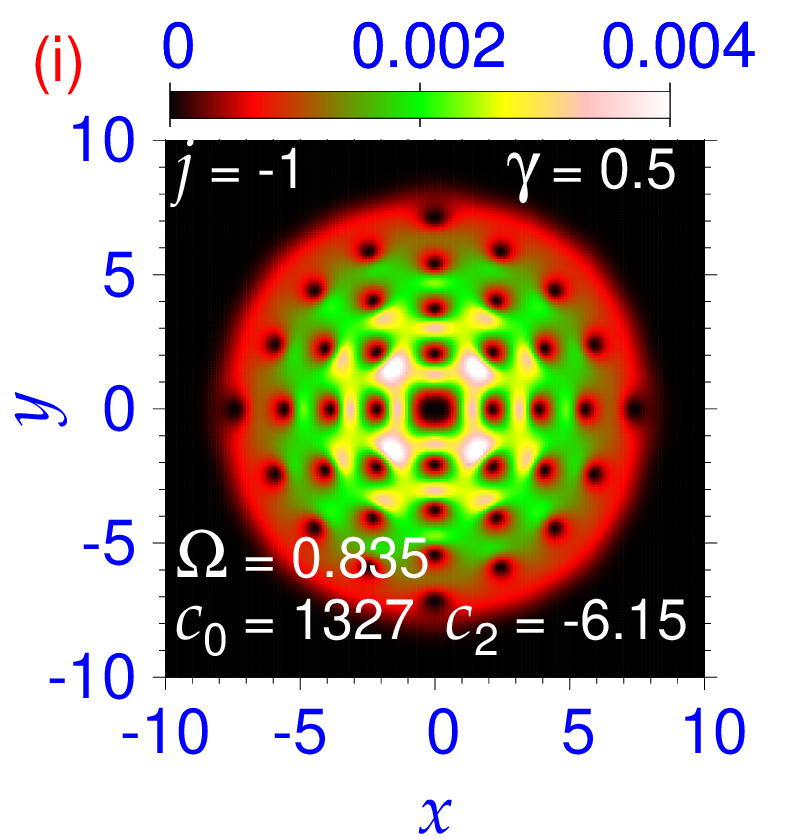}

\caption{(Color online) The same as in figure \ref{fig1} with approximate square symmetry for angular frequencies $\Omega = 0.4, 0.65,
$ and 0.835 in (a)-(c), (d)-(f), and  (g)-(i), respectively. The parameters of the ferromagnetic BEC are the same as in figure \ref{fig1}.}
\label{fig3}
\end{figure}

\begin{table}[!b]
\caption{(Color online) Energy of the different vortex-lattice  and anti-vortex-lattice 
states of hexagonal and approximate 
square symmetry  for a ferromagnetic and anti-ferromagnetic (polar) BEC
and the corresponding minimum energy state. In all cases the lattice has all concentric 
orbits with the maximum number of vortices.}
\label{tab1}
\begin{tabular}{ll|lll }
\hline
\hline
 & $\Omega$ & $E$  & $E$ & minimum     \\   
 &  & (hexagonal) & (square) & energy state  \\  
\hline
 & $+0.4$ & 12.262 &  12.207 & square \\ 
& $+0.65$ & 9.853 &  9.820& square \\  
ferro & $+0.835$ & 6.599 &  6.583  & square \\ 
  & $-0.39$ & 12.315& 12.264& square \\
 & $-0.62$ &10.203  & 10.192 & square\\
 & $-0.82$ & 6.921 &   6.915 & square\\
\hline
 & $+0.55$ &  6.751   & 6.621  &  square  \\
 & $+0.795$ & 4.539 &  4.512 &  square  \\  
 polar& $-0.55$&  6.751   & 6.715   &  square \\
  & $-0.79$&  4.598 &  4.575 & square \\
\hline
\end{tabular}
\end{table}

For the same sets of parameters as in figure \ref{fig1}, the vortex-lattice states  with an  approximate square symmetry 
are shown in figures \ref{fig3}(a)-(i) for angular frequencies $\Omega =0.4, 0.65, $ and 0.835, {
where vortices are arranged in approximate concentric  square orbits with 8, 12, and 16 vortices.}   The central spot in components $j=0$ and  $-1$  has a vortex of circulation  $+1$
and $+2$, respectively, whereas the same  in component $j=+1$ is vortex free.  In table \ref{tab1} we display the respective energies of the vortex-lattice states of figures \ref{fig1} and \ref{fig3} and find that 
the states of figure \ref{fig3} have smaller energies
as compared to the respective states in   figure \ref{fig1}.

\begin{figure}[!t]
\centering
 
\includegraphics[trim = 6mm 0mm 11mm 0mm,clip,width=.32\linewidth]{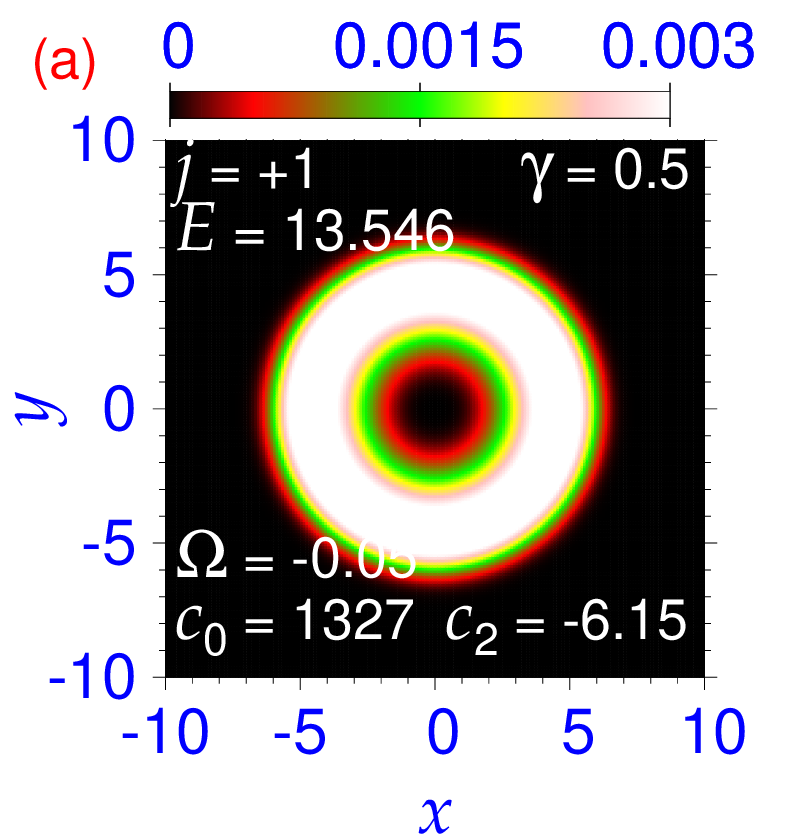}
\includegraphics[trim = 6mm 0mm 11mm 0mm,clip,width=.32\linewidth]{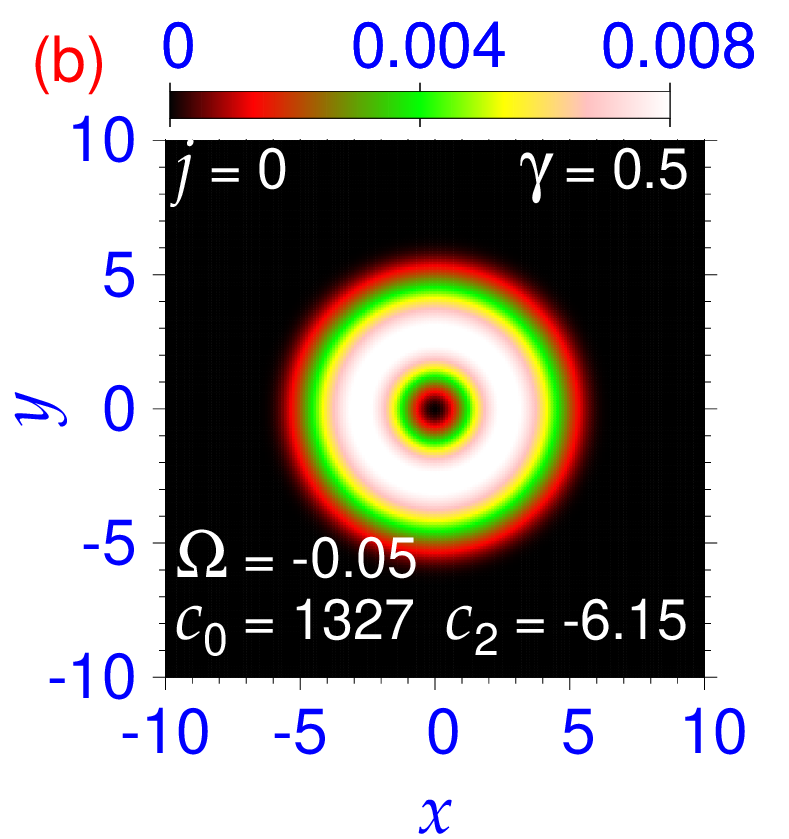}
\includegraphics[trim = 6mm 0mm 11mm 0mm,clip,width=.32\linewidth]{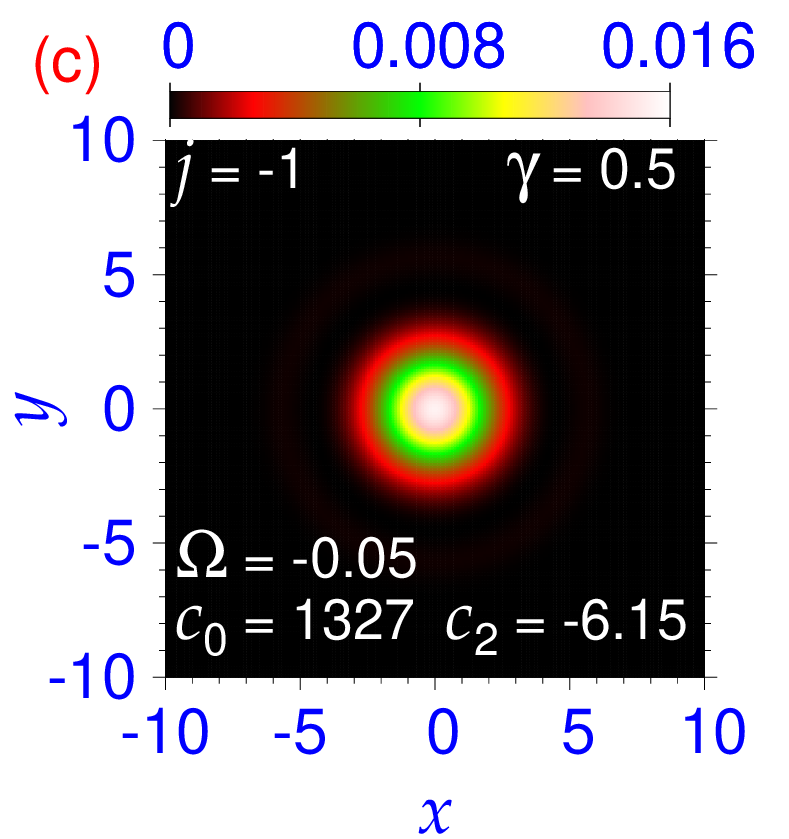}
\includegraphics[trim = 6mm 0mm 11mm 0mm,clip,width=.32\linewidth]{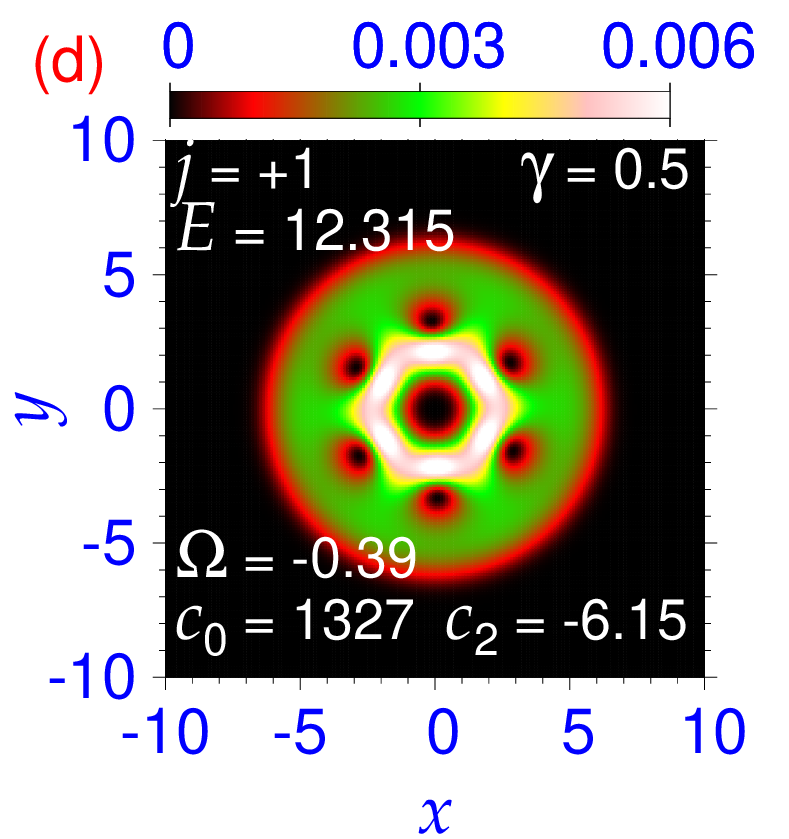} 
\includegraphics[trim = 6mm 0mm 11mm 0mm,clip,width=.32\linewidth]{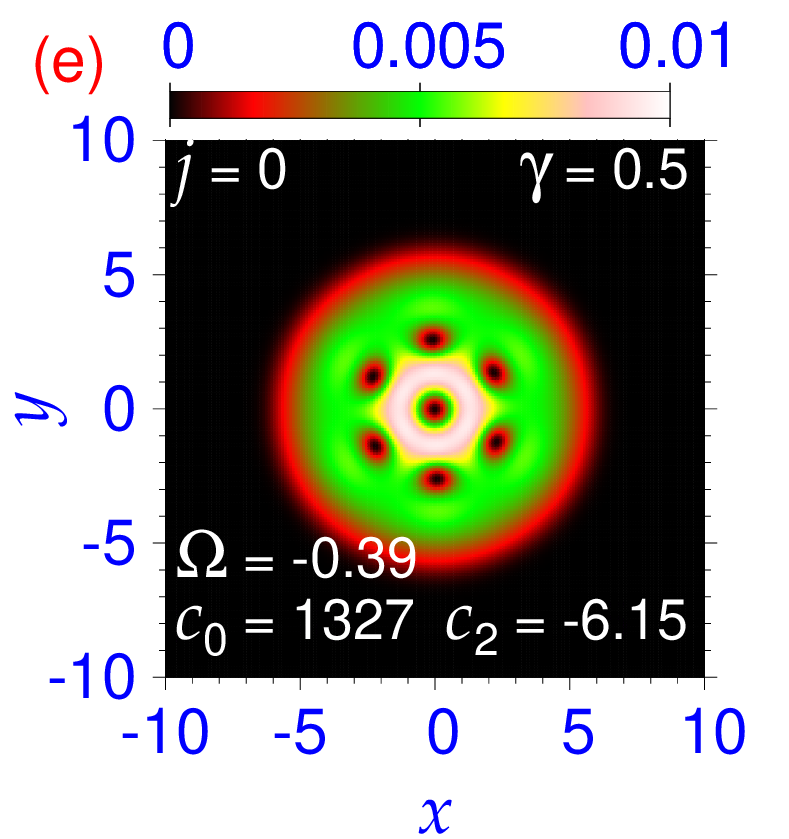}
\includegraphics[trim = 6mm 0mm 11mm 0mm,clip,width=.32\linewidth]{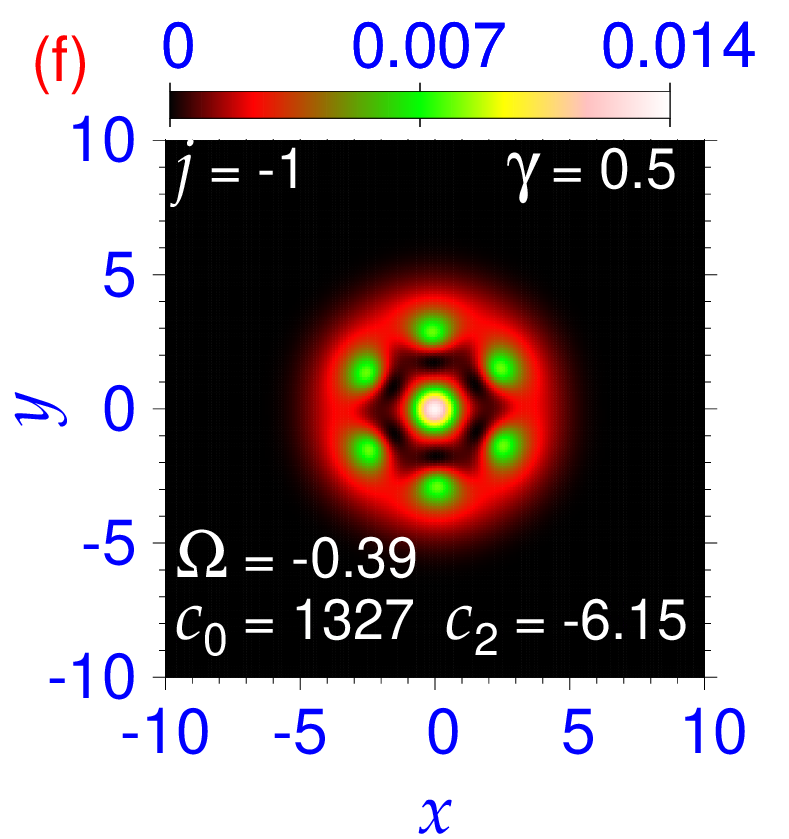} 
\includegraphics[trim = 6mm 0mm 11mm 0mm,clip,width=.32\linewidth]{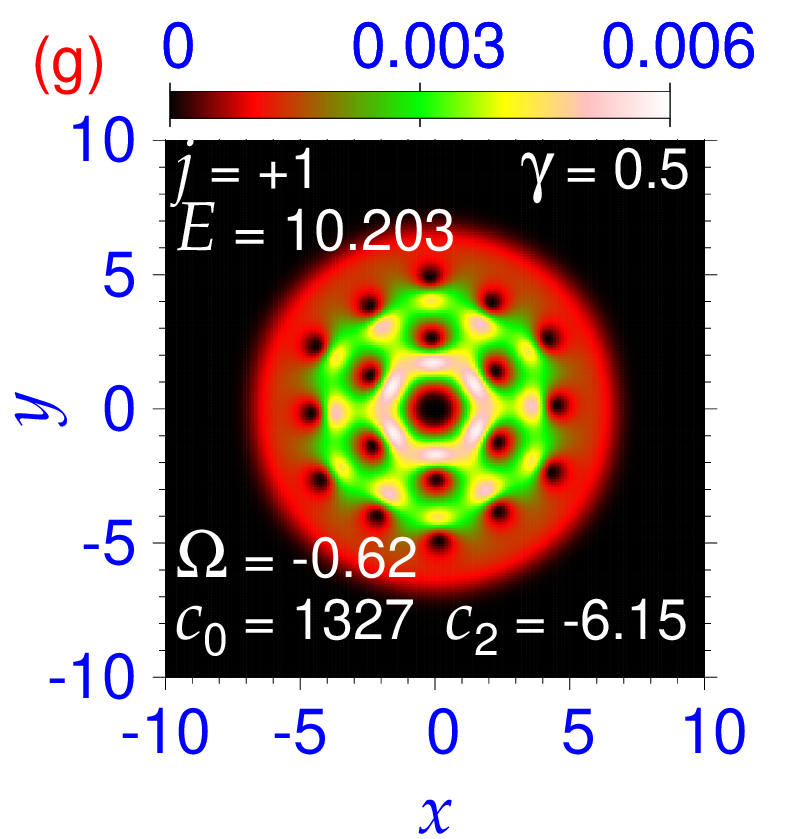} 
\includegraphics[trim = 6mm 0mm 11mm 0mm,clip,width=.32\linewidth]{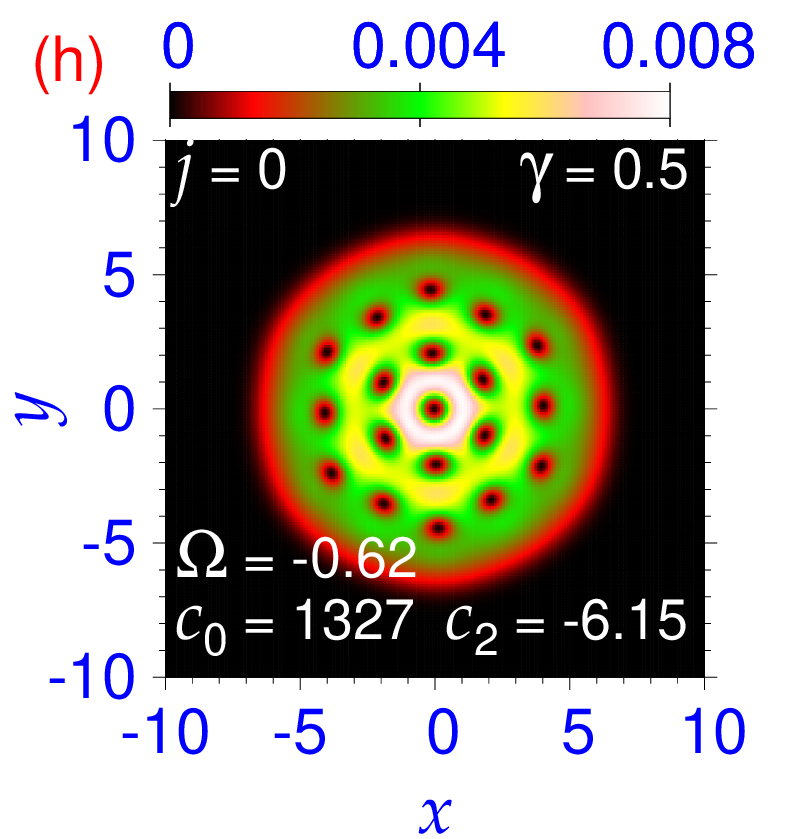}
\includegraphics[trim = 6mm 0mm 11mm 0mm,clip,width=.32\linewidth]{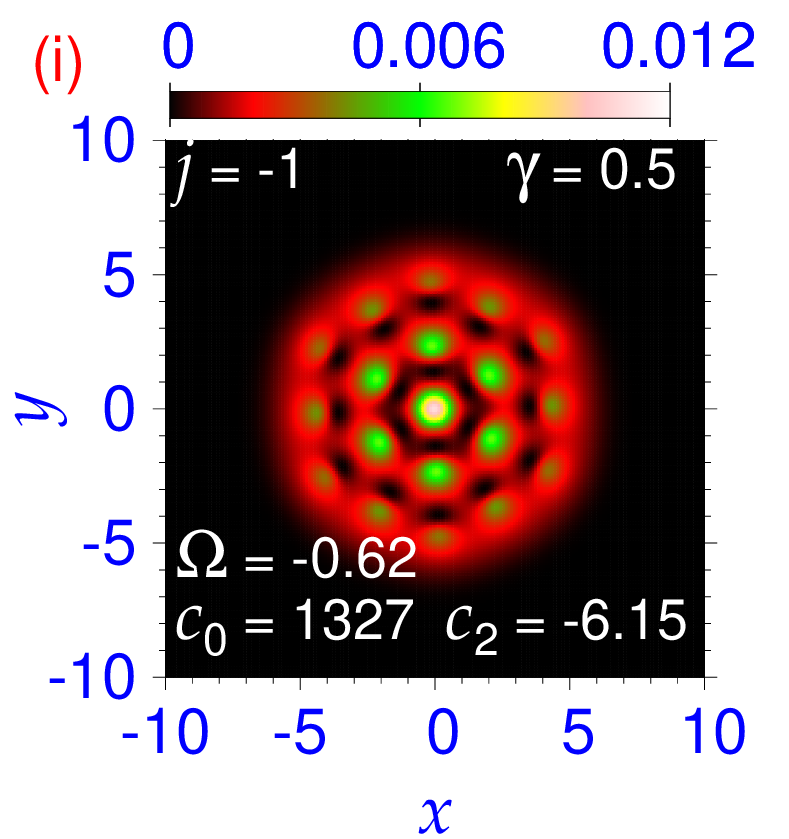} 
\includegraphics[trim = 6mm 0mm 11mm 0mm,clip,width=.32\linewidth]{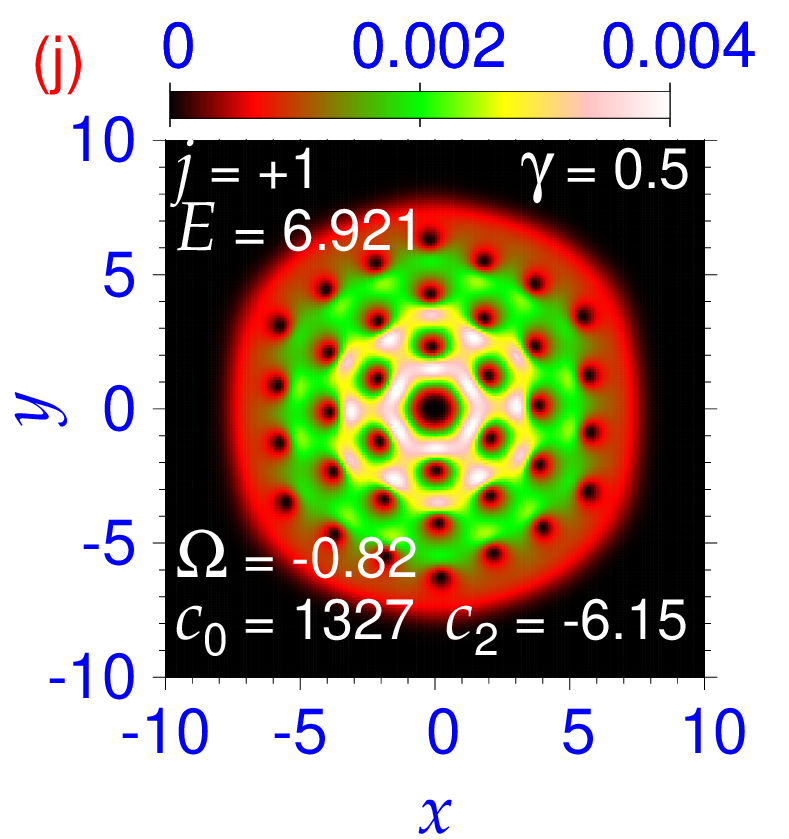} 
\includegraphics[trim = 6mm 0mm 11mm 0mm,clip,width=.32\linewidth]{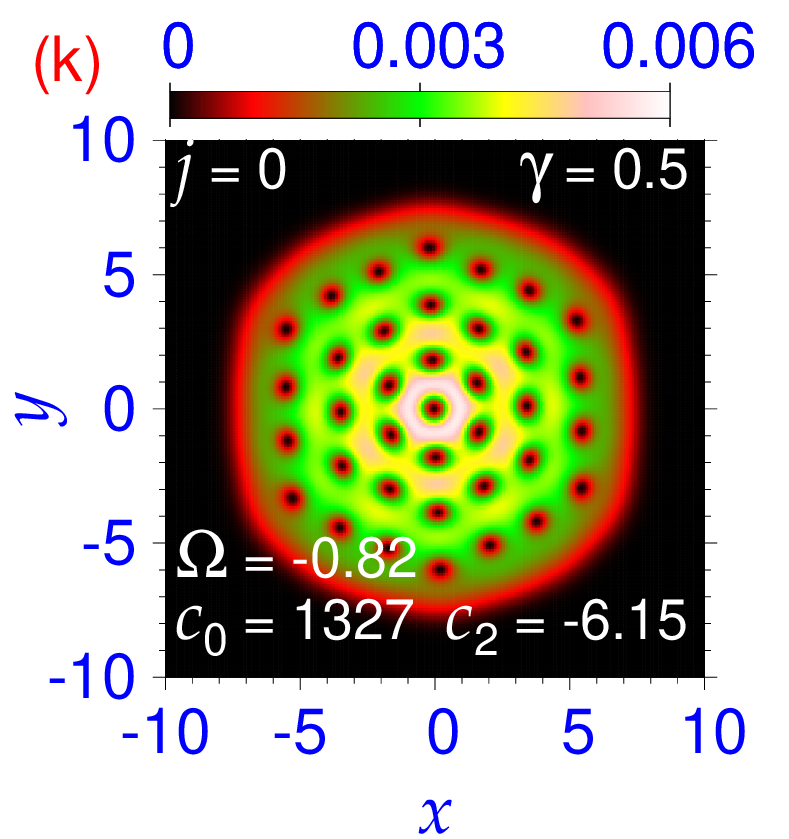}
\includegraphics[trim = 6mm 0mm 11mm 0mm,clip,width=.32\linewidth]{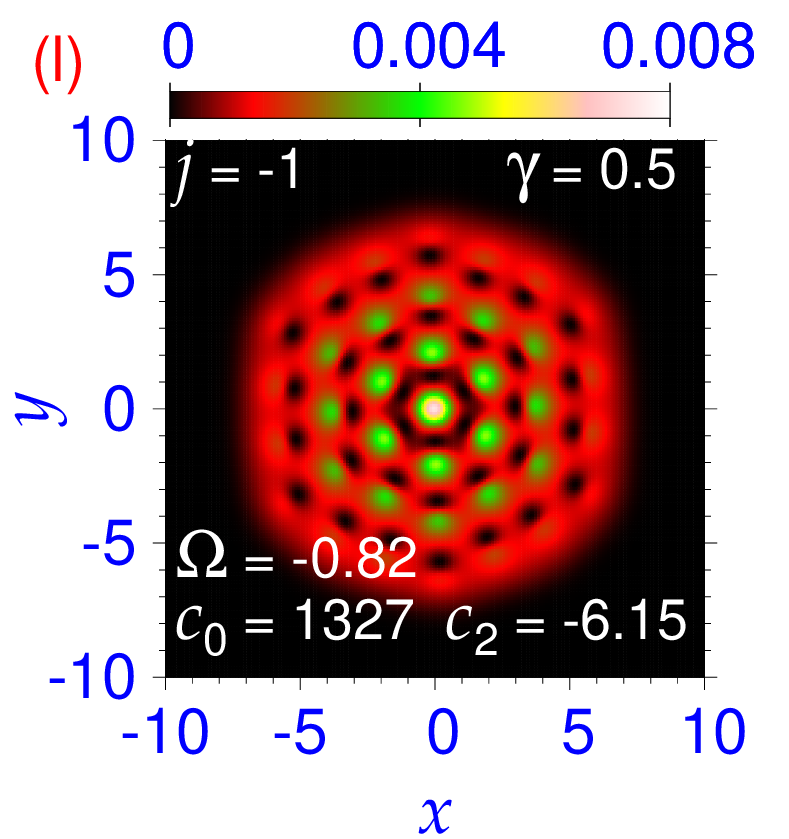} 

\caption{(Color online) The same as in figure \ref{fig1}  for angular frequencies $\Omega =-0.05, -0.39, -0.62,$ and $-0.82$ in plots (a)-(c), (d)-(f), (g)-(i), and (j)-(l), respectively.  The angular 
momentum of rotation is anti-parallel to the vorticity direction of the non-rotating state in figures \ref{fig1}(a)-(c).
The parameters of the ferromagnetic BEC are the same as in figure \ref{fig1}.
   }
\label{fig4}
\end{figure}

 \begin{figure}[!t]
\centering
\includegraphics[trim = 6mm 0mm 11mm 0mm,clip,width=.32\linewidth]{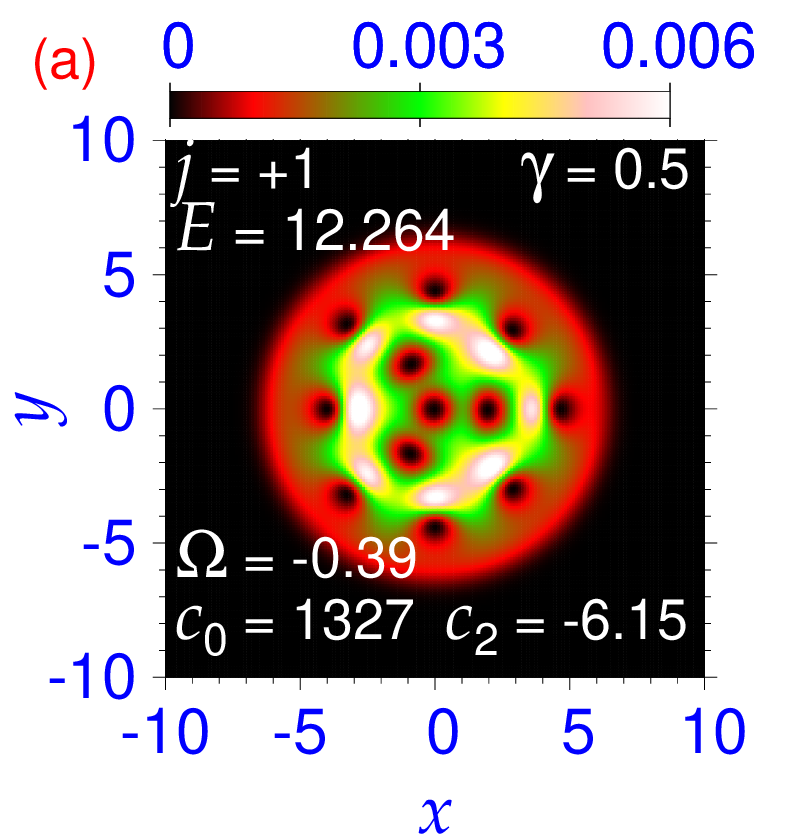}
\includegraphics[trim = 6mm 0mm 11mm 0mm,clip,width=.32\linewidth]{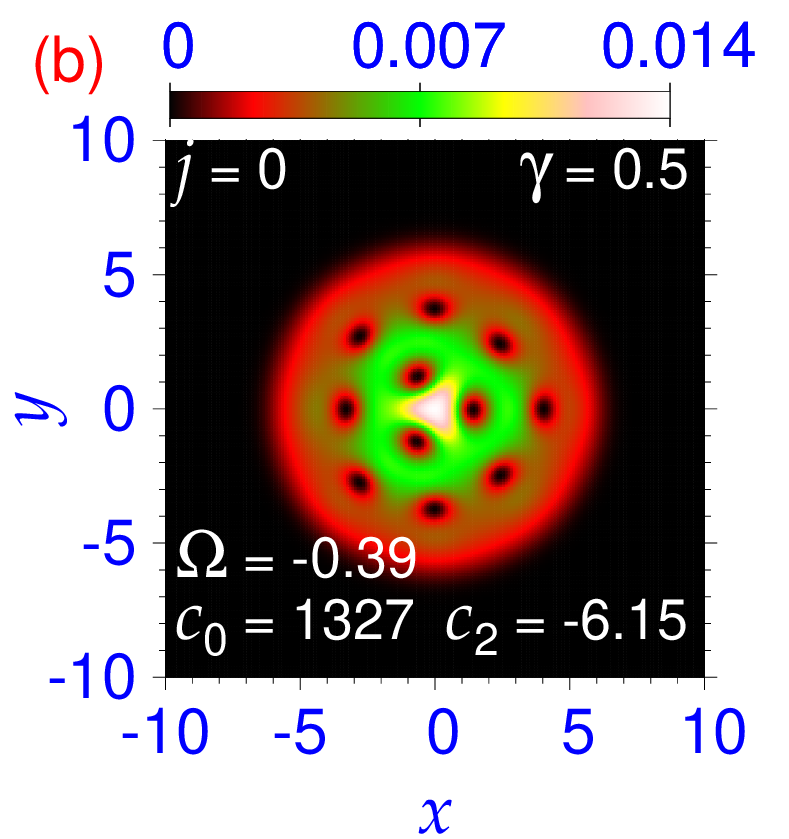}
 \includegraphics[trim = 6mm 0mm 11mm 0mm,clip,width=.32\linewidth]{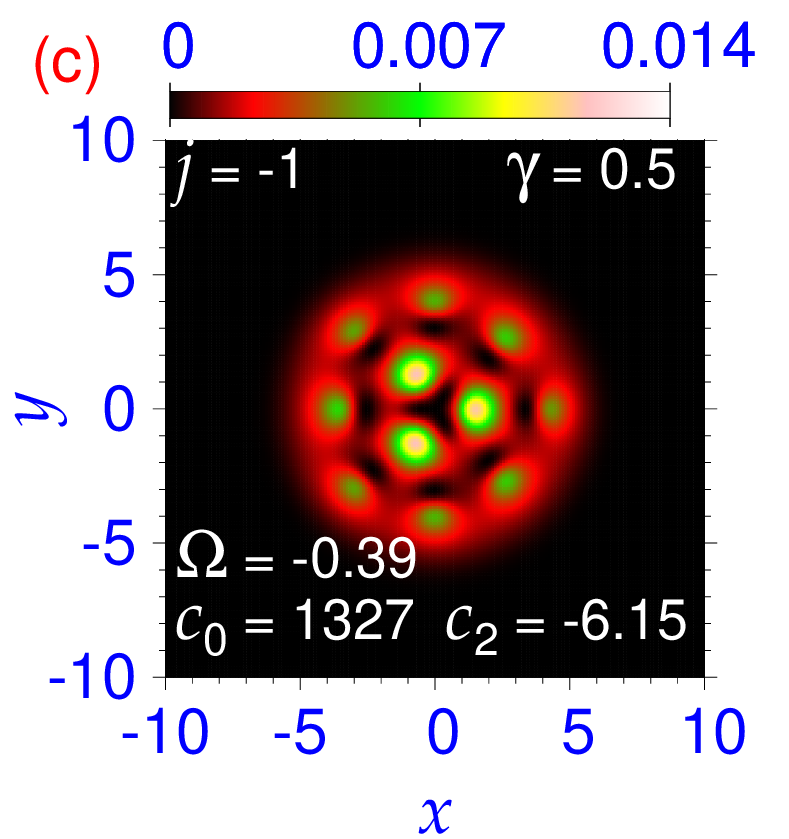}
\includegraphics[trim = 6mm 0mm 11mm 0mm,clip,width=.32\linewidth]{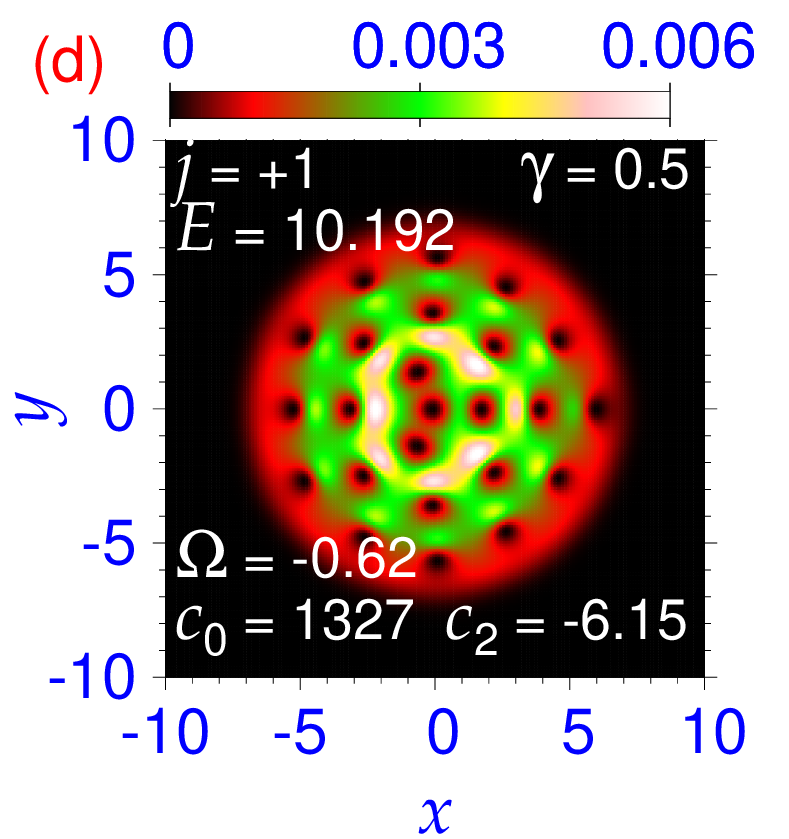}
\includegraphics[trim = 6mm 0mm 11mm 0mm,clip,width=.32\linewidth]{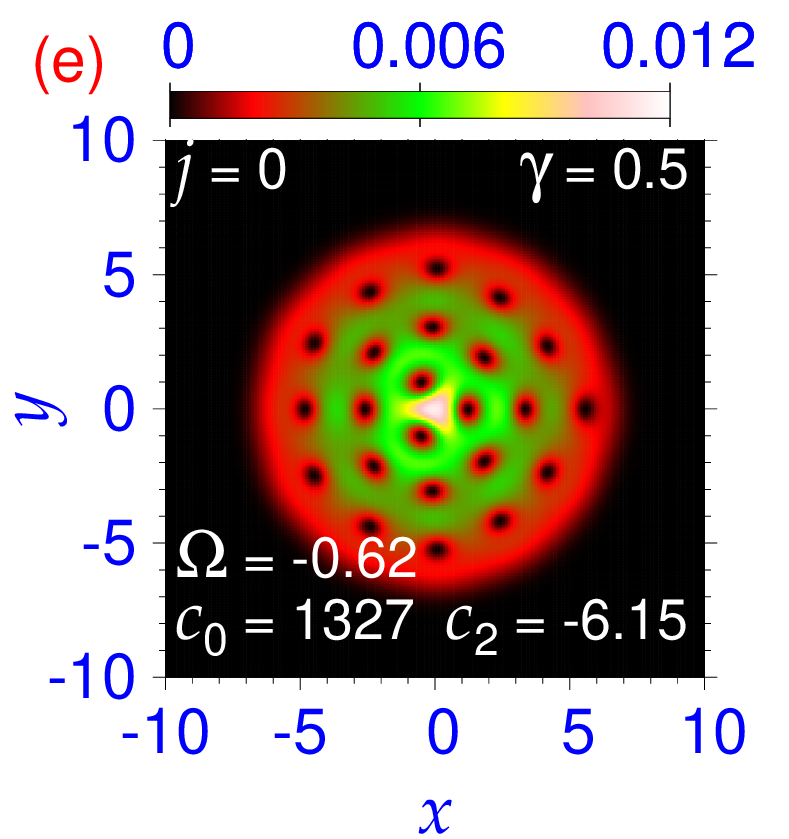}
 \includegraphics[trim = 6mm 0mm 11mm 0mm,clip,width=.32\linewidth]{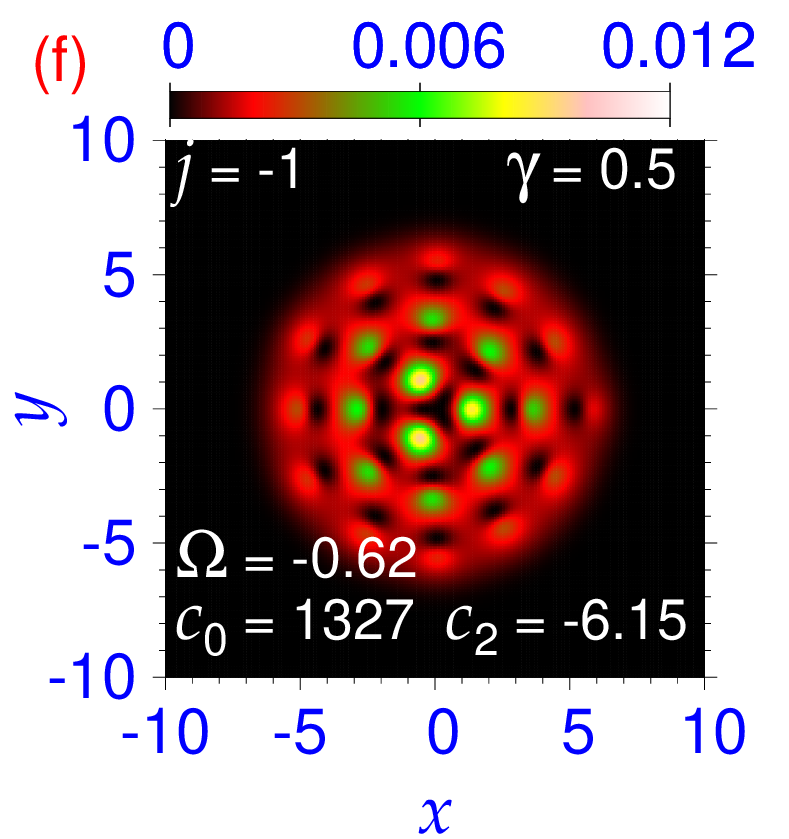}
\includegraphics[trim = 6mm 0mm 11mm 0mm,clip,width=.32\linewidth]{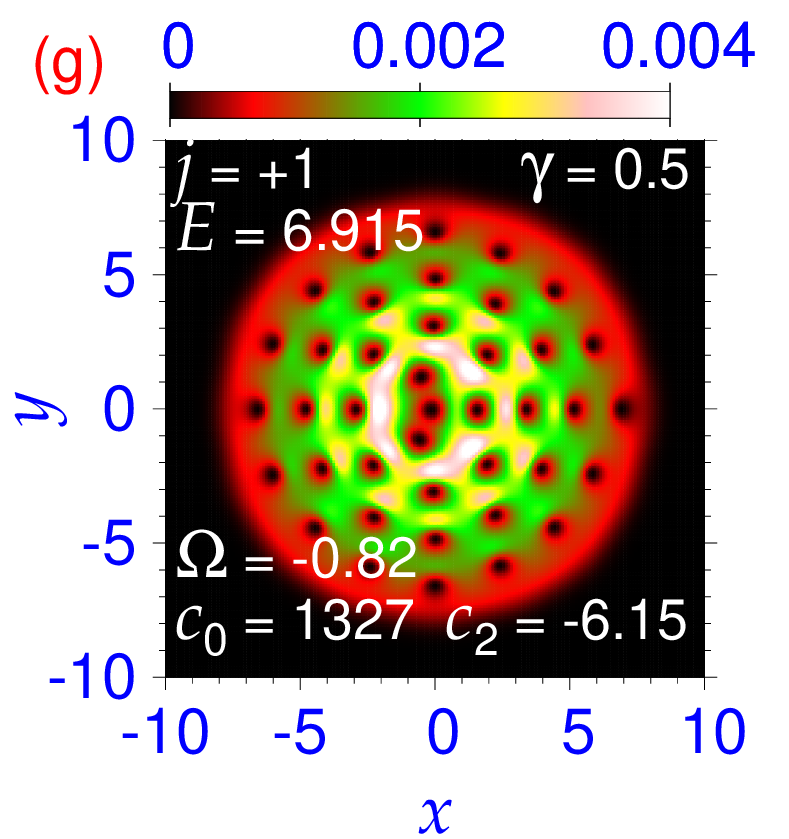}
\includegraphics[trim = 6mm 0mm 11mm 0mm,clip,width=.32\linewidth]{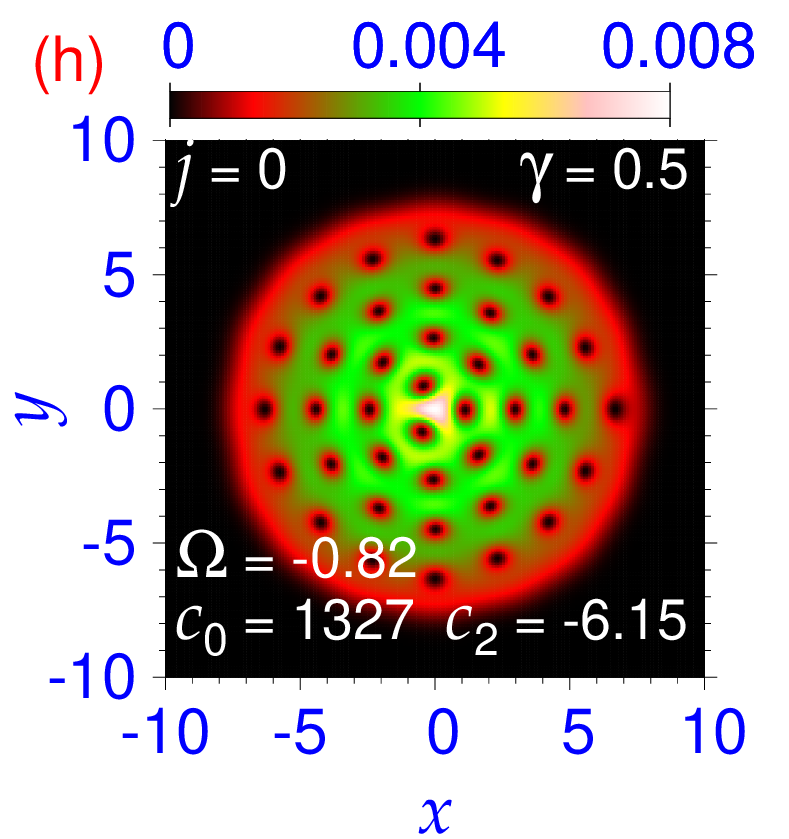}
 \includegraphics[trim = 6mm 0mm 11mm 0mm,clip,width=.32\linewidth]{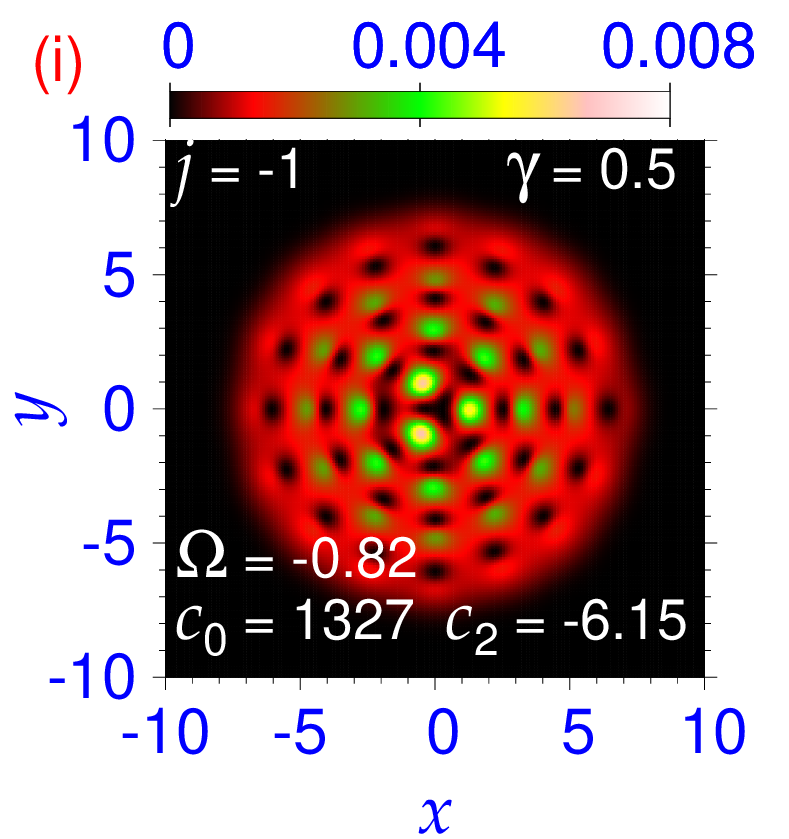}

\caption{(Color online) The same as in figure \ref{fig4} with approximate square symmetry for angular frequencies $\Omega = -0.39, -0.62,
$ and $-0.82$ in (a)-(c), (d)-(f), and  (g)-(i), respectively.  The angular 
momentum of rotation is anti-parallel to the vorticity direction of the non-rotating state in figures \ref{fig1}(a)-(c). The parameters of the ferromagnetic BEC are the same as in figure \ref{fig4}.}
\label{fig5}
\end{figure}

Next we consider the formation of vortex lattice in a quasi-2D Rashba SO-coupled ferromagnetic spin-1 spinor BEC upon rotation 
with angular momentum opposite to the intrinsic vorticity of the non-rotating BEC, denoted by negative values of 
angular frequency $\Omega$ in  (\ref{EQ1}) and  (\ref{EQ2}).  The contour plots of generated vortex lattices with hexagonal symmetry 
for angular frequencies $\Omega = -0.05,  -0.39, -0.62$ and $-0.82$ are shown in figures \ref{fig4}(a)-(c), 
(d)-(f), (g)-(i), (j)-(l), respectively. In all plots of  figure \ref{fig4} the circulation  of rotation has opposite 
sign compared to the same in figures \ref{fig1} and \ref{fig3}. Hence these vortices are anti-vortices.  Upon rotation of the state  of type $(0,+1,+2)$ of figures \ref{fig1}(a)-(c) 
with angular frequency $\Omega=-0.05$, with opposite vorticity, a state of type $(-2,-1,0)$ is generated as shown in 
figures \ref{fig4}(a)-(c).     The opposite vorticity 
of the vortices  in figures \ref{fig4}(a)-(c), as compared to those in figures \ref{fig1}(a)-(c),  was confirmed  from a plot of 
the corresponding phase  in figures \ref{fig2}(g)-(i). 
The phase drop upon a clockwise rotation of $2\pi$ in figures \ref{fig2}(g) (h) is $-4\pi$  ($-2\pi$) indicating an anti-vortex of
circulation  $-2$ ($-1$), viz. compare with vortices in figures \ref{fig2}(b)-(c). 
 As angular frequency $|\Omega|$ is increased,  an anti-vortex lattice   with hexagonal symmetry  is generated in the three components maintaining 
the anti-vortices  $(-2,-1,0)$   at the center, viz. figures  \ref{fig4}(d)-(f), \ref{fig4}(g)-(i), and \ref{fig4}(j)-(l), for $\Omega =-0.39, -0.62,$ and $-0.82$, respectively.  


 \begin{figure}[!t]
\centering
\includegraphics[trim = 6mm 0mm 11mm 0mm,clip,width=.32\linewidth]{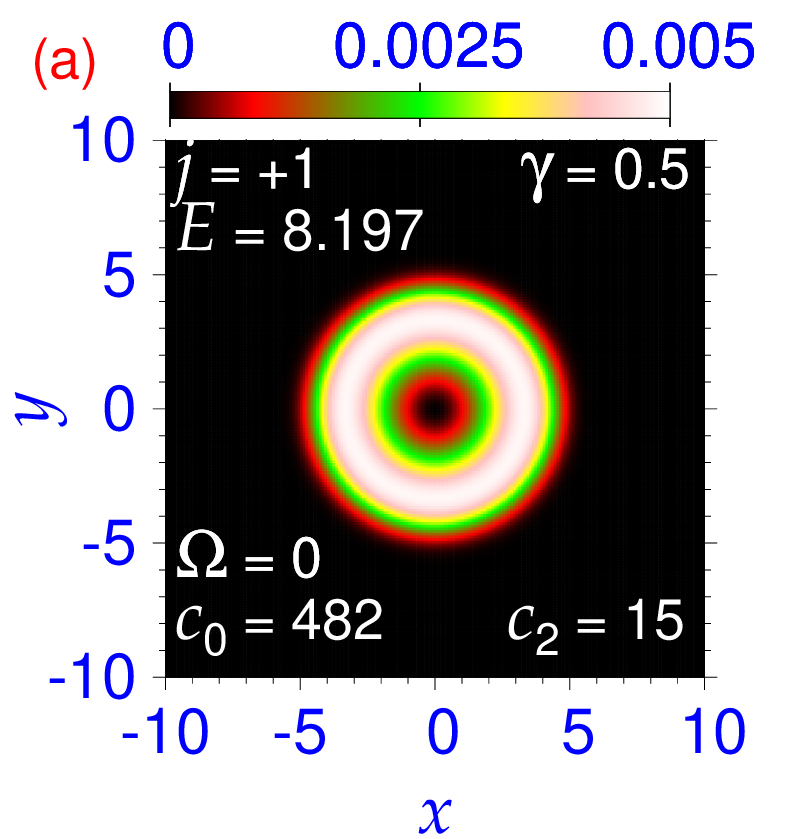}
\includegraphics[trim = 6mm 0mm 11mm 0mm,clip,width=.32\linewidth]{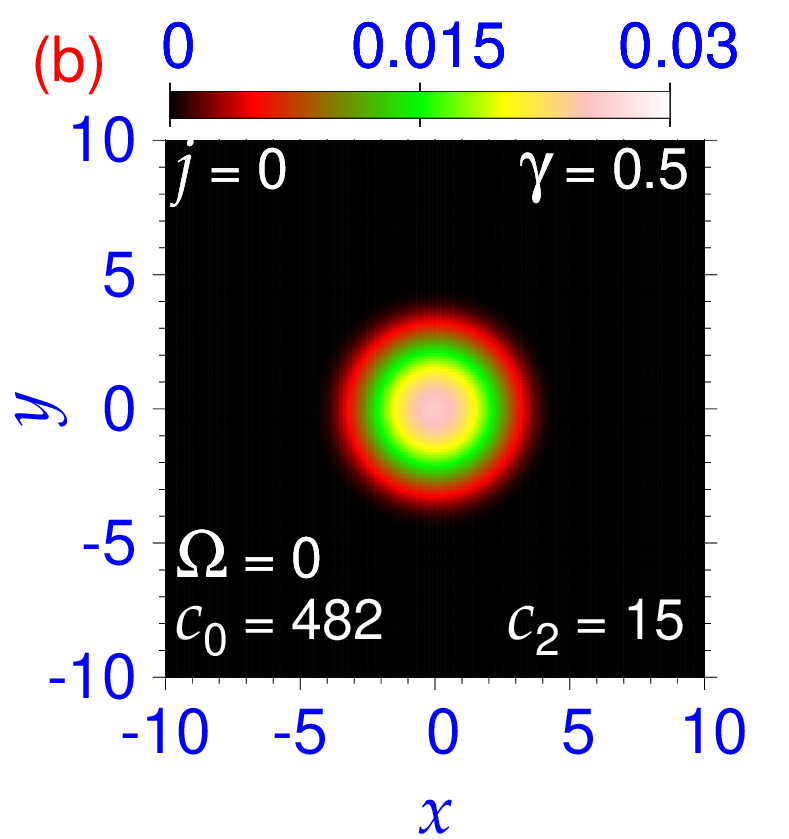}
 \includegraphics[trim = 6mm 0mm 11mm 0mm,clip,width=.32\linewidth]{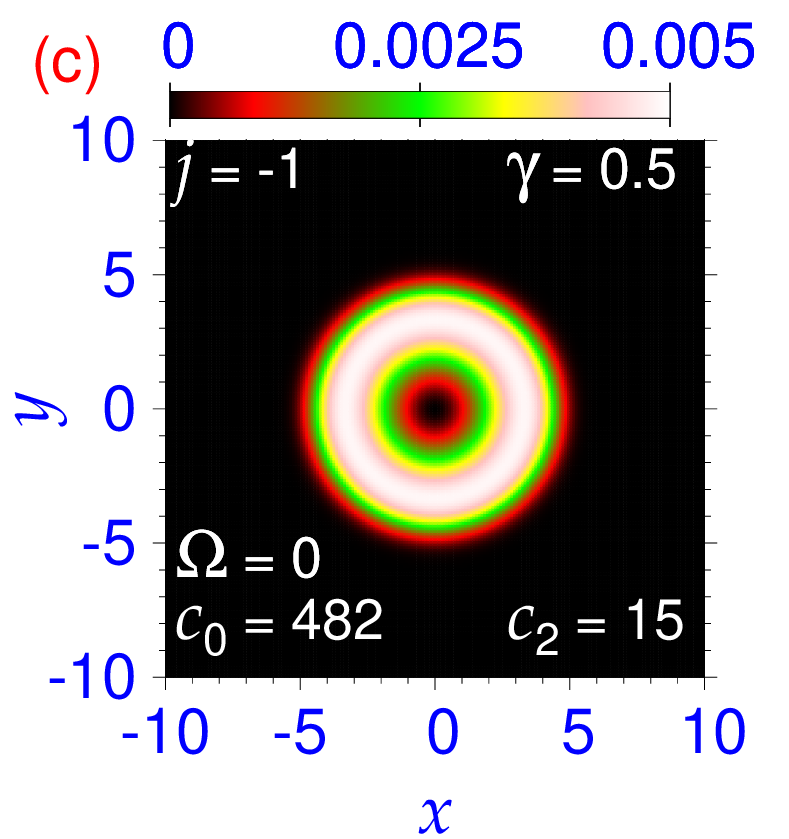}
\includegraphics[trim = 6mm 0mm 11mm 0mm,clip,width=.32\linewidth]{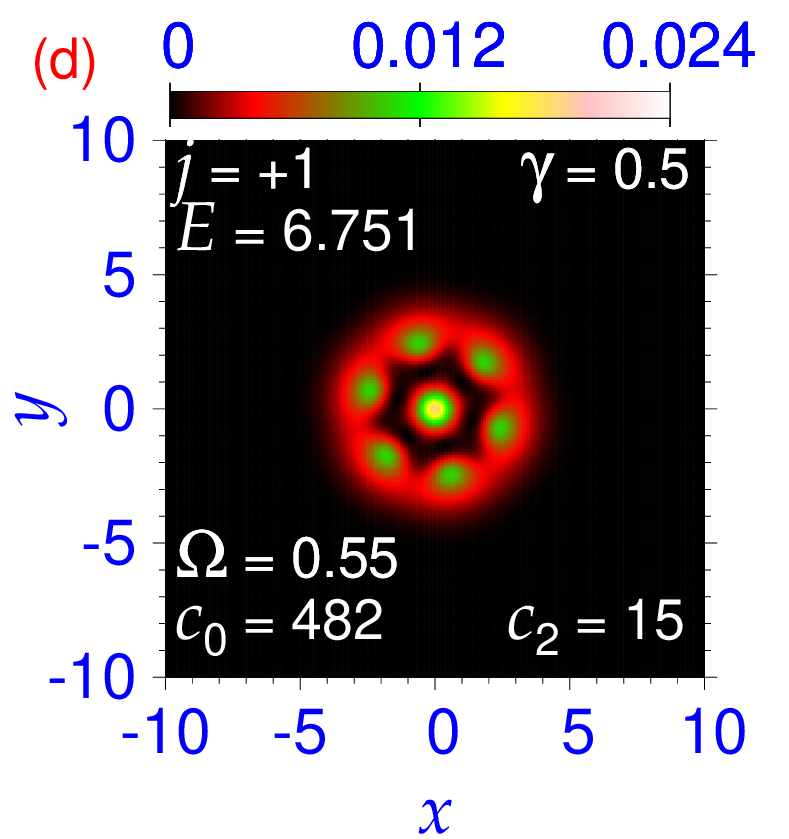}
\includegraphics[trim = 6mm 0mm 11mm 0mm,clip,width=.32\linewidth]{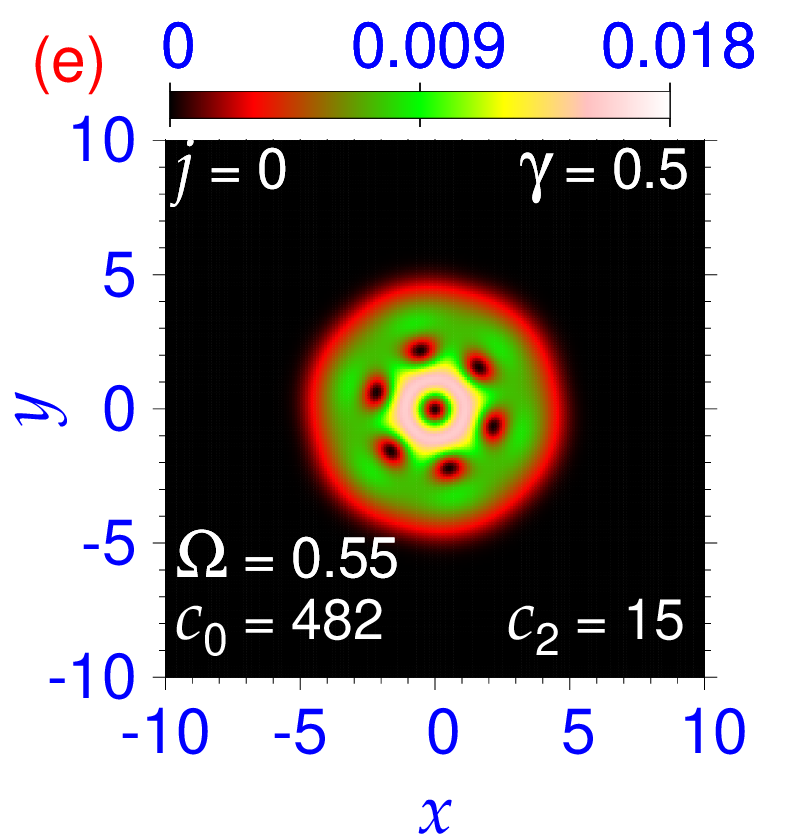}
 \includegraphics[trim = 6mm 0mm 11mm 0mm,clip,width=.32\linewidth]{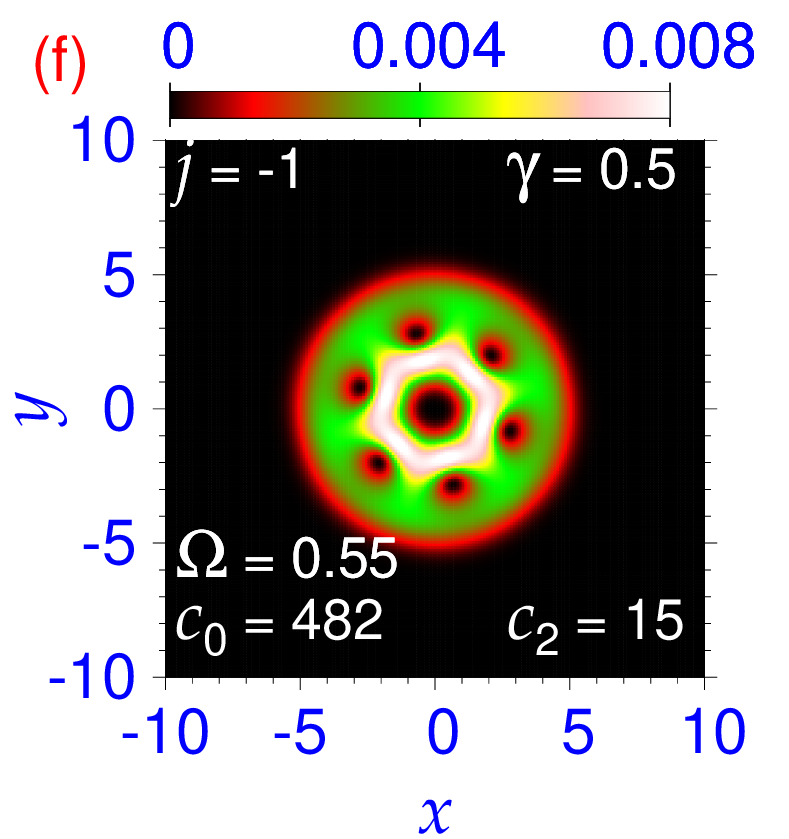}
\includegraphics[trim = 6mm 0mm 11mm 0mm,clip,width=.32\linewidth]{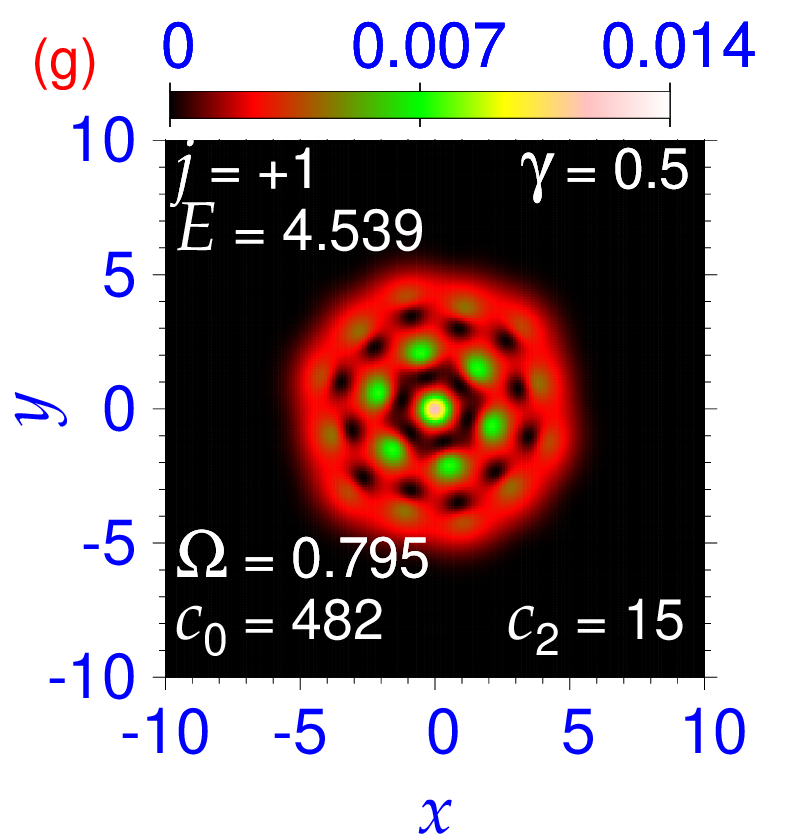}
\includegraphics[trim = 6mm 0mm 11mm 0mm,clip,width=.32\linewidth]{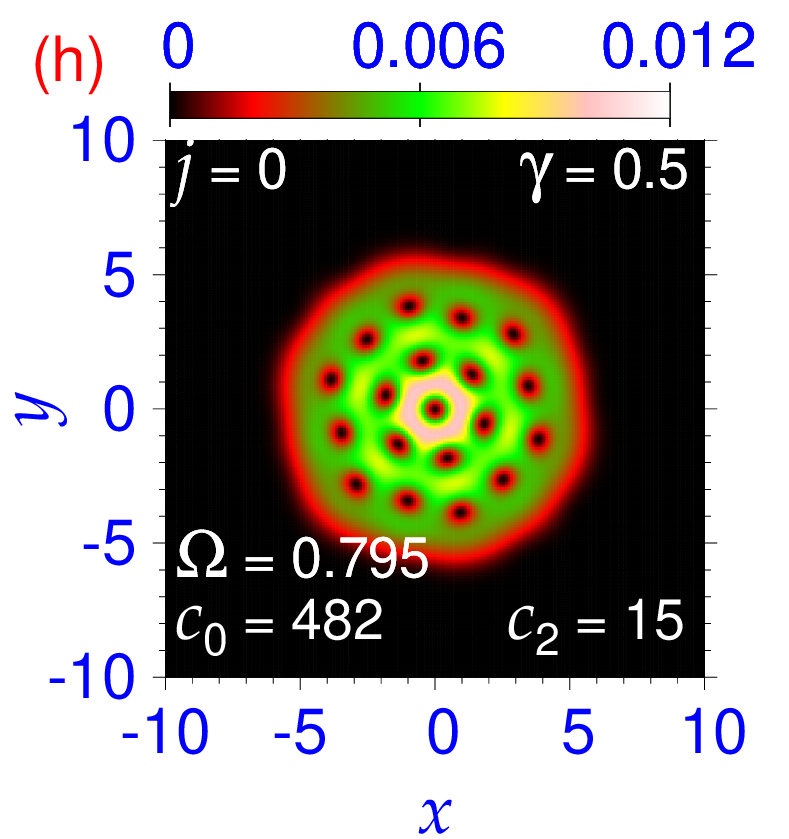}
 \includegraphics[trim = 6mm 0mm 11mm 0mm,clip,width=.32\linewidth]{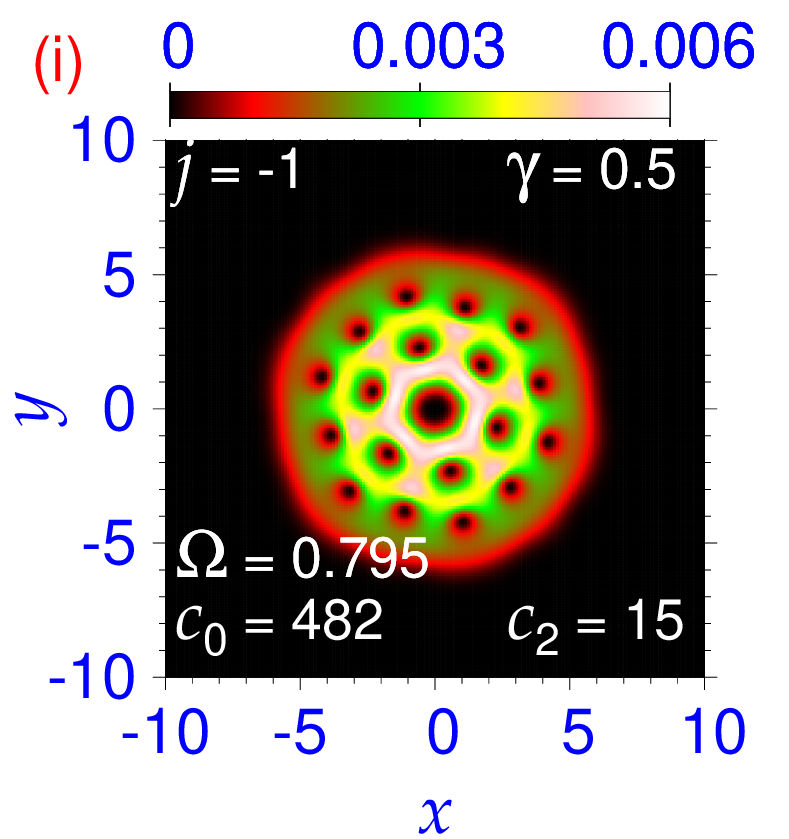}

\caption{(Color online) Contour plot of component densities $n_j(\boldsymbol \rho)\equiv | \psi _j(\boldsymbol \rho)|^2$
of vortex-lattice states of a rotating  Rashba SO-coupled anti-ferromagnetic spin-1 quasi-2D spinor BEC with hexagonal symmetry   for angular 
frequencies $\Omega =0, 0.55,$ and $ 0.795,$  in plots (a)-(c), (d)-(f), and (g)-(i), respectively.  The angular 
momentum of rotation is parallel to the vorticity direction of the vortex in component $j=-1$ in (c). The non-linearity parameters $c_0=482, c_2=15$,  and SO-coupling strength $\gamma =0.5$. }
\label{fig6}
\end{figure}

 Apart from the anti-vortex lattice with hexagonal symmetry of figure \ref{fig4}, one can  also have the same with approximate square 
symmetry for the same angular frequencies $\Omega= -0.39, -0.62$ and $-0.82$ as shown in figures \ref{fig5}(a)-(c), 
(d)-(f) and (g)-(i), respectively. The central part in figure \ref{fig5} has 4 and 3 anti-vortices of unit circulation in components $j=+1$ and 0, 
and an anti-vortex of circulation $-2$ in component $j=-1$.  
The net circulation of the anti-vortex at the center of component $j =-1$ was obtained from a consideration of the phase plot of the wave function for  $\Omega= -0.39$  as 
displayed in figures \ref{fig2}(j)-(l).
The phase drop upon a
clockwise rotation of $2\pi$ in figure  \ref{fig2}(l) is  $-4\pi$ 
indicating a circulation of $-2$  in  component  $j=-1$. 
The other anti-vortices of unit circulation in figures \ref{fig5}(a)-(c) can be identified in the phase plots in figures \ref{fig2}(j)-(l).
This central part is a superposition of the non-rotating state $(0,+1,+2)$ and the state $(-4,-4,-4)$ generated by rotation.
 The energies of the anti-vortex-lattice states with hexagonal and square symmetries of figure \ref{fig4} and \ref{fig5} are shown in table \ref{tab1} for angular frequencies
$\Omega = -0.39, -0.62$ and $-0.82$.

\begin{figure}[!t] 
\centering
\includegraphics[width=.32\linewidth]{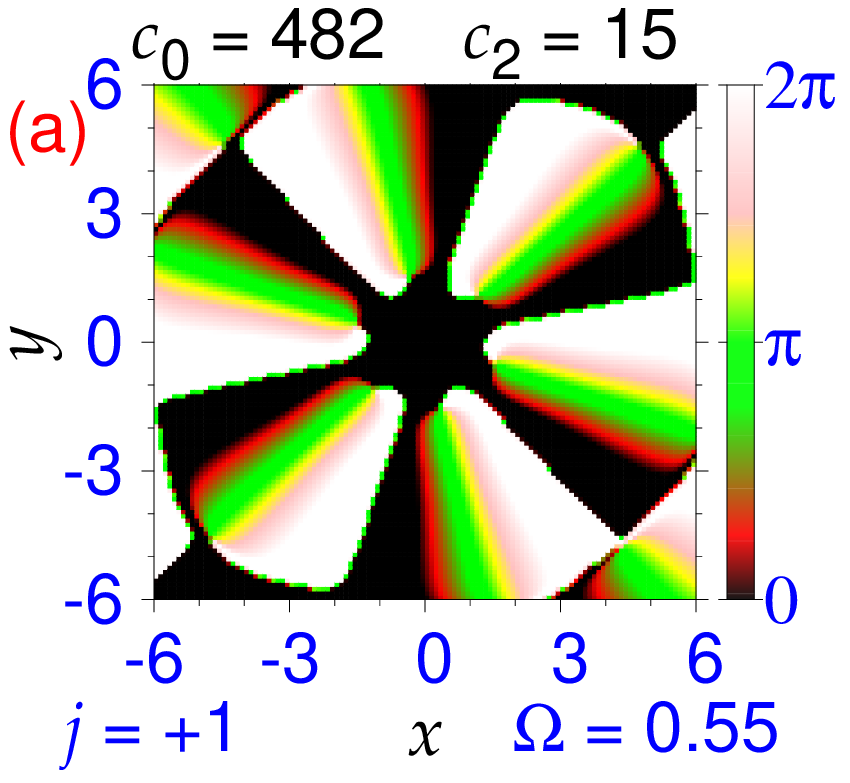} 
\includegraphics[width=.32\linewidth]{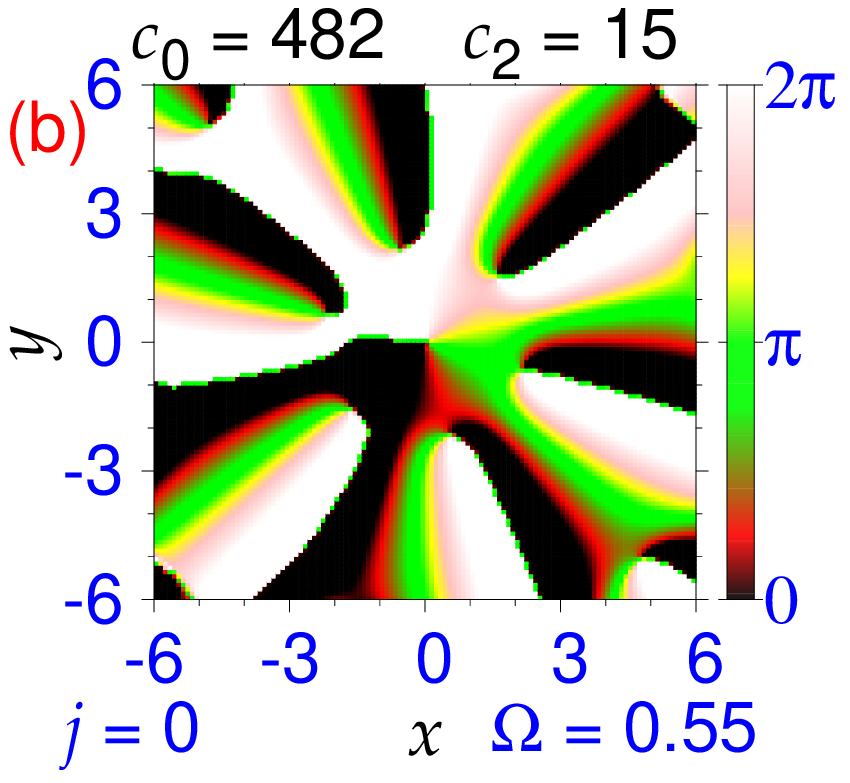}
\includegraphics[width=.32\linewidth]{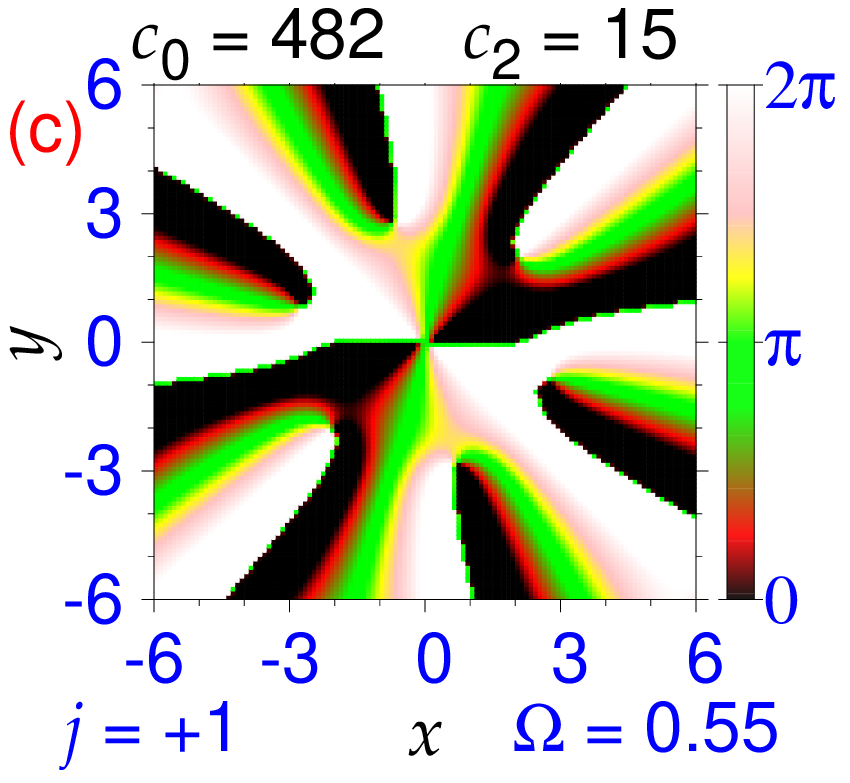} 
\includegraphics[width=.32\linewidth]{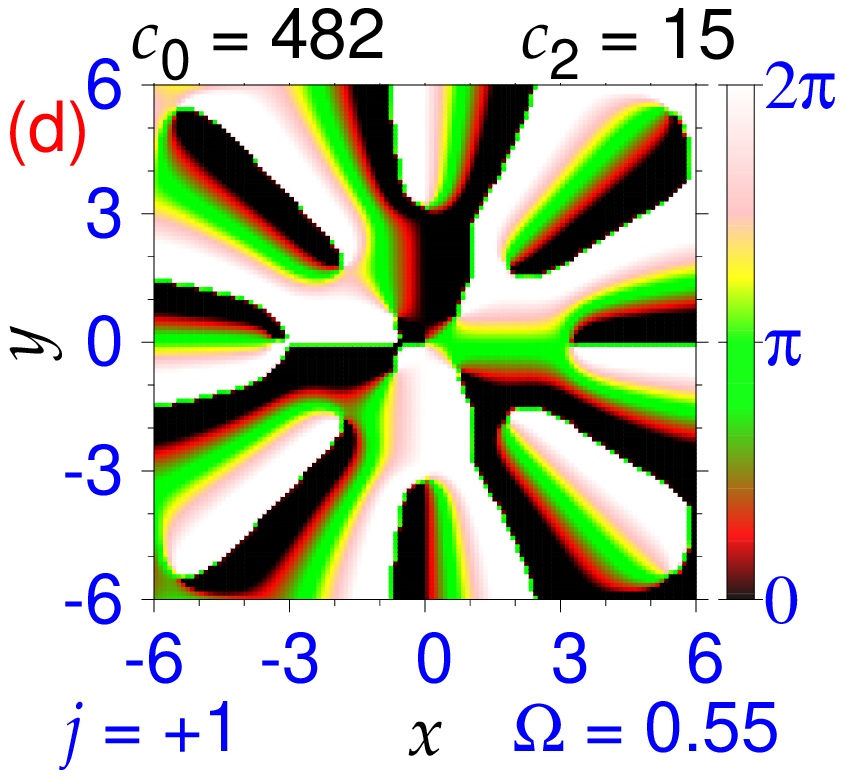} 
\includegraphics[width=.32\linewidth]{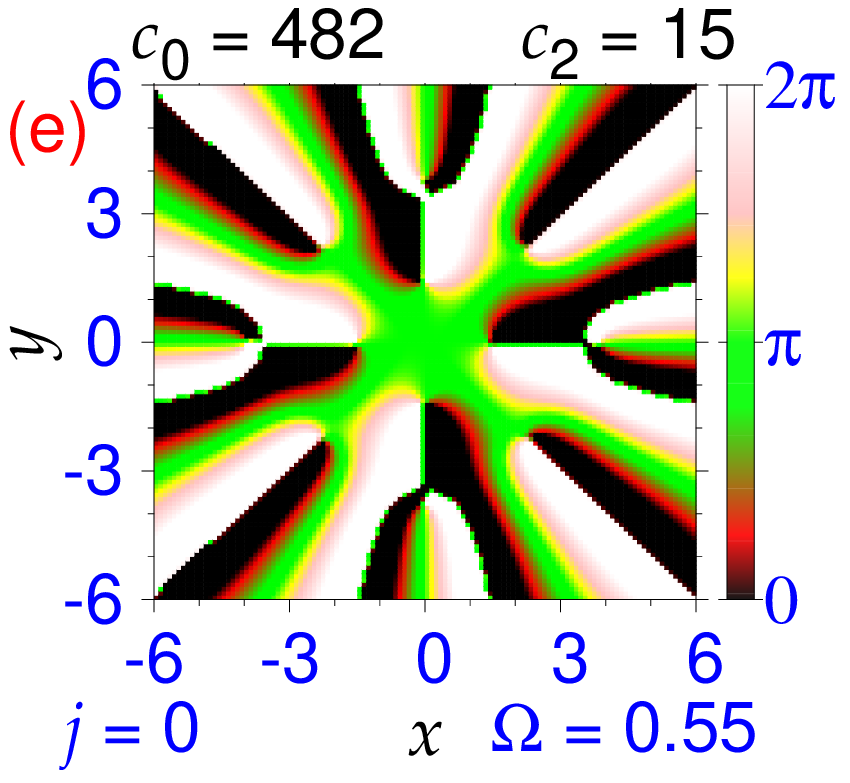}
\includegraphics[width=.32\linewidth]{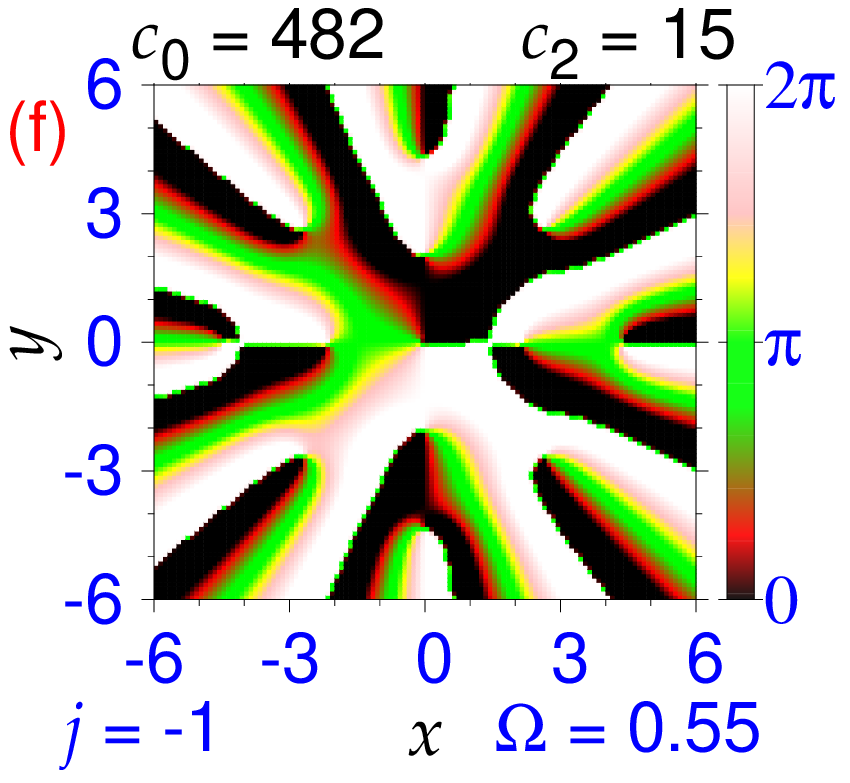}
 \includegraphics[width=.32\linewidth]{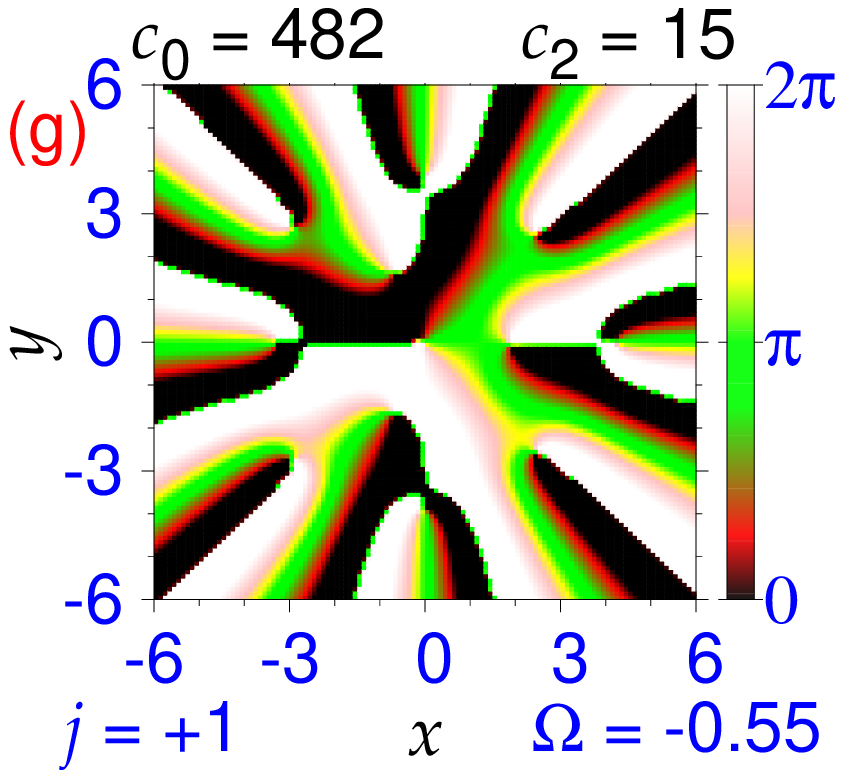} 
\includegraphics[width=.32\linewidth]{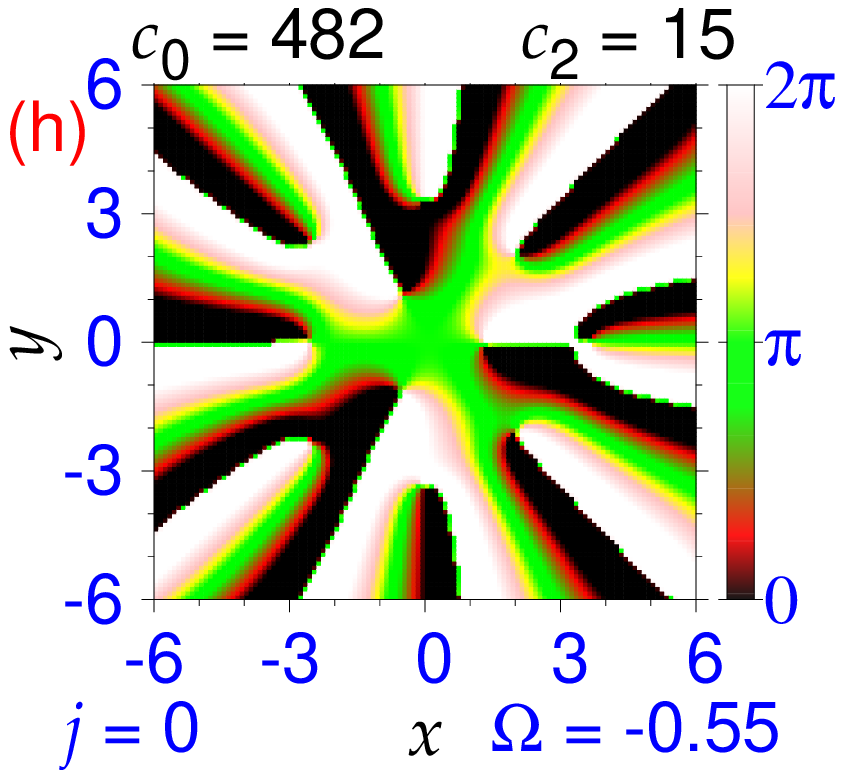}
\includegraphics[width=.32\linewidth]{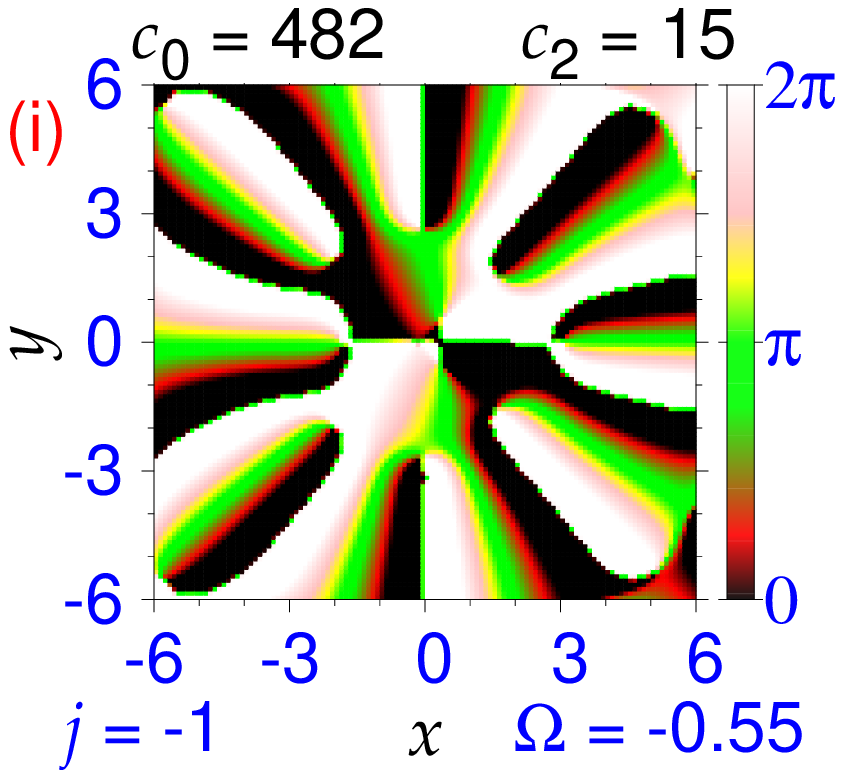} 
 
\caption{(Color online) (a)-(c) Contour plot  of the phase $\delta({\boldsymbol \rho})$ of the wave function
 of the rotating anti-ferromagnetic spinor BEC, with angular frequency $\Omega=0.55$,
 of  figures \ref{fig6}(d)-(f).
 (d)-(f) The same of the rotating anti-ferromagnetic spinor BEC, with angular frequency $\Omega=0.55$,
of  figures \ref{fig8}(a)-(c). (g)-(i) The same of the rotating anti-ferromagnetic spinor BEC, with angular frequency $\Omega=-0.55$,
 of  figures \ref{fig10}(a)-(c).  
 }
\label{fig7}

\end{figure}

\subsection{Anti-ferromagnetic condensate}

The vortex-lattice formation with hexagonal symmetry
in an anti-ferromagnetic rotating  Rashba SO-coupled quasi-2D  $^{23}$Na  spin-1  BEC 
with SO-coupling strength  $\gamma=0.5$,  and non-linearities $c_0=482$ and $c_2 =15$ is demonstrated in figure \ref{fig6}.
 The non-rotating state ($\Omega=0$), in this case, is of the $(-1,0,+1)$ type as shown in figures \ref{fig6}(a)-(c) through a contour plot of densities. The vortex and anti-vortex nature of the two states is  confirmed from the corresponding phase plot of the wave function (not shown here). 
 Upon rotation, the   $(-1,0,+1)$-type state, with the appearance of a vortex of circulation $+1$ in all components, transforms into a state of the  $(0,+1,+2)$ type.  These vortices of circulation $+1$ and $+2$ at the center of  components $j=0$ and $-1$ are maintained in the vortex lattice  with hexagonal symmetry of a rapidly rotating anti-ferromagnetic quasi-2D spin-1 spinor BEC as in the case of a  rapidly rotating ferromagnetic  spin-1 spinor BEC considered in figure \ref{fig1}. We display the formation of vortex lattice with hexagonal symmetry in the anti-ferromagnetic spinor BEC
  for angular frequencies  $\Omega = 0.55,$ and $ 0.795$  in figures \ref{fig6}(d)-(f), (g)-(i),  respectively. 
We checked the vorticity and circulation of
the components analyzing the phase plot of the wave
function of the BEC
displayed in figures \ref{fig6}(d)-(f),  viz.  Figs. \ref{fig7}(a)-(c). The phase drop
upon a clockwise rotation of $2\pi$ 
in figure \ref{fig7}(b)  (c) is $2\pi$
($4\pi$) indicating circulation $+1 (+2)$  at the center. The $j =-1$ component 
with circulation $+2$ has a larger vortex core than the
$j = 0$ component with circulation $+1$.
The hexagonal vortex lattices in figure \ref{fig6} for an  anti-ferromagnetic  quasi-2D spin-1 spinor BEC are quite similar to the vortex lattices in figure \ref{fig1} for a ferromagnetic spinor BEC.
 
\begin{figure}[!t]
\centering
\includegraphics[trim = 6mm 0mm 11mm 0mm,clip,width=.32\linewidth]{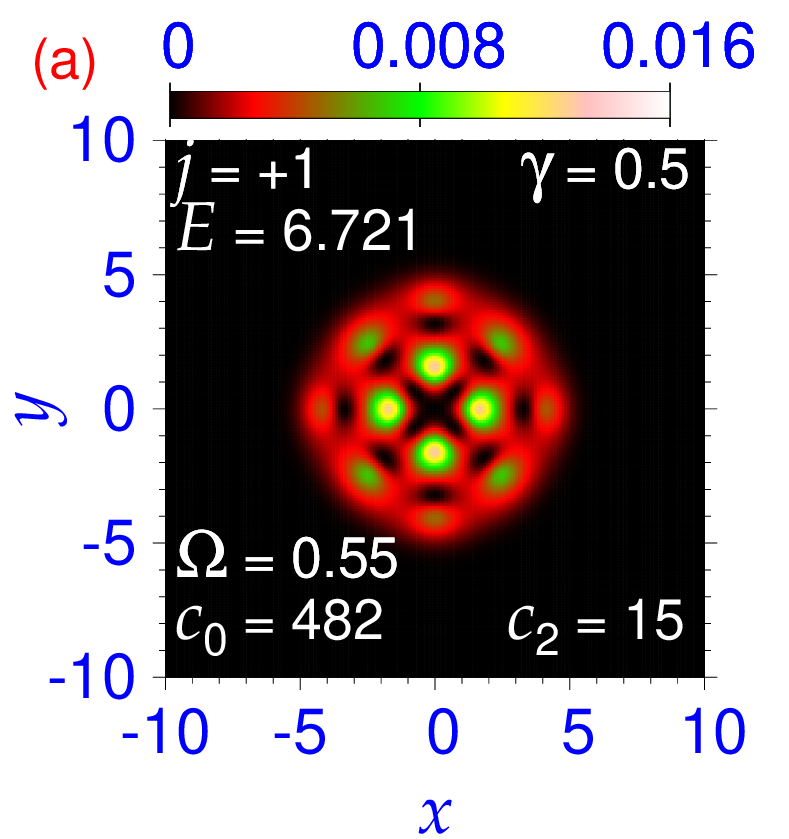}
\includegraphics[trim = 6mm 0mm 11mm 0mm,clip,width=.32\linewidth]{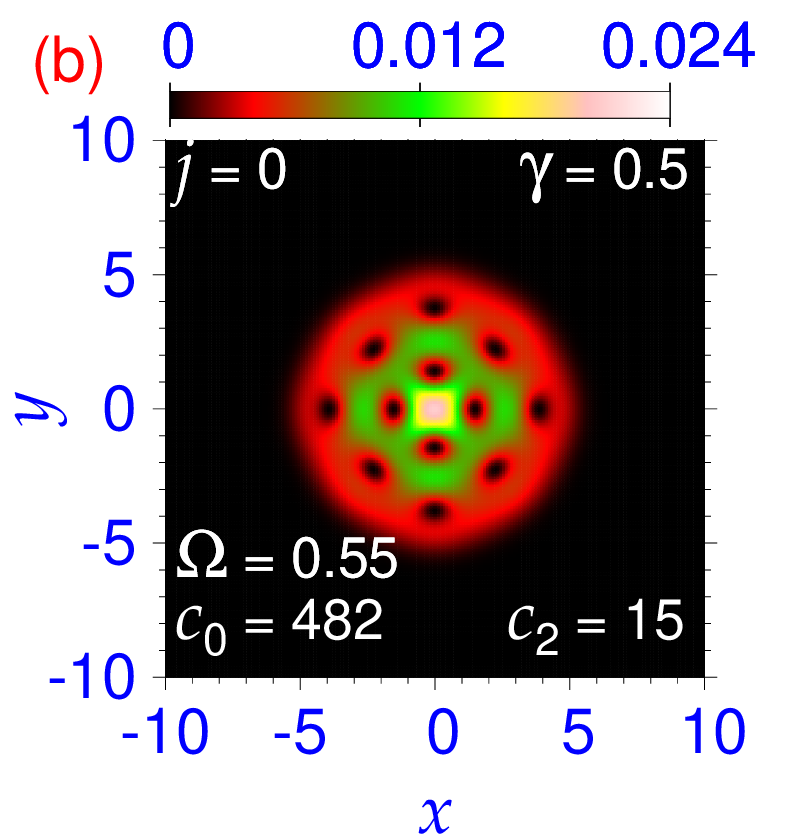}
 \includegraphics[trim = 6mm 0mm 11mm 0mm,clip,width=.32\linewidth]{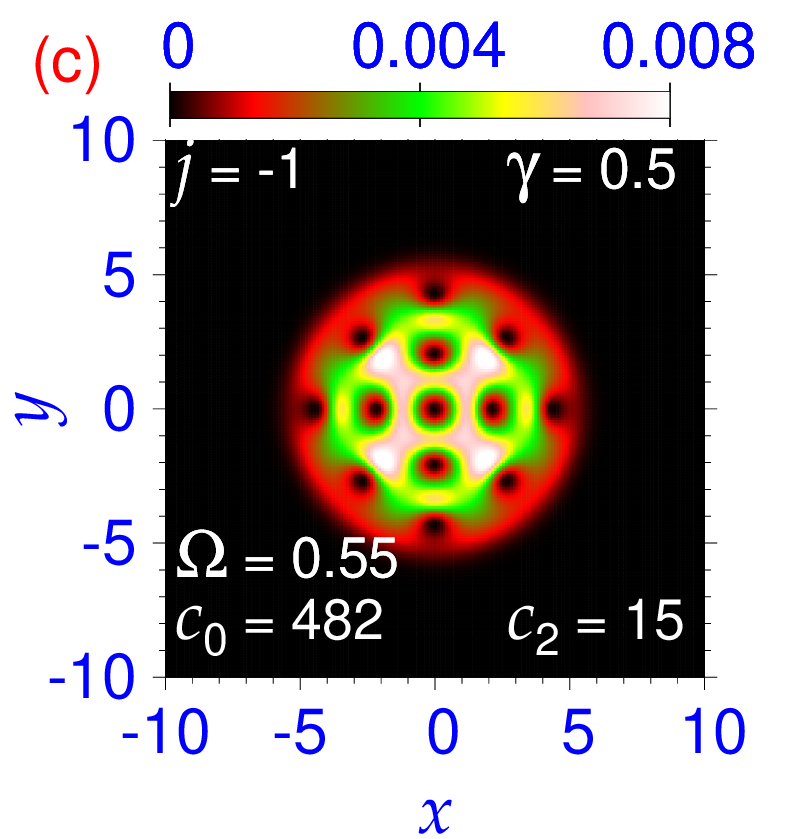}
\includegraphics[trim = 6mm 0mm 11mm 0mm,clip,width=.32\linewidth]{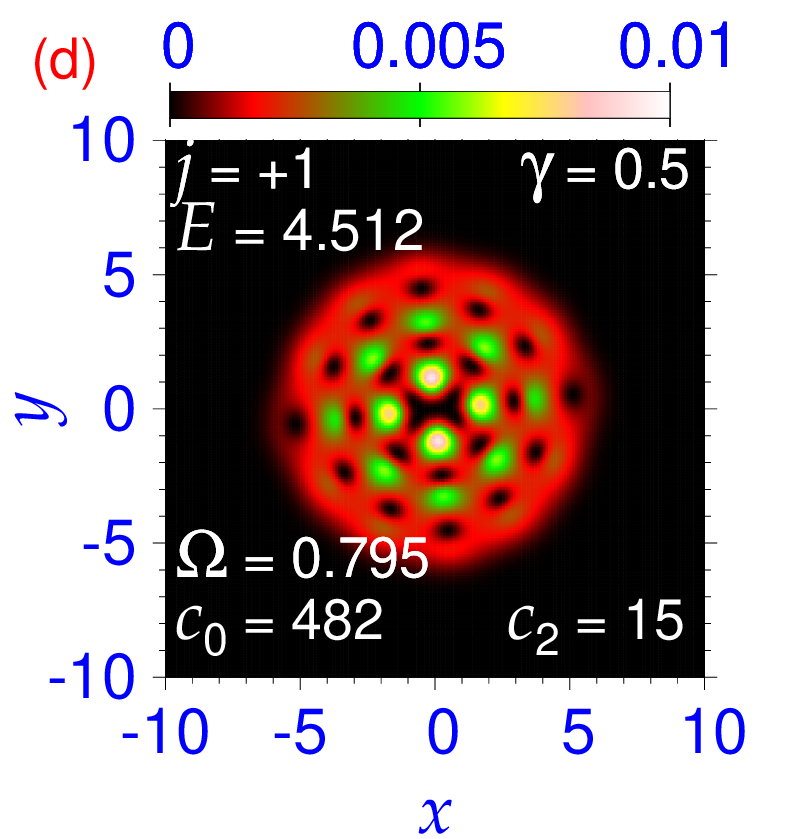}
\includegraphics[trim = 6mm 0mm 11mm 0mm,clip,width=.32\linewidth]{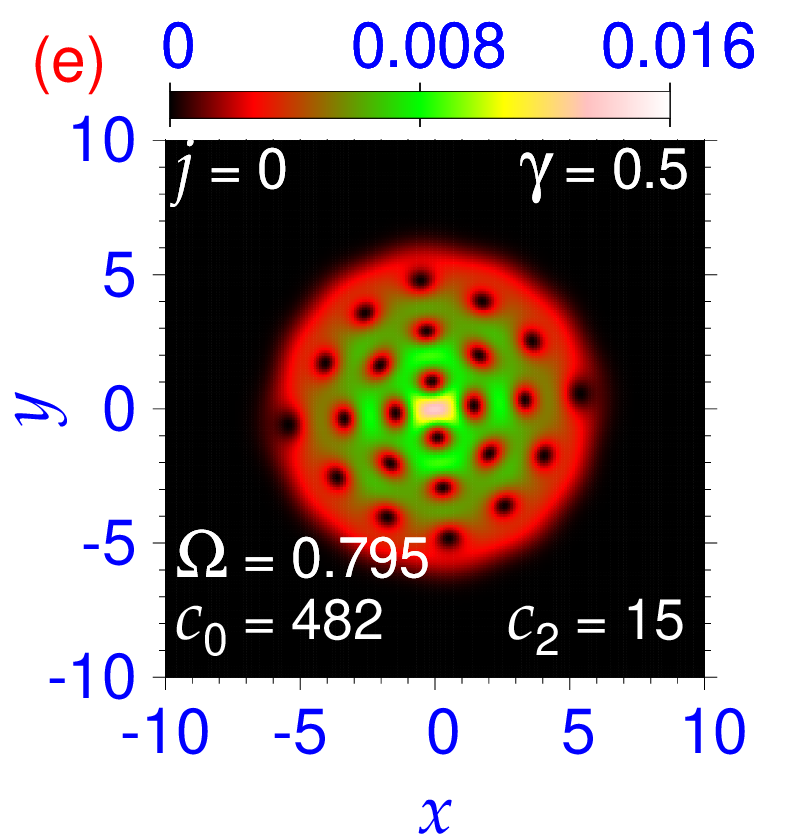}
 \includegraphics[trim = 6mm 0mm 11mm 0mm,clip,width=.32\linewidth]{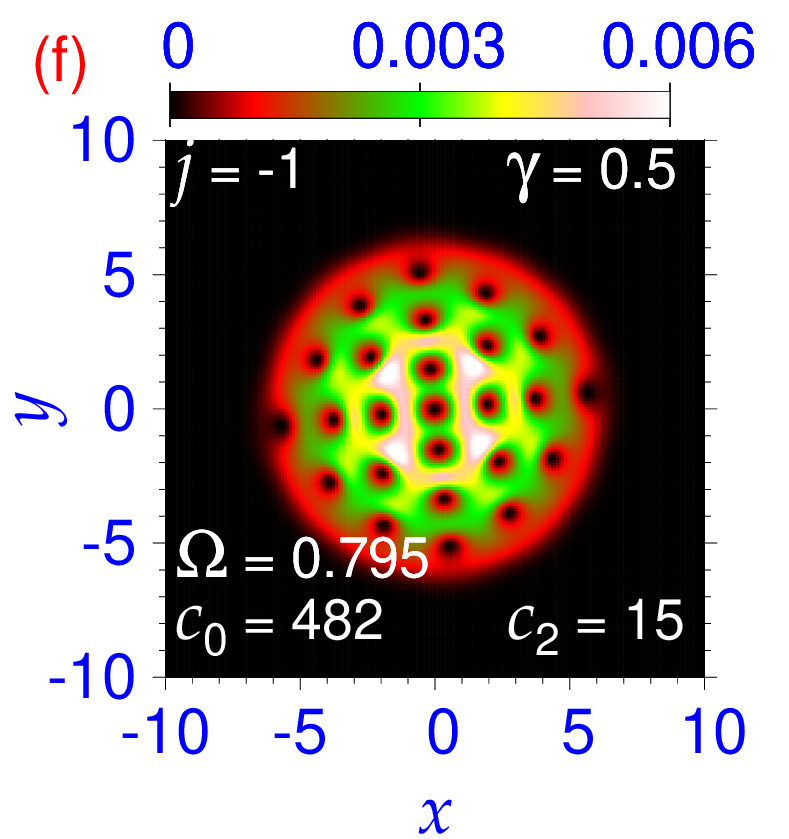}

\caption{(Color online) The same as in figure \ref{fig6} with approximate square symmetry with angular frequencies $\Omega = 0.55
$ and 0.795 in (a)-(c) and (d)-(f),  respectively.  The parameters of the anti-ferromagnetic BEC are the same as in figure \ref{fig6}.}
\label{fig8}
\end{figure}

Next we consider the formation of  vortex lattice with approximate square symmetry in a  Rashba SO-coupled  anti-ferromagnetic  quasi-2D spinor BEC upon rotation. The resultant vortex lattices, in this case, for angular frequencies $\Omega =0.55$ and 0.795
are displayed in figures \ref{fig8}(a)-(c) and (d)-(f), respectively.  The central region of the vortex lattice,  in this case, is different from  that in figure \ref{fig6}. The central region is formed by a superposition of the $(-1,0,+1)$-type state of the non-rotating BEC, viz. figures \ref{fig6}(a)-(c), with a $(+4,+4,+4)$-type state  formed by rotation, thus resulting in a state  with vortex of circulation +3 (+4, +5) in component $j=+1$ ($j=0, j=-1$).  The vortex of circulation +4 (+5) in component $j=0$ ($j=-1$)
breaks into 4 (5) vortices of unit circulation, whereas the $j=+1$ component contains a complex vortex structure of circulation +3, as can be seen from a phase plot of the wave function of the 
condensate displayed in  figures \ref{fig8}
(a)-(c), viz. figures \ref{fig7}(d)-(f).
In the outer region we have vortices arranged in concentric square orbits  with 
8 and 12 vortices, viz. figures \ref{fig8}(a)-(c) and (d)-(f).  Hence, although the vortex lattices with hexagonal 
symmetry in figures \ref{fig1} and \ref{fig6}   are quite similar, those with approximate square symmetry in  figures \ref{fig3} and \ref{fig8}   for ferromagnetic and anti-ferromagnetic spinor BECs are different in the central region
maintaining similarity in the outer region.

 \begin{figure}[!t]
\centering
\includegraphics[trim = 6mm 0mm 11mm 0mm,clip,width=.32\linewidth]{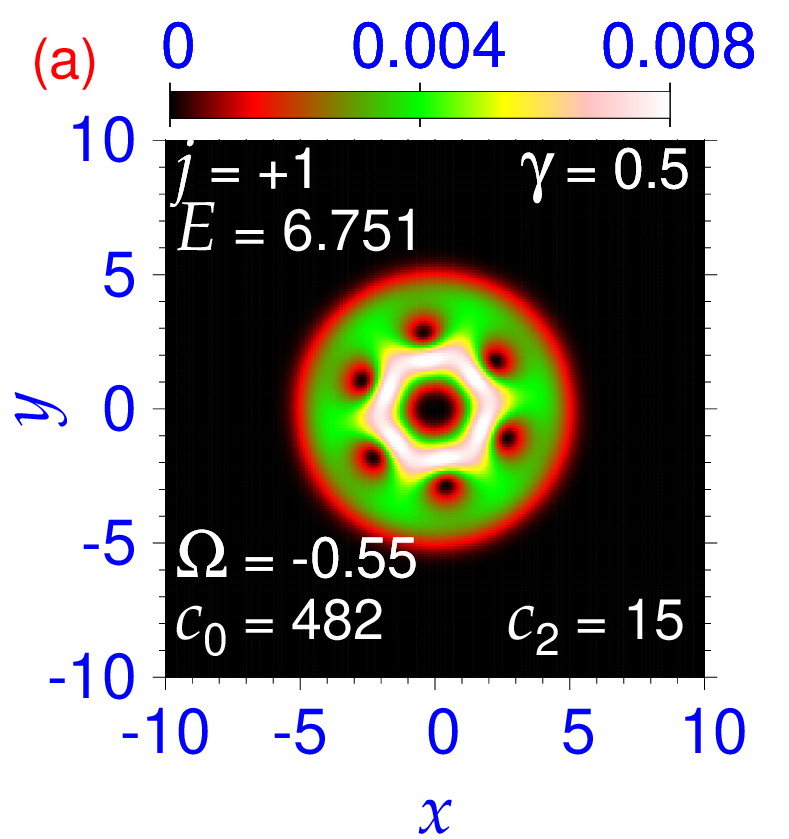}
\includegraphics[trim = 6mm 0mm 11mm 0mm,clip,width=.32\linewidth]{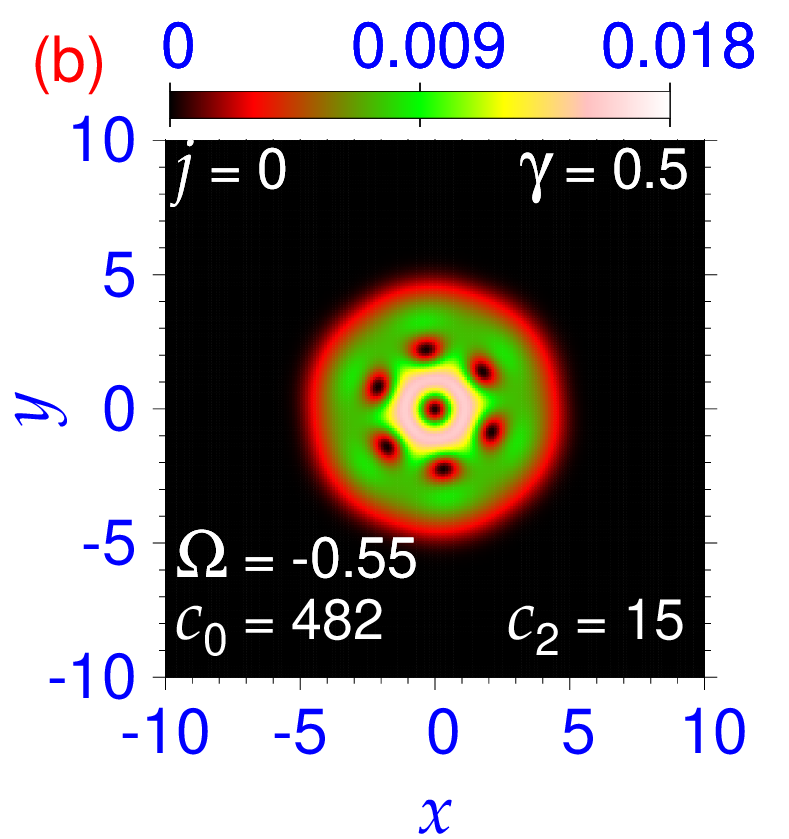}
 \includegraphics[trim = 6mm 0mm 11mm 0mm,clip,width=.32\linewidth]{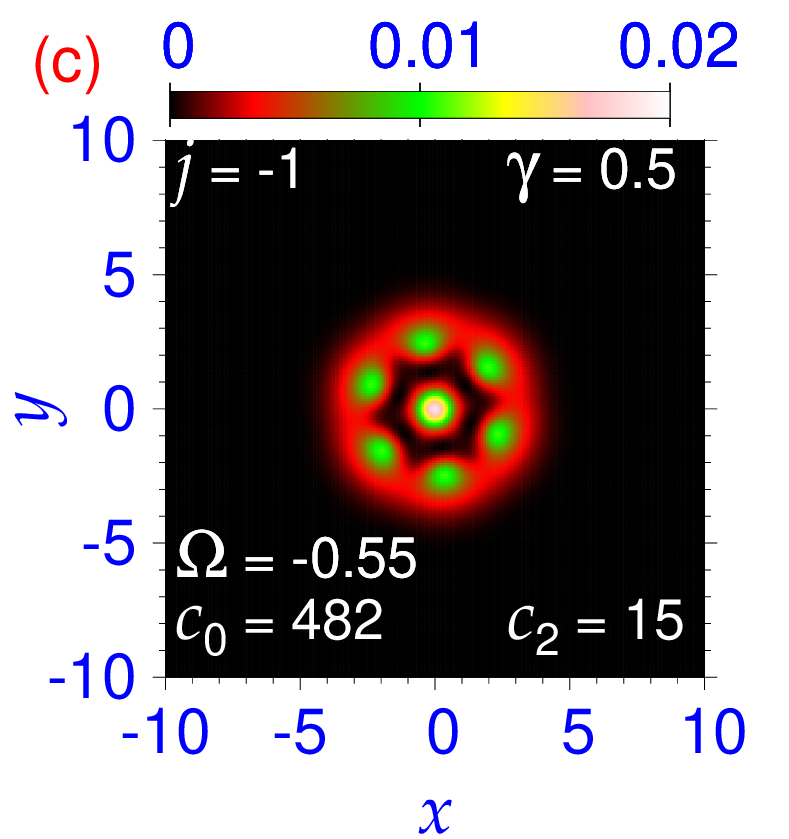}
\includegraphics[trim = 6mm 0mm 11mm 0mm,clip,width=.32\linewidth]{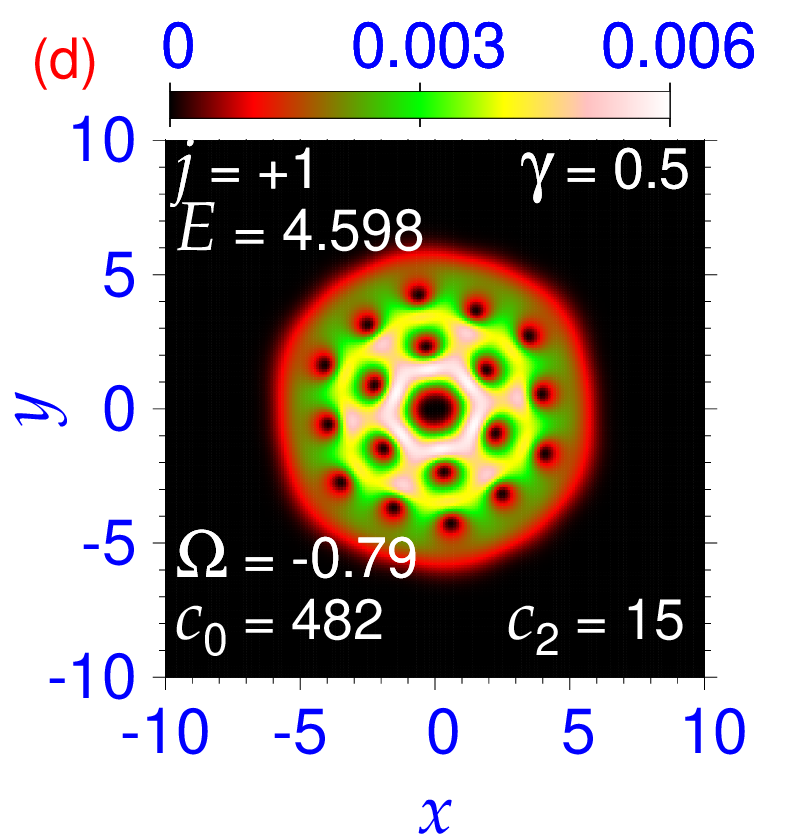}
\includegraphics[trim = 6mm 0mm 11mm 0mm,clip,width=.32\linewidth]{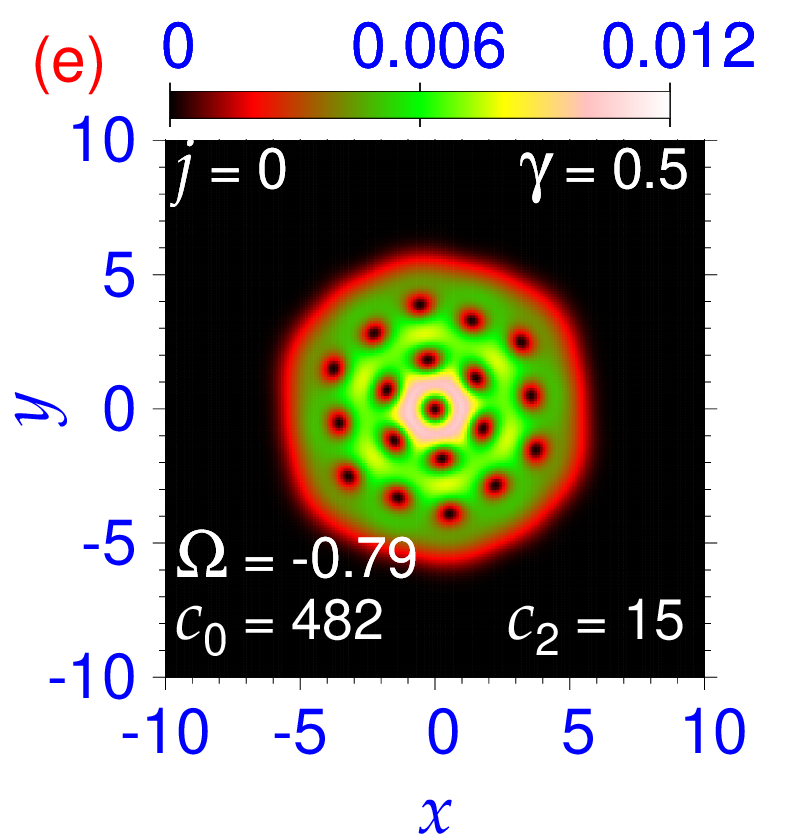}
 \includegraphics[trim = 6mm 0mm 11mm 0mm,clip,width=.32\linewidth]{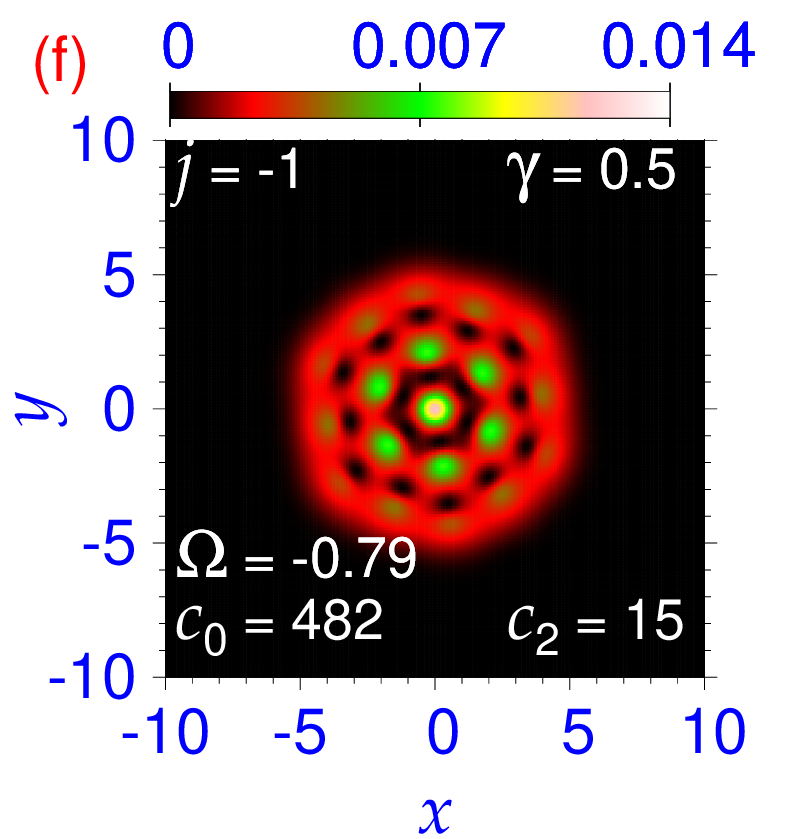}

\caption{(Color online)  The same as in figure \ref{fig6}  for angular frequencies $\Omega =-0.55,$ and $ -0.79,$  in  (a)-(c), and  (d)-(f),  respectively.   The angular 
momentum of rotation is anti-parallel to the vorticity direction of the vortex in component $j=-1$ in figure \ref{fig6}
(c). 
The parameters of the anti-ferromagnetic BEC are the same as in figure \ref{fig6}. }
\label{fig9}
\end{figure}

 \begin{figure}[!t]
\centering
\includegraphics[trim = 6mm 0mm 11mm 0mm,clip,width=.32\linewidth]{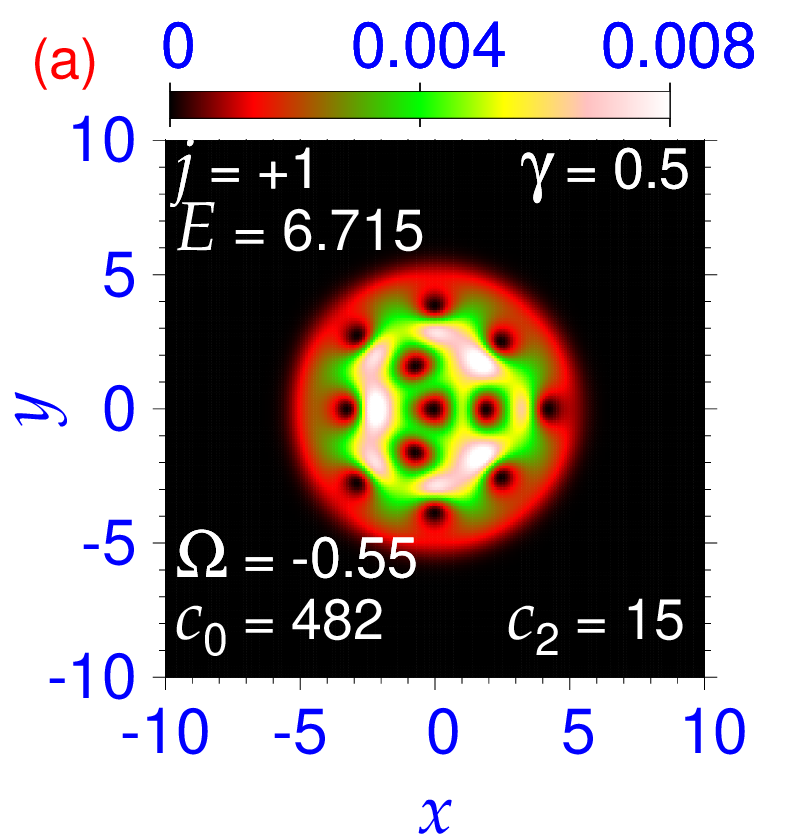}
\includegraphics[trim = 6mm 0mm 11mm 0mm,clip,width=.32\linewidth]{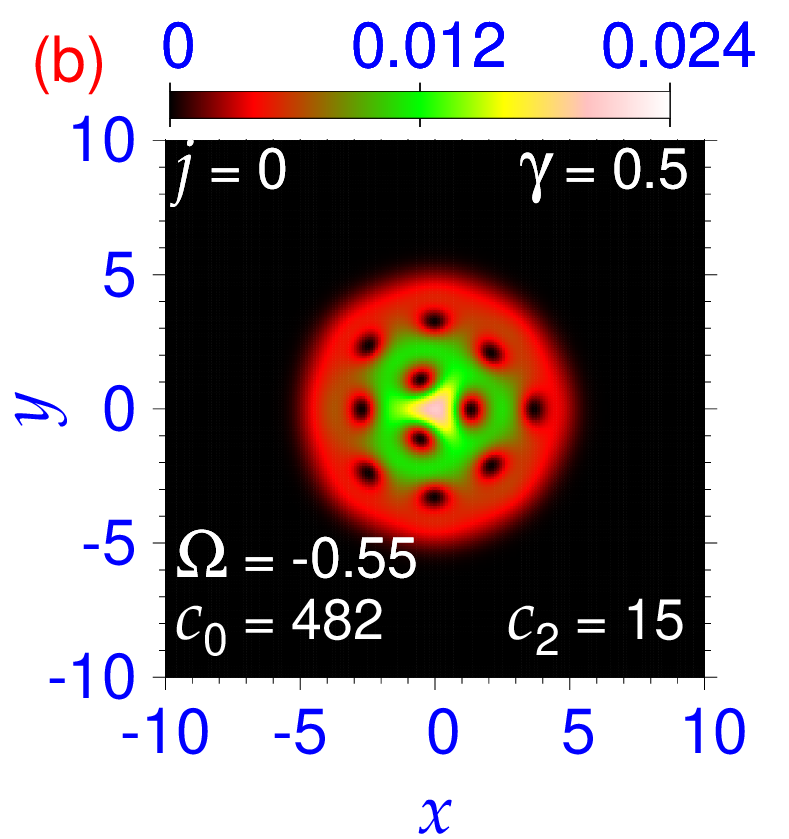}
 \includegraphics[trim = 6mm 0mm 11mm 0mm,clip,width=.32\linewidth]{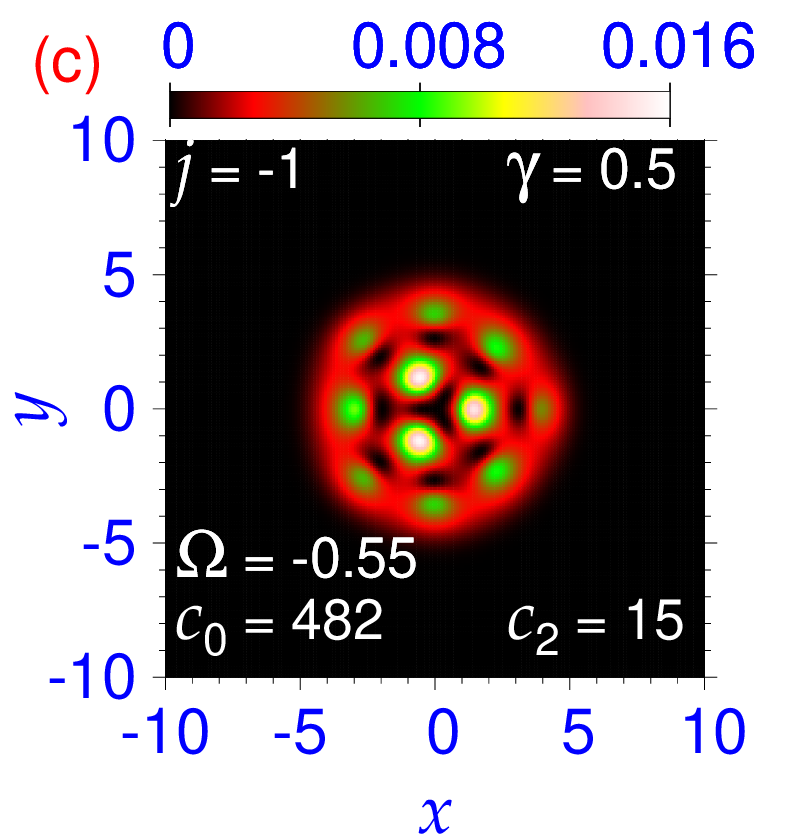}
\includegraphics[trim = 6mm 0mm 11mm 0mm,clip,width=.32\linewidth]{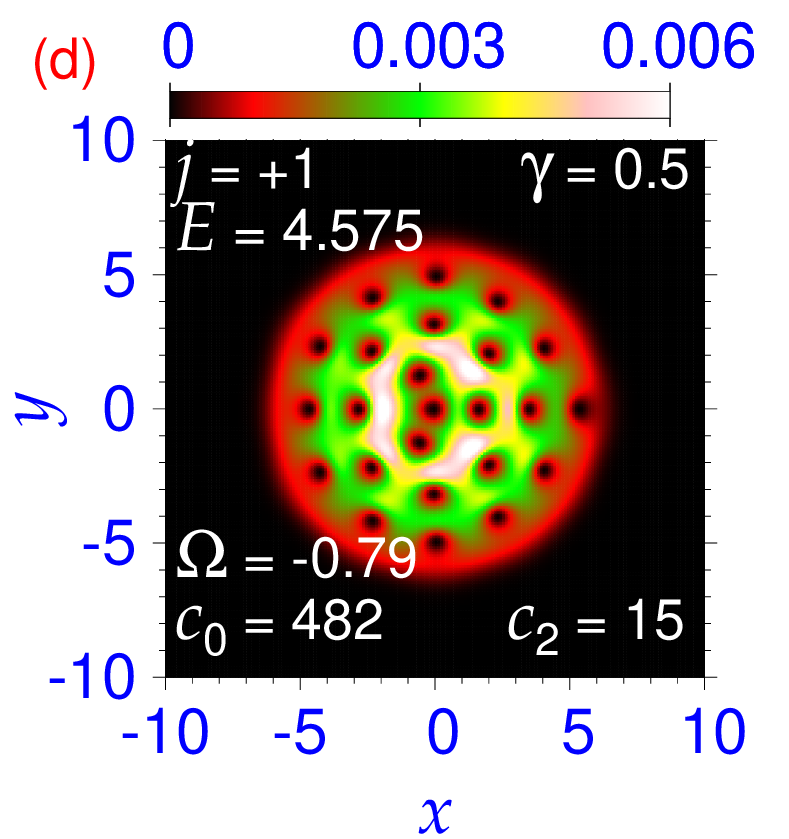}
\includegraphics[trim = 6mm 0mm 11mm 0mm,clip,width=.32\linewidth]{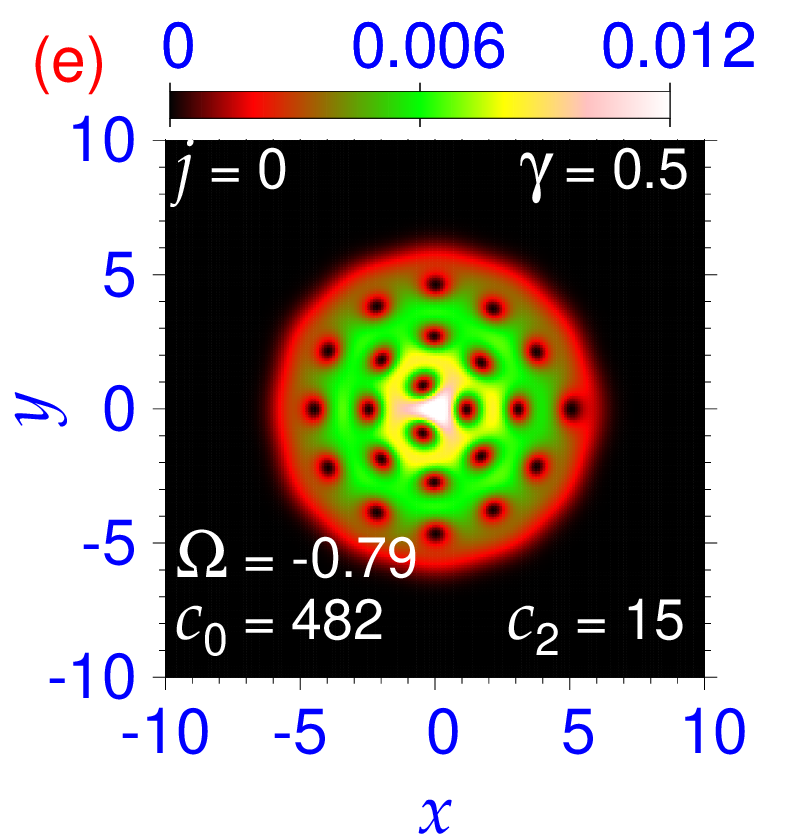}
 \includegraphics[trim = 6mm 0mm 11mm 0mm,clip,width=.32\linewidth]{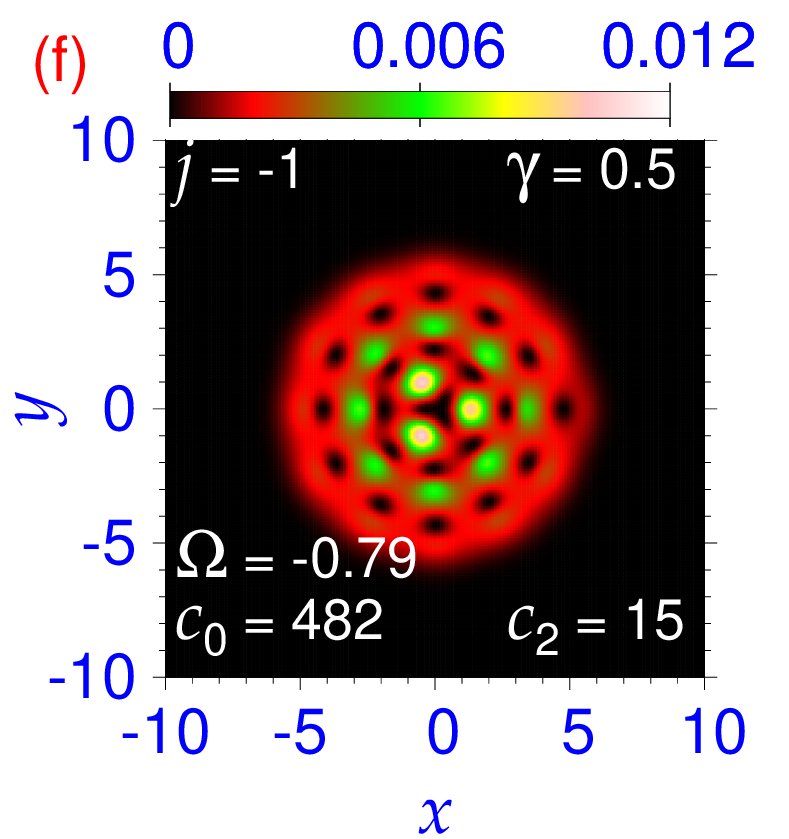}  

\caption{(Color online) The same as in figure \ref{fig6} with approximate square symmetry with angular frequencies $\Omega = -0.55,$  and $ -0.79,
$ in (a)-(c),  and (d)-(f) respectively.  The angular 
momentum of rotation is anti-parallel to the vorticity direction of the vortex in component $j=-1$ in figure \ref{fig6}
(c).  The parameters of the anti-ferromagnetic BEC are the same as in figure \ref{fig6}.}
\label{fig10}
\end{figure}

 Let us next consider the formation of anti-vortex lattice with hexagonal  symmetry in a rotating  Rashba SO-coupled anti-ferromagnetic spinor BEC  with the angular momentum of rotation opposite to the vorticity direction of the vortex in component $j=-1$ of  figure \ref{fig6}(c).  The non-rotating state in figures \ref{fig6}(a)-(c) is of the $(-1, 0, +1)$ type.  For small angular frequency of rotation, 
 an anti-vortex is generated in all three components in the form of a $(-1,-1,-1)$-type state, which when superposed on the 
$(-1, 0, +1)$-type state leads to a $(-2,-1,0)$-type state at the center.  The generated anti-vortex lattice in this case maintains this scenario in the central region, e.g., one (two) anti-vortex of  circulation  $-1$ in component $j=0$ ($j=+1$) and none in component $j=-1$, around which the hexagonal vortex lattice is formed. This is illustrated by a plot of contour density of  components $j=+1,0,-1$   for angular frequencies $\Omega=-0.55,$ and $ -0.79 $  in figures \ref{fig9}(a)-(c), and 
(d)-(f), respectively.  
The generated anti-vortex lattice states  in figure \ref{fig9} for the anti-ferromagnetic phase are  identical  to those figure \ref{fig4} for the ferromagnetic phase.

Finally, we consider  the formation of anti-vortex lattice with square  symmetry in a rotating SO-coupled
 anti-ferromagnetic spinor BEC  with the angular momentum of rotation opposite to the vorticity direction of the vortex in component $j=-1$, viz. figure \ref{fig6}(c). 
In this case, the generated anti-vortex lattice displayed in figure \ref{fig10}  for $\Omega =-0.55$ and $-0.79$ 
is quite similar to the anti-vortex lattice in the case of a ferromagnetic spinor BEC presented in figure \ref{fig5}.  The vortices for   $\Omega =-0.55$ can be identified from a phase plot of the wave function in
figures \ref{fig7}(g)-(i). 
The anti-vortex lattice with square  symmetry  for   $ \Omega =-0.55,$ and $-0.79$   is presented in   figures \ref{fig10}(a)-(c), and (d)-(f)  respectively. In table \ref{tab1}  we also display the energies of  anti-vortex-lattice states of
hexagonal and square symmetries
 of figures \ref{fig9} and \ref{fig10}.   The energies of the hexagonal anti-vortex-lattice {states are larger than} the corresponding states with square symmetry in all cases.




\section{Discussion}

{ We now compare the present results of vortex-lattice formation in a Rashba 
SO-coupled spin-1 BEC   with previous results \cite{yy,yy2,yy1} for vortex-lattice   formation in a Rashba 
SO-coupled pseudo spin-1/2 BEC. A Rashba SO-coupled pseudo spin-1/2 BEC supports  a half-quantum vortex state
with an unit vortex in one component and zero  vortex in the other, which is a $(+1,0)$-type state with intrinsic 
vorticity. This state should be compared with $(-1,0,+1)$- and $(0,+1,+2)$-type states in the present Rashba SO-coupled spin-1 BEC.
The vortex lattice for the pseudo spin-1/2 BEC has a vortex of unit circulation at  the center of one component, whereas the center of the  other component has no vortex in  analogy with vortices of circulation 0, $+1$, and $+2$ at the centers of the three components, viz. figures \ref{fig1}(d)-(f), in the ferromagnetic case.
 The presence of  the (+1,0)-type state with intrinsic vorticity in a pseudo spin-1/2 BEC
breaks the symmetry between rotation with vorticity along the $z$ and $-z$ axes and thus   might generate
different  vortex-lattice and anti-vortex-lattice states in a rotating SO-coupled quasi-2D pseudo spin-1/2 BEC 
for these two  types of rotation. However,
the previous studies  \cite{yy,yy1} did not  explore this possibility. The detailed numerical study \cite{yy2}  for
 vortex-lattice 
formation in the pseudo spin-1/2 case  confirmed  the formation of   lattice with hexagonal symmetry only. In this study on  the spin-1 BEC,   in addition, we also demonstrate   the formation of lattice with square symmetry.    }

 \begin{figure}[!t]
\centering
\includegraphics[width=.99\linewidth]{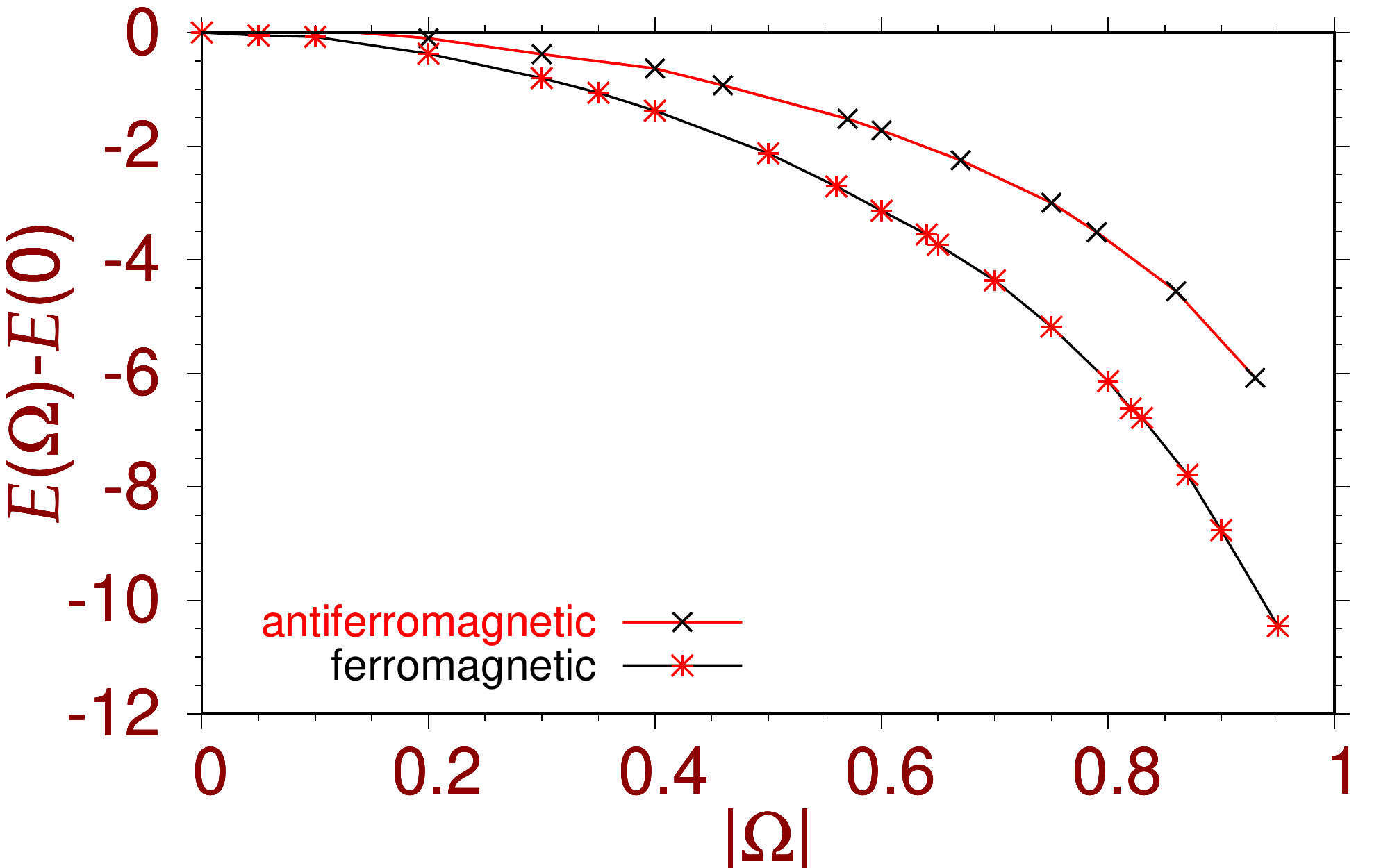}

\caption{(Color online) The rotational energy in  the rotating frame $[E(\Omega)-E(0)]$ versus angular frequency of rotation for the ferromagnetic and anti-ferromagnetic BEC. 
The points are numerically calculated whereas the lines are to guide the eye.}
\label{fig11}
\end{figure}

The rotational energy of a scalar BEC $[E(\Omega)-E(0)]$ in the rotating frame
is the energy of rigid-body rotation $\sim -I\Omega^2/2$ where $I$ is the moment 
of inertia of the condensate. This energy is  proportional to the square of the angular 
frequency \cite{fetter}, where $E(\Omega)$ is given by  (\ref{energy}). In the case of 
spin-1 spinor ferromagnetic and anti-ferromagnetic BECs, a similar relation also holds.  
We illustrate  in  figure \ref{fig11} the rotational energy of the vortex- and anti-vortex-lattice
states of both  
ferromagnetic and anti-ferromagnetic BECs with square and hexagonal symmetry 
as a function of angular frequency $\Omega$, where we plot 
the energy of the minimum-energy state versus $|\Omega|$. 
The energies of ferromagnetic and anti-ferromagnetic BECs lie on two distinct 
lines showing similar qualitative behavior.
The energy decreases with increasing angular frequency of rotation as the contribution of
the rotational energy $-\Omega L_z$
in the expression for energy (\ref{energy})
is negative for large $|\Omega|$. For small $\Omega$, in the perturbative limit, this
contribution is linearly proportional to $\Omega$. Hence as $\Omega \to 0$ ($\Omega \lessapprox 0.1$), the rotational energy
for positive (negative) values of $\Omega$ is positive (negative).  For $\Omega \gtrapprox 0.1$  rotational  energies 
for both positive and negative  $\Omega$ are negative. For $\Omega  \gtrapprox 0.2$
 these two energies are practically equal and negative. The difference between the energies for $\pm \Omega$ is small and for clarity of the plot, this detail is not displayed in figure   \ref{fig11} and an average of the two energies are exhibited for small $|\Omega|$.
But as  $|\Omega|$ increases,    
  the rotational energy behaves as $\sim -\Omega^2$ \cite{fetter}.

\section{Summary}
\label{IV}

We studied the formation of  vortex lattice in a  quasi-2D Rashba SO-coupled spin-1 spinor BEC in the $x-y$ plane,  
under rapid rotation, using the mean-field GP equation in the rotating frame, where the generated 
vortex-lattice state is a stationary state. In the case of a scalar BEC, the generated vortex lattice for rotation 
with vorticity along $z$ and $-z$ axes  are the same. 
The { lowest-energy circularly-symmetric} state of a non-rotating ferromagnetic Rashba SO-coupled spin-1 spinor BEC
is  of the $(0,+1,+2)$ type, whereas the same for an anti-ferromagnetic BEC is of the type $(-1,0,+1)$.  The intrinsic vorticity 
of these two states make the  rotation 
with vorticity along $z$ and $-z$ axes conceptually different for an  SO-coupled spin-1 spinor BEC. Consequently,
 different from a scalar BEC, the generated vortex-lattice structure for a quasi-2D Rashba SO-coupled  spin-1 spinor BEC 
 for rotation 
with vorticity along $z$ and $-z$ axes are different. For rotation with vorticity along $z$ direction, a vortex lattice is formed and for rotation with vorticity along $-z$ direction an anti-vortex lattice is formed. Two types of vortex and anti-vortex 
lattices were found to be formed predominantly:
a hexagonal lattice and an approximate square lattice.  For rotation with vorticity along $z$ direction,
the hexagonal lattice has vortices arranged in closed concentric orbits which accommodate the following maximum number  of vortices: 6, 12, 18  etc., whereas the square lattice  has vortices arranged in closed concentric orbits with the maximum numbers 8, 12, 16 etc. 
We illustrated, for different angular frequencies, the formation of vortex lattices with closed concentric orbits of vortices 
while all orbits accommodate the allowed maximum number  of vortices.  
The central region in both cases is occupied by a complex structure  of vortices,  often accommodating vortices of circulation (angular momentum) greater than unity. In case of a scalar BEC, all vortices in a vortex lattice are of unit circulation.   For rotation with vorticity along $-z$ axis,  similar lattice structure emerges 
but with anti-vortices replacing vortices, although the central region of the lattice may have a different  distribution
of vortices  from the case of    rotation with vorticity along $z$ axis. Such a lattice structure is termed an anti-vortex lattice as opposed to a vortex lattice. If, for a fixed angular frequency of rotation, both a  square and a hexagonal vortex or anti-vortex lattice  can
 be formed, with closed concentric orbits,  the square vortex-lattice structure is found to possess the smaller energy as shown in table \ref{tab1}. This strongly suggests  the square lattice states to be the { lowest-energy} state.
For larger values of $\gamma$ (not considered in this paper), we could not find the hexagonal lattice states; only square lattice states were found. 
The rotational energy of the generated vortex or anti-vortex lattice for both ferromagnetic and anti-ferromagnetic BECs is found to be proportional to the square of angular frequency $\Omega$ as displayed in figure \ref{fig11},  consistent with a theoretical suggestion by Fetter for a scalar BEC
 \cite{fetter}.


\section*{Acknowledgments}
\noindent

S.K.A. acknowledges support by the CNPq (Brazil) grant 301324/2019-0, and by the ICTP-SAIFR 
FAPESP (Brazil) grant 2016/01343-7.

\end{document}